\documentclass[12pt]{article}
\usepackage{float}
\usepackage{amsmath}
\usepackage{amssymb}
\usepackage{graphicx}
\usepackage{setspace}
\usepackage[authoryear]{natbib}
\usepackage{appendix}
\usepackage{grffile}
\usepackage{rotating}
\usepackage{tikz}
\usepackage{multirow}
\usepackage{hyperref}

\hypersetup{
    colorlinks=true,
    linkcolor=blue,
    citecolor=blue,
    filecolor=magenta,      
    urlcolor=cyan,
}

\newtheorem{example}{Example}

\newtheorem{theorem}{Theorem}
\newtheorem{lemma}{Lemma}
\newtheorem{definition}{Definition}
\newtheorem{assumption}{Assumption}
\allowdisplaybreaks
\input{tcilatex}
\begin{document}

\title{Time series experiments and causal estimands: \\
exact randomization tests and trading\thanks{%
We thank Edo Airoldi, Joshua Angrist, Guillaume Basse, Stephen Blyth, Peng
Ding, Pierre Jacob, Guido Kuersteiner, Anthony Ledford, Daniel Lewis,
Fabrizia Mealli, Xiao-Li Meng, Luke Miratrix, Susan Murphy, David Parkes,
Mikkel Plagborg-Moller, James M. Robins, Donald B. Rubin, Jim Stock and
Panos Toulis for various suggestions, and AHL Partners LLP (London, UK) for
giving us the financial data we use. \ }}
\author{Iavor Bojinov \\
\textit{Department of Statistics,}\\
\textit{Harvard University}\\
\texttt{bojinov@fas.harvard.edu} \and Neil Shephard \\
\textit{Department of Economics and }\\
\textit{Department of Statistics,}\\
\textit{Harvard University}\\
\texttt{shephard@fas.harvard.edu}}
\maketitle

\begin{abstract}
We define causal estimands for experiments on single time series, extending
the potential outcome framework to dealing with temporal data. \ Our
approach allows the estimation of a broad class of these estimands and exact
randomization based $p$-values for testing causal effects, without imposing
stringent assumptions. \ We further derive a general central limit theorem
that can be used to conduct conservative tests and build confidence
intervals for causal effects. \ Finally, we provide three methods for
generalizing our approach to multiple units that are receiving the same
class of treatment, over time. \ We test our methodology on simulated
\textquotedblleft potential autoregressions,\textquotedblright which have a
causal interpretation. \ Our methodology is partially inspired by data from
a large number of experiments carried out by a financial company who
compared the impact of two different ways of trading equity futures
contracts. \ We use our methodology to make causal statements about their
trading methods.

\end{abstract}

\baselineskip=20pt

\graphicspath{ {./figures/} }

\section{Introduction}

In longitudinal experiments with time-varying exposure causal estimands are
traditionally defined as averages over a population. \ Population averaging
usually requires that the number of units in the experiment exceed the
duration of the study. \ However, there are applications where the length of
the experiment is greater than the number of units. \ The most extreme case
is when there is only one unit that is being experimented on over time. \ We
refer to this as a ``time series experiment.'' \ For time series
experiments, the usual population estimands fail to capture the personalized
nature of the problem and are virtually impossible to estimate without
imposing strong, often unrealistic, assumptions. \ 

In this paper, we generalize the standard one-period experimental treatments
on multiple units setup to a multiple-period experimental treatment path
carried out on a single unit. \ Our time series experiments have at their
heart potential outcome paths, allowing us to define unit level causal
estimands. \ We define a broad class of causal effects and show how to
estimate several key examples under a relatively weak non-anticipating
treatments assignment assumption. \ For these causal effects, we derive two
non-parametric inferential strategies based on the randomization alone. \
One of these strategies delivers an exact randomization test of no causality
in time series. \ We then generalize our results to the setting with
multiple units. \ 

Our methods are partially inspired by our analysis of a database of
experiments carried out by AHL Partners LLP (London, UK), a large
quantitative hedge fund, who have been executing orders through both human
traders and computer algorithms. \ To quantify the relative performance of
these two methods they have been running experiments, randomly allocating
jobs to the human and the computer. \ We will use our causal methods to make
inference on the causal effect of these treatments on the relative costs of
trading. \ The data covers a year of experiments on 10 futures markets on
equity indexes. \ 

Our approach embraces the potential outcomes phrasing of experiments and
causal inference, which has its origins in \cite{Neyman(23)}, \cite%
{Kempthorne(55)}, \cite{Cox(58book)}, \cite{rubin(74)} and \cite%
{robins1986new}. \ In Section \ref{S:otherwork} we will be precise about how
our work relates to other studies of using temporal data to learn about
causal effects. \ In particular, we will discuss the work of, for example,
Robins, et al, Murphy et al and Angrist, Kuersteiner et al as well as the
many papers which have been published on impulse response functions, Granger
causality, highly structured time series models and \textquotedblleft
natural\textquotedblright\ experiments. \ \ 

This paper's structure is as follows. \ In Section \ref{sect:treatment path}%
, we define the potential outcome paths, the outcome path and the
non-anticipating treatment path. \ In Section \ref{sect:causal effects}, we
define what we mean by causal effects and estimands. \ In Section \ref%
{sect:estimation}, we discuss experiments and using the results to estimate
causal effects. In Section \ref{sec:inference}, we propose how to conduct
inference on the causal effects using just the randomization in the
treatment. \ These methods are exact for any time series being treated so
long as the treatment regime is non-anticipating. \ \ In Section \ref%
{S:otherwork}, we compare our causal effects and inference approaches with
those familiar in the literature. In Section \ref{Muliple units}, we extend
our time series results on a single unit to the case of multiple units each
recorded through time. \ In Section \ref{sect:simulation study}, we conduct
some simulation experiments to see the effectiveness of our procedures
focusing on what we call a potential autoregression. \ In Section \ref%
{sect:empirical example}, we give our empirical illustration, measuring the
causal effect of trading. \ In Section \ref{sect:conclusion}, we give our
concluding remarks. \ The Appendix gathers some of the more technical proofs
for the theorems and lemmas discussed in the main body of the text. The Web
Appendix contains a practical description of how to conduct a randomization
test, some more general theorems and extra figures. \ 


\section{The treatment path and potential outcome paths\label{sect:treatment
path}}


At each time step, $t=1,\dots ,T$, we will expose a single unit to either
treatment, $W_{t}=1$, or control, $W_{t}=0$, and subsequently measure an
outcome. \ We focus on binary treatments; however, our results generalize to
multiple treatments. \ The random \textquotedblleft treatment
path\textquotedblright\ is 
\begin{equation*}
W_{1:t}=(W_{1},\dots ,W_{t}).
\end{equation*}%
We denote a realization of $W_{1:t}$ by $w_{1:t}$. \ 
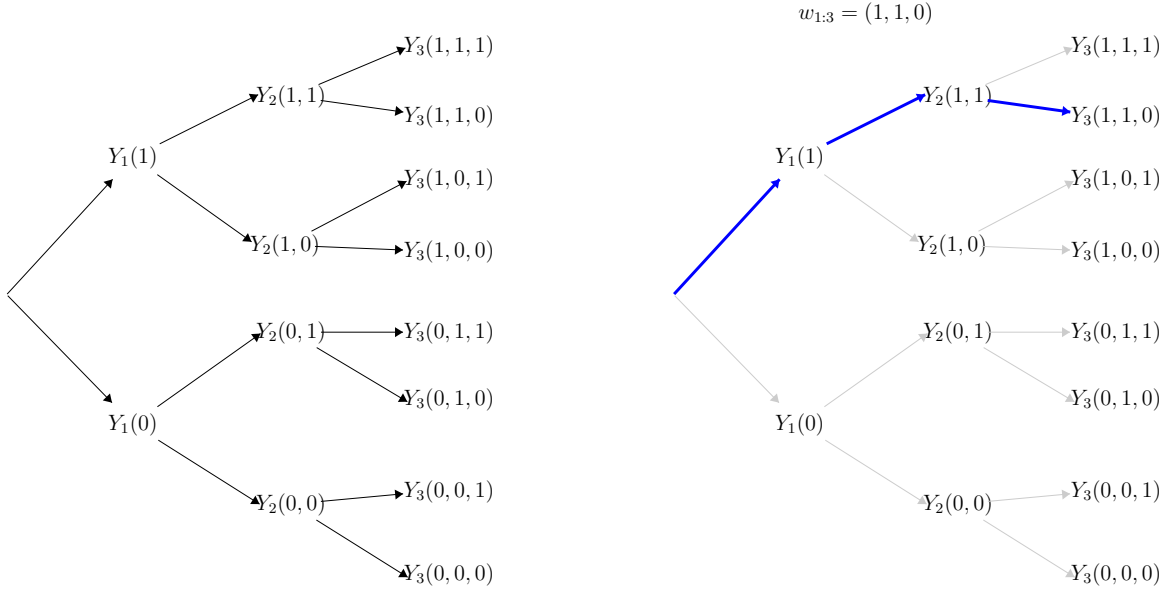
\begin{figure}[t]
\label{F:Potential_Outcomes}
\par
\begin{center}
\scalebox{0.7}{\begin{tikzpicture}[scale=0.19]
    \tikzstyle{every node}+=[inner sep=0pt]
      \draw (23.4,-14.4) node {$Y_1(1)$};
      \draw (23.4,-41.1) node {$Y_1(0)$};
      \draw (39.2,-8.3) node {$Y_2(1,1)$};
      \draw (38.6,-23.2) node {$Y_2(1,0)$};
      \draw (39.2,-31.9) node {$Y_2(0,1)$};
      \draw (39.2,-49.1) node {$Y_2(0,0)$};
      \draw (55,-3.3) node {$Y_3(1,1,1)$};
      \draw (55,-10.3) node {$Y_3(1,1,0)$};
      \draw (55,-16.6) node {$Y_3(1,0,1)$};
      \draw (55,-23.8) node {$Y_3(1,0,0)$};
      \draw (55,-31.9) node {$Y_3(0,1,1)$};
      \draw (55,-38.6) node {$Y_3(0,1,0)$};
      \draw (55,-56.1) node {$Y_3(0,0,0)$};
      \draw (55,-47.8) node {$Y_3(0,0,1)$};
      \draw [black] (10.83,-28.09) -- (21.37,-16.61);
      \fill [black] (21.37,-16.61) -- (20.46,-16.86) -- (21.2,-17.54);
      \draw [black] (41.77,-33.44) -- (50.83,-38.86);
      \fill [black] (50.83,-38.86) -- (50.4,-38.02) -- (49.88,-38.88);
      \draw [black] (42.19,-48.83) -- (50.41,-48.07);
      \fill [black] (50.41,-48.07) -- (49.57,-47.65) -- (49.66,-48.64);
      \draw [black] (41.74,-50.69) -- (50.86,-56.41);
      \fill [black] (50.86,-56.41) -- (50.45,-55.56) -- (49.91,-56.41);
      \draw [black] (41.96,-7.13) -- (50.64,-3.47);
      \fill [black] (50.64,-3.47) -- (49.71,-3.32) -- (50.09,-4.24);
      \draw [black] (42.17,-8.72) -- (50.43,-9.88);
      \fill [black] (50.43,-9.88) -- (49.71,-9.27) -- (49.57,-10.27);
      \draw [black] (41.25,-21.8) -- (50.75,-16.8);
      \fill [black] (50.75,-16.8) -- (49.81,-16.73) -- (50.27,-17.61);
      \draw [black] (41.6,-23.32) -- (50.4,-23.68);
      \fill [black] (50.4,-23.68) -- (49.62,-23.15) -- (49.58,-24.15);
      \draw [black] (26.08,-13.06) -- (35.92,-8.14);
      \fill [black] (35.92,-8.14) -- (34.98,-8.05) -- (35.42,-8.95);
      \draw [black] (25.84,-16.14) -- (35.26,-22.86);
      \fill [black] (35.26,-22.86) -- (34.9,-21.99) -- (34.32,-22.8);
      \draw [black] (25.85,-39.36) -- (36.15,-32.04);
      \fill [black] (36.15,-32.04) -- (35.21,-32.09) -- (35.79,-32.91);
      \draw [black] (25.94,-42.7) -- (36.06,-49.1);
      \fill [black] (36.06,-49.1) -- (35.65,-48.25) -- (35.12,-49.09);
      \draw [black] (10.89,-28.25) -- (21.31,-38.95);
      \fill [black] (21.31,-38.95) -- (21.11,-38.03) -- (20.39,-38.73);
      \draw [black] (42.2,-31.9) -- (50.4,-31.9);
      \fill [black] (50.4,-31.9) -- (49.6,-31.4) -- (49.6,-32.4);
    \end{tikzpicture} } \hspace{2cm} 
\scalebox{0.7}{\begin{tikzpicture}[scale=0.19]
        \tikzstyle{every node}+=[inner sep=0pt]
        \draw (30, 0) node {$w_{1:3} = (1,1,0)$};
        \draw (23.4,-14.4) node {$Y_1(1)$};
        \draw (23.4,-41.1) node {$Y_1(0)$};
        \draw (39.2,-8.3) node {$Y_2(1,1)$};
        \draw (38.6,-23.2) node {$Y_2(1,0)$};
        \draw (39.2,-31.9) node {$Y_2(0,1)$};
        \draw (39.2,-49.1) node {$Y_2(0,0)$};
        \draw (55,-3.3) node {$Y_3(1,1,1)$};
        \draw (55,-10.3) node {$Y_3(1,1,0)$};
        \draw (55,-16.6) node {$Y_3(1,0,1)$};
        \draw (55,-23.8) node {$Y_3(1,0,0)$};
        \draw (55,-31.9) node {$Y_3(0,1,1)$};
        \draw (55,-38.6) node {$Y_3(0,1,0)$};
        \draw (55,-56.1) node {$Y_3(0,0,0)$};
        \draw (55,-47.8) node {$Y_3(0,0,1)$};
        \draw [ultra thick,blue] (10.83,-28.09) -- (21.37,-16.61);
        \fill [ultra thick,blue] (21.37,-16.61) -- (20.46,-16.86) -- (21.2,-17.54);
        \draw [gray!40] (41.77,-33.44) -- (50.83,-38.86);
        \fill [gray!40] (50.83,-38.86) -- (50.4,-38.02) -- (49.88,-38.88);
        \draw [gray!40] (42.19,-48.83) -- (50.41,-48.07);
        \fill [gray!40] (50.41,-48.07) -- (49.57,-47.65) -- (49.66,-48.64);
        \draw [gray!40] (41.74,-50.69) -- (50.86,-56.41);
        \fill [gray!40] (50.86,-56.41) -- (50.45,-55.56) -- (49.91,-56.41);
        \draw [gray!40] (41.96,-7.13) -- (50.64,-3.47);
        \fill [gray!40] (50.64,-3.47) -- (49.71,-3.32) -- (50.09,-4.24);
        \draw [ultra thick,blue] (42.17,-8.72) -- (50.43,-9.88);
        \fill [ultra thick,blue] (50.43,-9.88) -- (49.71,-9.27) -- (49.57,-10.27);
        \draw [gray!40] (41.25,-21.8) -- (50.75,-16.8);
        \fill [gray!40] (50.75,-16.8) -- (49.81,-16.73) -- (50.27,-17.61);
        \draw [gray!40] (41.6,-23.32) -- (50.4,-23.68);
        \fill [gray!40] (50.4,-23.68) -- (49.62,-23.15) -- (49.58,-24.15);
        \draw [ultra thick,blue] (26.08,-13.06) -- (35.92,-8.14);
        \fill [ultra thick,blue] (35.92,-8.14) -- (34.98,-8.05) -- (35.42,-8.95);
        \draw [gray!40] (25.84,-16.14) -- (35.26,-22.86);
        \fill [gray!40] (35.26,-22.86) -- (34.9,-21.99) -- (34.32,-22.8);
        \draw [gray!40] (25.85,-39.36) -- (36.15,-32.04);
        \fill [gray!40] (36.15,-32.04) -- (35.21,-32.09) -- (35.79,-32.91);
        \draw [gray!40] (25.94,-42.7) -- (36.06,-49.1);
        \fill [gray!40] (36.06,-49.1) -- (35.65,-48.25) -- (35.12,-49.09);
        \draw [gray!40] (10.89,-28.25) -- (21.31,-38.95);
        \fill [gray!40] (21.31,-38.95) -- (21.11,-38.03) -- (20.39,-38.73);
        \draw [gray!40] (42.2,-31.9) -- (50.4,-31.9);
        \fill [gray!40] (50.4,-31.9) -- (49.6,-31.4) -- (49.6,-32.4);
      \end{tikzpicture}}
\end{center}
\caption{The left figure shows all the potential outcome paths for $T=3$.
The right figure shows the observed outcome path $Y_{1:3}(w_{1:3})$ for $%
w_{1:3}=(1,1,0)$, indicated by the thick blue line. The gray arrows indicate
the missing data. }
\label{fig: potential outcome path}
\end{figure}

In classical causal inference the treatment path is of length $1$, and each
unit only has $2^{1}$ potential outcomes (the outcome that would be observed
if the unit receives control and the outcome that would be observed if the
unit receives treatment). \ In a time series experiment, we follow a unit
over time and administer $T$ different treatments. \ At each time point $t$
the treatment path is of length $t$ and therefore there must be $%
2+2^{2}+...+2^{t}=2(2^{t}-1)$ potential outcomes up to time $t$ representing
the $2^{t}$ different treatment paths that could be traversed.

\begin{example}
\label{Example T=3}Suppose $T=3$. When $t=1$ there are $2^{1}=2$ potential
outcomes $Y_{1}(0),Y_{1}(1)$, as the unit can either receive treatment $0$
or $1$. At time $t=2$, there are $2^{2}=4$ potential outcomes $%
Y_{2}(0,0),Y_{2}(0,1),Y_{2}(1,0),Y_{2}(1,1)$, as the unit can receive $%
\{0,0\}$, $\{0,1\}$, $\{1,0\}$ or $\{1,1\}$. \ When $t=3$ there are $2^{3}=8$
potential outcomes, $Y_{3}(0,0,0)$, $Y_{3}(0,1,0)$, $Y_{3}(0,0,1)$, $%
Y_{3}(0,1,1)$, $Y_{3}(1,0,0)$, $Y_{3}(1,0,1)$, $Y_{3}(1,1,0)$, $Y_{3}(1,1,1)$%
, corresponding to all the possible values the treatment path could have
taken. \ The left hand side of Figure \ref{fig: potential outcome path}
depicts this. \ 
\end{example}

The set of $2^{t}$ potential outcomes at time $t$, denoted by $Y_{t}(\bullet
)$, are defined by 
\begin{equation*}
Y_{t}(\bullet )=\{Y_{t}(w_{1:t}):w_{1:t}\in \{0,1\}^{t}\},
\end{equation*}%
so the potential outcomes at time $t$ only functionally depend upon current
and past treatments. \ This stops current potential outcomes depending upon
future treatments. \ 

The collection of all potential paths up to time $t$ is written as 
\begin{equation*}
Y_{1:t}(\bullet )=\left\{ Y_{1}(\bullet ),Y_{2}(\bullet ),...,Y_{t}(\bullet
)\right\} ,
\end{equation*}%
containing $2(2^{t}-1)$ random variables. The potential path for the
treatment path $w_{1:t}$ is 
\begin{equation*}
Y_{1:t}(w_{1:t})=\left\{
Y_{1}(w_{1}),Y_{2}(w_{1:2}),...,Y_{t}(w_{1:t})\right\} .
\end{equation*}

Formulating treatments and potential outcomes as paths was introduced into
longitudinal analysis by \cite{robins1986new}. \ For binary time series it
was used by \cite{RobinsGreenlandHu(99)} in their Section 7, for state space
models by \cite{BondersenGallusserKoehlerRemyScott(15)} and generally by 
\cite{BlackwellGlynn(16)}. \ 

We do not make any assumptions on the dimension of $Y_{t}(w_{1:t})$. \ In
particular, nothing changes if the outcome contains both primary and
secondary outcomes of interest. \ Therefore without any loss of generality,
we can assume that any covariates that are measured are unaffected by the
experiment; otherwise, they would be considered as secondary outcomes. \ 

\begin{example}
(continuing Example \ref{Example T=3}). The number of potential outcomes up
to time $3$ is $14$.\ There are $2^{3}=8$ potential paths, e.g. $%
Y_{1:3}(1,1,1)=\left\{ Y_{1}(1),Y_{2}(1,1),Y_{3}(1,1,1)\right\} $. In the
experiment we observe one of these potential paths, the other seven will be
missing.
\end{example}

Our framework is model free, but it helps intuition to consider an example.\ 

\begin{example}
\label{defn potential autoregression}A first-order (vector)
\textquotedblleft potential autoregression\textquotedblright\ $%
Y_{1:T}(\bullet )$ holds when, for every permissible treatment path $w_{1:T}$%
, the potential outcomes follow the dynamic 
\begin{equation*}
Y_{t}(w_{1:t})=\mu (w_{1:t})+\phi \left( w_{1:t}\right)
Y_{t-1}(w_{1:t-1})+\sigma (w_{1:t})\varepsilon _{t},\quad t=1,2,...,T,
\end{equation*}%
where $\mu (\bullet )$, $\phi \left( \bullet \right) $ and $\sigma (\bullet
) $ are non-stochastic. The treatment-invariant $\varepsilon
_{1},...,\varepsilon _{T}$ couple the potential paths. When $\phi \left(
w_{1:t}\right) =\phi $, $\mu (w_{1:t})=\mu +\sigma \mu (w_{t})$ and $\sigma
(w_{1:t})=\sigma $ we label this an \textquotedblleft impulse potential
autoregression\textquotedblright . \ Then $\mu (w_{1:t})+\sigma
(w_{1:t})\varepsilon _{t}=\mu +\sigma \left\{ \varepsilon _{t}+\mu
(w_{t})\right\} $. A first order \textquotedblleft potential moving
average\textquotedblright\ $Y_{1:T}(\bullet )$ holds when 
\begin{equation*}
Y_{t}(w_{1:t})=\mu (w_{1:t})+\sigma (w_{1:t})\varepsilon _{t}+\theta
(w_{1:t-1})\sigma (w_{1:t-1})\varepsilon _{t-1},\quad t=1,2,...,T,
\end{equation*}%
where $\mu (\bullet )$, $\theta (\bullet )$ and $\sigma (\bullet )$ are
non-stochastic. When $\theta (w_{1:t-1})=\theta $, $\mu (w_{1:t})=\mu
+\sigma \mu ^{\ast }(w_{t})+\sigma \mu ^{\ast }(w_{t-1})$ and $\sigma
(w_{1:t})=\sigma $, then we label this an \textquotedblleft impulse moving
average\textquotedblright\ $Y_{t}(w_{1:t})=\mu +\sigma \left\{ \varepsilon
_{t}+\mu ^{\ast }(w_{t})\right\} +\theta \sigma \left\{ \varepsilon
_{t-1}+\mu ^{\ast }(w_{t-1})\right\} $. \ The right hand side of this only
depends upon $w_{t-1:t}$, a special case of the m-dependent potential
process discussed in Web Appendix \ref{sect:m dependent}. \ \ The
autoregressive and moving average impulse models have a treatment which acts
as an \textquotedblleft impulse\textquotedblright\ for the innovation in the
model \`{a} la \cite{Sims(80)}.
\end{example}

The challenge of causal inference on time series is that we only observe one
treatment path, $w_{1:T}^{\text{obs}}=(w_{1}^{\text{obs}},\dots ,w_{T}^{%
\text{obs}})$, since we can only administer one treatment at each time
step.\ After administering the treatment path the outcomes are $y_{1:T}^{%
\text{obs}}=(y_{1}^{\text{obs}},\dots ,y_{T}^{\text{obs}})$.

To link the treatment and the outcome paths, we assume that there is only
one version of the treatment, the treatment assigned is the treatment the
unit is offered and receives. Then the observed path is 
\begin{equation}
y_{1:t}^{\text{obs}}=y_{1:t}(w_{1:T}^{\text{obs}})=\sum_{w\in
\{0,1\}^{t}}1_{w_{1:t}^{\text{obs}}=w}y_{1:t}(w),\ t=1,\dots ,T,
\label{E:observed_outcome_path}
\end{equation}%
while the potential path at time $t$ is, for the $w_{1:t}^{\text{obs}}$
treatment path, 
\begin{equation*}
Y_{1:t}^{\text{obs}}=Y_{1:t}(w_{1:t}^{\text{obs}})=\sum_{w\in
\{0,1\}^{t}}1_{w_{1:t}^{\text{obs}}=w}Y_{1:t}(w),\ t=1,\dots ,T.
\end{equation*}

\begin{example}
(continuing Example \ref{Example T=3}). The right hand side of Figure \ref%
{F:Potential_Outcomes} visualizes this setup when $T=3$, with $w_{1:3}^{%
\text{obs}}=(1,1,0)$. \ Then we will see the path $Y_{1:t}(w_{1:3}^{\text{obs%
}})$ (3 random variables: $Y_{1}(1)$, $Y_{2}(1,1)$ and $Y_{3}(1,1,0)$) while
the others are missing.
\end{example}

To design time series experiments and extract causal effects we view our
treatments as causing contemporaneous or subsequent movements in the outcome
variables. Therefore we must, a priori, rule out the opposite, that the
treatments are themselves influenced by future values of potential outcomes
(e.g. apply Axiom A of \cite{Granger(80cause)}). We encapsulate this
\textquotedblleft non-anticipation\textquotedblright\ by the following
Assumption.

\begin{assumption}
\label{Assum:non-anticipating gen} For each $t$ and for all $%
w_{1:t-1},Y_{1:T}(\bullet )$, 
\begin{equation*}
\Pr (W_{t}=w_{t}|W_{1:t-1}=w_{1:t-1},Y_{1:T}(\bullet ))=\Pr
(W_{t}=w_{t}|W_{1:t-1}=w_{1:t-1},Y_{1:t-1}(\bullet )).
\end{equation*}
\end{assumption}

The time series nomenclature for this assumption is that $Y_{t:T}(\bullet )$
does not Granger cause $W_{t}$ (e.g. \cite{Kursteiner(10)}, \cite{Sims(72)}, 
\cite{Chamberlain(83)}, \cite{Lechner(11)}). \ Non-anticipating treatments
are thus \textquotedblleft latent ignorable\textquotedblright\ (\cite%
{FrangakisRubin(99)}) and are the same as the \textquotedblleft latent
sequential ignorability\textquotedblright\ longitudinal assumption of \cite%
{RicciardiMatteiMealli(16)} when $T=2$.

A stronger assumption is that the only potential outcomes which matter are
those which have followed the treatment path. \ We formalize this below. \ 

\begin{assumption}[Non-anticipating treatment]
\label{Assum: non-anticipating} For each $t$ and for all $%
w_{1:t-1},Y_{1:T}(\bullet )$, 
\begin{equation}
\Pr (W_{t}=w_{t}|W_{1:t-1}=w_{1:t-1},Y_{1:T}(\bullet ))=\Pr
(W_{t}=w_{t}|W_{1:t-1}=w_{1:t-1},Y_{1:t-1}(w_{1:t-1})).  \notag
\label{non-anticipating}
\end{equation}
\end{assumption}

The class of treatments which only depend upon past observables is at the
heart of the \textquotedblleft sequential randomization\textquotedblright\
assumption of Robins (e.g. \cite{Robins(94)}, \cite{RobinsGreenlandHu(99)}, 
\cite{AbbringBerg(03)} and \cite{Lok(08)}) in his longitudinal studies%
\footnote{%
We make assumptions about $\Pr
(W_{t}=w_{t}|W_{1:t-1}=w_{1:t-1},Y_{1:T}(\bullet ))$, which conditions on $%
Y_{1:T}(\bullet )$. \ \cite{RobinsGreenlandHu(99)}, for example, focuses on
identifying marginal effects and requires that for all $u_{1:T}\in
\{0,1\}^{T}$, $\Pr (W_{t}=w_{t}|W_{1:t-1}=w_{1:t-1},Y_{1:T}(u_{1:T}))=\Pr
(W_{t}=w_{t}|W_{1:t-1}=w_{1:t-1},Y_{1:t-1}(w_{1:t-1}))$. \ To produce our
randomization test we need to be able to condition on the full $%
Y_{1:T}(\bullet )$. \ }. This is attractive as the person assigning
treatment will often not have access to the lagged unobserved potential
outcomes. \ A slight generalization is to condition on a filtration (or
information set), that at least contains $w_{1:t-1},Y_{1:t-1}(w_{1:t-1})$. \
This allows us to naturally include any observed covariates in the
conditional distribution of the treatment assignment.


\section{Causal effects\label{sect:causal effects}}

Time series causal effects are defined as a comparison between the potential
outcomes at a fixed point in time. \ The primary object of interest is the
temporal average of these causal effects. \ We will not invoke super
population arguments (averaging over units) or any properties of the
potential path processes (e.g. averaging over time by assuming stationarity
or by averaging with respect to a model). \ The only source of randomness in
our formulation is the randomization of the treatment.

\subsection{General causal effects}

Any comparison of potential outcomes, at a fixed point in time, has a causal
interpretation, e.g. $Y_{1}(1)-Y_{1}(0)$ and $Y_{2}(1,0)-Y_{2}(0,1)$. \
After $T$ time steps, a single unit has $2(2^{T}-1)$ potential outcomes; we
can, therefore, define a large number causal estimands.

\begin{definition}
(General Causal Effects) For paths $w_{1:t}$ and $w_{1:t}^{\prime }$, the $t$%
-th causal effect is 
\begin{equation*}
\tau _{t}(w_{1:t},w_{1:t}^{\prime })=Y_{t}(w_{1:t})-Y_{t}(w_{1:t}^{\prime }).
\end{equation*}%
The temporal average treatment effect of the paths $w_{1:T}$ and $%
w_{1:T}^{\prime }$ is 
\begin{equation*}
\bar{\tau}(w_{1:T},w_{1:T}^{\prime })=\frac{1}{T}\sum_{t=1}^{T}\tau
_{t}(w_{1:t},w_{1:t}^{\prime }).
\end{equation*}
\end{definition}

Think of the $t$-th causal effect in a similar way as in the classical
setting where the causal estimands are defined as comparisons between the
unit level potential outcomes. We are mainly interested in the temporal
average treatment effect.

\begin{example}
\label{ex:total cause}The causal effect that has received the most attention
in the literature is $\tau _{t,\text{Total}}=Y_{t}(1,\dots ,1)-Y_{t}(0,\dots
,0)$. This asks how a unit performs at time $t$ if under constant treatment
compared with constant control.
\end{example}

The general causal effect directly compares two different paths. \ A special
case of this, focuses on measuring the causal effect on the outcome at time $%
t$ of treatment compared to control at time $t-p$, when $p\geq 0$. \ The
main way of measuring this effect is 
\begin{equation}
Y_{t}(\underset{(t-p-1)\times 1}{w^{\dag }},1,\underset{p\times 1}{w})-Y_{t}(%
\underset{(t-p-1)\times 1}{w^{\dag }},0,\underset{p\times 1}{w}),
\label{lagged tau crude}
\end{equation}%
for a particular choice of $w^{\dag }\in \left\{ 0,1\right\} ^{t-p-1}$ and $%
w\in \left\{ 0,1\right\} ^{p}$. To make the notation clearer, we will
sometimes use the under-set script to specify the length of a vector or a
matrix. \ Effect (\ref{lagged tau crude}) can be generalized to an average
over the paths 
\begin{equation}
\tau _{t,p}^{\dag }(\{1\},\{0\})=\sum_{\substack{ w^{\dag }\in \left\{
0,1\right\} ^{t-p-1}  \\ w\in \left\{ 0,1\right\} ^{p}}}a_{w^{\dag
},w}\left\{ Y_{t}(\underset{(t-p-1)\times 1}{w^{\dag }},1,\underset{p\times 1%
}{w})-Y_{t}(\underset{(t-p-1)\times 1}{w^{\dag }},0,\underset{p\times 1}{w}%
)\right\} ,  \label{lagged tau average}
\end{equation}%
with non-stochastic weights $\sum_{w^{\dag },w}a_{w^{\dag },w}=1$ and $%
a_{w^{\dag },w}\geq 0$, e.g. setting $a_{w^{\dag },w}\propto 1$ leads to
uniform weights. \ The averaging removes the dependence on a particular path
and incorporates all of the $2^{t}$ potential outcomes.

The weights $a$ can down-weight certain paths which are less probable and
exclude ones which are not possible. \ For example, in a randomized
experiment where after a unit receives the treatment they are considered to
be in the treatment group until the conclusion of the study, at time step $t$
there are only $1+t$ potential outcomes. \ The causal effect is then defined
by setting to zero the weights for the outcomes paths that can not be
observed.

\subsection{$p\geq 0$ lag causal effect of treatment on outcome}

Under our set up we observe one outcome path, and therefore without strong
assumptions we can not estimate (\ref{lagged tau average}). \ However, by
defining the weights as a function of parts of the observed treatment we can
define a class of causal estimands that can be estimated from one
experimental unit. \ The price we pay is that the estimand changes as a
function of the observed past treatment path.

Setting $w^{\dag }=w_{1:t-p-1}^{\text{obs}}$, $a_{w^{\dag },w}=1_{w^{\dag
}=w_{1:t-p-1}^{\text{obs}}}a_{w}$, where\quad $\sum_{w\in \left\{
0,1\right\} ^{p}}a_{w}=1\ $and\quad $a_{w}\geq 0$, leads to the following
special case of (\ref{lagged tau average}).

\begin{definition}[$p\geq 0$ lag causal effect of treatment on outcome]
\label{defn:lag cause effect}Let $a_{w}$ be non-stochastic weights, then let
the $p\geq 0$ lag causal effect of treatment on outcome be 
\begin{equation*}
\tau _{t,p}(\{1\},\{0\})=\sum_{w\in \left\{ 0,1\right\} ^{p}}a_{w}\left\{
Y_{t}(w_{1:t-p-1}^{\text{obs}},1,\underset{p\times 1}{w})-Y_{t}(w_{1:t-p-1}^{%
\text{obs}},0,\underset{p\times 1}{w})\right\} ,
\end{equation*}%
where $\sum_{w}a_{w}=1$ and $a_{w}\geq 0$. The temporal average $p$ lag
treatment effect is 
\begin{equation*}
\overline{\tau }_{p}(\{1\},\{0\})=\frac{1}{T-p}\sum_{t=p+1}^{T}\tau
_{t,p}(\{1\},\{0\})\text{.}
\end{equation*}%
When $p=0$ the (weight free) \textquotedblleft contemporaneous causal
effect\textquotedblright\ is 
\begin{equation*}
\tau _{t,0}(\{1\},\{0\})=Y_{t}(w_{1:t-1}^{\text{obs}},1)-Y_{t}(w_{1:t-1}^{%
\text{obs}},0).
\end{equation*}%
This measures the instant effect of administering treatment at time $t$.
\end{definition}

In order to keep the notation simpler, most of our theoretical results will
be stated for the case when the weights are uniform.

\begin{example}
\label{Ex:2lag} Let $p=1$ and $a_{w}=1/2$, then $\tau _{t,1}(\{1\},\{0\})=%
\frac{1}{2}[\{Y_{t}(w_{1:t-2}^{\text{obs}},1,0)-Y_{t}(w_{1:t-2}^{\text{obs}%
},0,0)\}+\{Y_{t}(w_{1:t-2}^{\text{obs}},1,1)-Y_{t}(w_{1:t-2}^{\text{obs}%
},0,1)\}]$. This is, for the observed path up to time $t-2$, the average
difference in the potential outcomes between assigning the unit to treatment
as opposed to control at time $t-1$, averaged over a uniformly randomized
treatment assignment at time $t$.
\end{example}

\subsection{Causal effects and treatment path}

Defining causal effects conditional on the observed treatments may seem like
a departure from the classical causal inference setting where the causal
estimands are only a function of the potential outcomes. \ There are three
reasons why this is a reasonable approach to take. \ 

Firstly, by making the estimands dependent on the observed treatment path we
define a wide class of causal estimands that can be estimated without
imposing further assumptions. \ Secondly, this is similar to focusing on the
average effect on the treated which is defined as a conditional estimand and
is widely accepted as a valid causal estimand (see \cite{ImbensRubin(15)}).
\ Thirdly, although the average $p$ lag treatment effect depends on the
observed path, it satisfies a central limit theorem (see Theorem \ref{T:CLT}%
); therefore, inference drawn conditional on one path is close to the
inference we would have drawn if we had observed a different treatment path.

To reduce the impact of the observed treatment path on the definition of the
causal estimand we propose using a \textquotedblleft stepping
approach.\textquotedblright\ \ For fixed $p\geq 0$ and $q\geq 0$ the $q$
step lag $p$ causal effect is, 
\begin{equation*}
\tau _{t,p}^{(q)}(\{1\},\{0\})=\sum_{\substack{ w\in \left\{ 0,1\right\}
^{p}  \\ w^{\dag }\in \left\{ 0,1\right\} ^{q}}}a_{w^{\dag },w}\left\{
Y_{t}(w_{1:t-p-q-1}^{\text{obs}},\underset{q\times 1}{w^{\dag }},1,\underset{%
p\times 1}{w})-Y_{t}(w_{1:t-p-q-1}^{\text{obs}},\underset{q\times 1}{w^{\dag
}},0,\underset{p\times 1}{w})\right\} ,
\end{equation*}%
which averages over possible treatment paths $w^{\dag }\in \left\{
0,1\right\} ^{q}$, at time $t-p-q$ to $t-p-1$. \ Again $a_{w^{\dag },w}$ are
non-negative weights which sum to one. \ As $q$ increases the dependence on
the observed treatment path in the definition of the $q$ step $p$ lag causal
effect decreases. \ At the extreme, when $q=t-p-1$ we obtain (\ref{lagged
tau average}) and when $q=0$, $\tau _{t,p}^{(0)}(\{1\},\{0\})=\tau
_{t,p}(\{1\},\{0\})$. For values of $t\in \lbrack p+1,\dots ,p+q+1]$ the $q$
step $p$ lag causal effect is $\tau _{t,p}^{(q)}=\tau _{t,p}^{(t-p+1)}$. In
practice, as $q$ increases the variance of the estimator we propose in the
subsequent section will also increase.

The temporal average $q$ step $p$ lag causal effect is defined as 
\begin{equation}
\bar{\tau}_{p}^{(q)} = \frac{1}{T-p} \sum_{t=p+1}^T \tau
_{t,p}^{(q)}(\{1\},\{0\}).
\end{equation}

\begin{example}
\label{Ex:2step} Let $q=1$ and $p=0$, then for each $t$, let $a_{w^{\dag
},w}=1/2$, then 
\begin{equation*}
\tau _{t,0}^{(1)}(\{1\},\{0\})=\frac{1}{2}\left[ \left\{ Y_{t}(w_{1:t-2}^{%
\text{obs}},1,1)-Y_{t}(w_{1:t-2}^{\text{obs}},1,0)\right\} +\left\{
Y_{t}(w_{1:t-2}^{\text{obs}},0,1)-Y_{t}(w_{1:t-2}^{\text{obs}},0,0)\right\} %
\right] .
\end{equation*}%
This is, for the observed path up to time $t-2$, the average difference in
the potential outcomes between assigning the unit to treatment as opposed to
control at time $t$, averaged over a uniformly randomized treatment
assignment at time $t-1$.
\end{example}

Comparing examples \ref{Ex:2lag} and \ref{Ex:2step}, we see that in general $%
\tau _{t,0}^{(1)}(\{1\},\{0\})\neq \tau _{t,1}(\{1\},\{0\})$.


\section{Experiments and estimation\label{sect:estimation}}

There are multiple ways of estimating $\bar{\tau}_{p}$ and $\bar{\tau}%
_{p}^{(q)}$. \ Here we use a Horvitz-Thompson type estimators. \ We show
these estimators are unbiased, over the randomization distribution; compute
their variances, which depends on the potential outcomes; and derive
unbiased estimators of an upper bound to these variances. \ 

\subsection{Probabilistic treatment}

\label{S:Inference}

Assume that at each time point we randomly administer a treatment $W_{t}=1$
or control $W_{t}=0$ with some probability 
\begin{equation*}
p_{t}(w_{t})=\Pr (W_{t}=w_{t}|\mathcal{F}_{T,t-1}),\quad w_{t}=0,1,
\end{equation*}%
which we will label the \textquotedblleft adapted propensity
score.\textquotedblright\ We use the array notation $\mathcal{F}_{T,t}$ for
the filtration (or information set) to remind us we always condition on $%
Y_{1:T}(\bullet )$ \footnote{%
There are two approaches to causal inference. \ The first conditions on the
potential outcomes (or equivalently treats them as fixed), and assumes that
the only randomness comes from the treatment assignment. \ The other
averages the potential outcomes over some model. In this paper we only focus
on the first.}. $\mathcal{F}_{T,t}$ contains at least $w_{1:t}$ and so
implicitly $Y_{1:t}(w_{1:t})$ and obeys the nesting property $\mathcal{F}%
_{T,t}\subseteq \mathcal{F}_{T,t+1}$ for all $t\leq T$ and $T\geq 1$. Under
Assumption \ref{Assum: non-anticipating}, the adapted propensity score
simplifies to 
\begin{equation*}
p_{t}(w_{t})=\Pr (W_{t}=w_{t}|W_{1:t-1}=w_{1:t-1},Y_{1:t-1}(w_{1:t-1})).
\end{equation*}%
Generalizing this notation helps. \ For the $p\geq 0$ lag, the
\textquotedblleft adapted path propensity score\textquotedblright\ is 
\begin{equation*}
p_{t}(\underset{\left( p+1\right) \times 1}{w_{t-p:t}})=\Pr
(W_{t-p:t}=w_{t-p:t}|\mathcal{F}_{T,t-p-1}),\quad \text{for }w_{t-p:t}\in
\{0,1\}^{p+1}.
\end{equation*}

To study the properties of our proposed estimators, we will compute their
moments with respect to $W_{1:T}|Y_{1:T}(\bullet )$, the randomization of
the treatment, holding fixed all the potential outcomes. \ Such
randomization driven conditional means, variances and covariances will be
noted by $\mathrm{E}^{R}$, $\mathrm{Var}^{R}$ and $\mathrm{Cov}^{R}$
respectively. \ To ensure that inference can be performed using the
randomization distribution we assume that treatment assignments are
probabilistic.

\begin{assumption}
\label{Prob assign} (Probabilistic Treatment Assignment). For every $t\geq 1$
and $\mathcal{F}_{T,t-1}$, 
\begin{equation}
0<\Pr (W_{t}=1|\mathcal{F}_{T,t-1})<1.  \label{E:pathuncon}
\end{equation}
\end{assumption}

This assumption implies that $p_{t}(w_{t-p:t})\in (0,1)$ for all $t\geq 1$, $%
\mathcal{F}_{T,t-p-1}$ and $w_{t-p:t}$.


\subsection{The $p$ lag causal effect and estimation}

Using the observed data, we can estimate $\bar{\tau}_{p}$ without imposing
any further assumptions. Recall that the $p$ lag causal effect with uniform
weights is, 
\begin{equation*}
\tau _{t,p}(\{1\},\{0\})=\sum_{w\in \{0,1\}^{p}}a_{w}\left\{
Y_{t}(w_{1:t-p-1}^{\text{obs}},1,w)-Y_{t}(w_{1:t-p-1}^{\text{obs}%
},0,w)\right\} .
\end{equation*}%
A feasible estimator of this causal effect is 
\begin{eqnarray*}
\hat{\tau}_{t,p} &=&\sum_{w\in \{0,1\}^{p}}a_{w}\left\{ \frac{1_{w_{t-p:t}^{%
\text{obs}}=(1,w)}}{p_{t}(1,w)}Y_{t}(w_{1:t-p-1}^{\text{obs}},1,w)-\frac{%
1_{w_{t-p:t}^{\text{obs}}=(0,w)}}{p_{t}(0,w)}Y_{t}(w_{1:t-p-1}^{\text{obs}%
},0,w)\right\} \\
&=&a_{w_{t-p+1:t}^{\text{obs}}}\frac{Y_{t}(w_{1:t}^{\text{obs}%
})(-1)^{1-w_{t-p}^{\text{obs}}}}{p_{t}(w_{t-p:t}^{\text{obs}})}.
\end{eqnarray*}%
The lead case of this is uniform weights $a_{w}=1/2^{p}$. The following
theorem shows that this estimator is unbiased and derives its variance, over
the randomization. \ 

\begin{theorem}[Properties of $p$ lag estimators]
\label{T:p-lag} Define $u_{t-p,p}=\hat{\tau}_{t,p}-\tau _{t,p}(\{1\},\{0\})$
as the estimation error. \ Then $\mathrm{E}^{R}(u_{t,p}|\mathcal{F}%
_{T,t-1})=0$, and $\mathrm{Var}^{R}(u_{t,p}|\mathcal{F}_{T,t-1})$ equals%
\begin{eqnarray*}
&&\sum_{w\in \{0,1\}^{p}}a_{w}^{2}\left( \frac{Y_{t+p}(w_{1:t-1}^{\text{obs}%
},1,w)^{2}}{p_{t+p}(1,w)}+\frac{Y_{t+p}(w_{1:t-1}^{\text{obs}},0,w)^{2}}{%
p_{t+p}(0,w)}\right) \\
&&-\underset{w^{\prime }\in \{0,1\}^{p}}{\sum_{w\in \{0,1\}^{p}}}%
a_{w}a_{w^{\prime }}\left\{ Y_{t+p}(w_{1:t-1}^{\text{obs}},1,w^{\prime
})-Y_{t+p}(w_{1:t-1}^{\text{obs}},0,w^{\prime })\right\} \left\{
Y_{t+p}(w_{1:t-1}^{\text{obs}},1,w)-Y_{t+p}(w_{1:t-1}^{\text{obs}%
},0,w)\right\} .
\end{eqnarray*}
Conditioning on $Y_{1:T}(\bullet )$, then $\mathrm{E}^{R}(\left\vert
u_{t,p}\right\vert )<\infty $, $\mathrm{E}^{R}(u_{t,p})=0$, $\mathrm{Cov}%
^{R}(u_{t,p},u_{s,p})=0$, $s\neq t$.
\end{theorem}

Proof. The proof is given in Appendix \ref{ss:proof_thrm_kstep}. \newline

The Theorem shows that the $\hat{\tau}_{t,p}$ are unbiased estimators of the 
$p$ lag causal effect. Moreover, the randomization ensures that $\left\{
u_{t,p}\right\} $ is a martingale difference array with respect to $\left\{ 
\mathcal{F}_{T,t-1}\right\} $, and hence $\hat{\tau}_{t,p}$ is conditionally
(on $Y_{1:T}(\bullet )$) unbiased and the errors are conditionally
uncorrelated through time.\ 

From now on, in the rest of the paper for simplicity of exposition, we will
always use uniform weights $a_{w}\propto 1$. \ 

\begin{example}
Assume $Y_{t}$ is scalar and $p=0$, then 
\begin{equation}
\hat{\tau}_{t,0}=\frac{1_{W_{t}=1}Y_{t}(w_{1:t-1}^{\text{obs}},1)}{p_{t}(1)}-%
\frac{1_{W_{t}=0}Y_{t}(w_{1:t-1}^{\text{obs}},0)}{p_{t}(0)}=\frac{%
Y_{t}(w_{1:t}^{\text{obs}})(-1)^{1-w_{t}^{\text{obs}}}}{p_{t}(w_{t}^{\text{%
obs}})},  \label{E:1-step_estimator}
\end{equation}%
while 
\begin{equation*}
\mathrm{Var}^{R}(u_{t,0}|\mathcal{F}_{T,t-1})=\frac{\left\{ Y_{t}(w_{1:t-1}^{%
\text{obs}},1)p_{t}(0)+Y_{t}(w_{1:t-1}^{\text{obs}},0)p_{t}(1)\right\} ^{2}}{%
p_{t}(1)p_{t}(0)}.
\end{equation*}%
In the Bernoulli randomized experiment, where $p_{t}(1)=p_{t}(0)=\frac{1}{2}$%
, for all $t$, then, $\mathrm{Var}^{R}(u_{t,0}|\mathcal{F}_{T,t-1})=\left\{
Y_{t}(w_{1:t-1}^{\text{obs}},1)+Y_{t}(w_{1:t-1}^{\text{obs}},0)\right\} ^{2}$%
, conditioning on past treatments and potential outcomes, while 
\begin{equation*}
\mathrm{Var}^{R}(u_{t,0})=\mathrm{E}^{R}\left[ \left\{
Y_{t}(W_{1:t-1},1)+Y_{t}(W_{1:t-1},0)\right\} ^{2}|Y_{1:t}(\bullet )\right] =%
\frac{1}{2^{t-1}}\sum_{w}\left\{ Y_{t}(w,1)+Y_{t}(w,0)\right\} ^{2},
\end{equation*}%
averaging over all possible treatment paths up to time $t-1$ with the
potential outcomes fixed. \ \ 
\end{example}

The variance of $\hat \tau_{t,p}$ is a function of the potential outcomes as
well as the treatment assignment probabilities. \ Since we never observe all
of the potential outcomes we can not estimate the variance without imposing
further assumption. \ Instead, we derive an upper bound; the following Lemma
contains the details and provides an unbiased estimator for the upper bound.
\ This upper bound is different to the usual Neyman-style upper bound for
the variance; so far we have been unable to establish a connection between
the two. \ The upper bound is only attained if the potential outcomes are
all equal and the variance is 0.

\begin{lemma}
\label{L:var_bound_p} Under non-anticipating treatments Assumption \ref%
{Assum: non-anticipating} and probabilistic assignment Assumption \ref{Prob
assign}, the variance of $u_{t,p}$, and in turn $\hat{\tau}_{t,p}$, is
bounded above by 
\begin{equation*}
\mathrm{Var}^{R}(u_{t,p}|\mathcal{F}_{T,t-1})\leq \sum_{w\in \{0,1\}^{p+1}}%
\frac{Y_{t+p}^{2}(w_{1:t-1}^{\text{obs}},w)\left[ 1+2p_{t+p}(w)(2^{p-1}-1)%
\right] }{p_{t+p}(w)}=\sigma _{t+p,p}^{2}.
\end{equation*}%
Moreover, this upper bound can be estimated by, 
\begin{equation}
\widehat{\sigma }_{t+p,p}^{2}=\sum_{w\in \{0,1\}^{p}}\frac{%
1_{W_{t:t+p}=w}Y_{t+p}(w_{1:t-1}^{\text{obs}},w)^{2}\left[
1+2p_{t+p}(w)(2^{p-1}-1)\right] }{p_{t+p}^{2}(w)}
\end{equation}%
and is conditionally unbiased, i.e. $\mathrm{E}^{R}(\widehat{\sigma }%
_{t+p,p}^{2}|\mathcal{F}_{T,t-1})=\sigma _{t+p,p}^{2}$.
\end{lemma}

Proof. The first part is proved in Appendix \ref{appendix:proof of lemma},
the unbiasedness is straightforward.\newline

\begin{example}
When $p=0$, then 
\begin{equation*}
\mathrm{Var}^{R}(u_{t,0}|\mathcal{F}_{T,t-1})\leq \frac{%
Y_{t}(w_{1:t-1},1)^{2}}{p_{t}(1)}+\frac{Y_{t}(w_{1:t-1},0)^{2}}{p_{t}(0)}%
=\sigma _{t,0}^{2}.
\end{equation*}%
which can be estimated by, 
\begin{equation}
\frac{1_{W_{t}=1}}{p_{t}(1)^{2}}Y_{t}(w_{1:t-1},1)^{2}+\frac{1_{W_{t}=0}}{%
p_{t}(0)^{2}}Y_{t}(w_{1:t-1},0)^{2}=\left( \hat{\tau}_{t,0}\right) ^{2}.
\label{E:var_bound}
\end{equation}
\end{example}

The average $p$ lag causal effect is $\overline{\tau }_{p}=\frac{1}{T-p}%
\sum_{t=p+1}^{T}\tau _{t,p}(\{1\},\{0\})$, which we estimate by $\hat{\bar{%
\tau}}_{p}=\frac{1}{T-p}\sum_{t=p+1}^{T}\hat{\tau}_{t,p}$. \ The causal
estimand depends on the observed treatment path and in general two different
treatment paths will lead to distinct estimates of $\bar{\tau}_{p}$ (which
itself is a function of the observed treatment path). \ The variance of $%
\hat{\bar{\tau}}_{p}$ is a combination of the variances of $\hat{\tau}_{t,p}$%
, and its properties are given in the following key Theorem.

\begin{theorem}[Properties of average $p$ lag estimator]
\label{T:average_p-step} Let $\widehat{\gamma }_{p}=\frac{1}{(T-p)^{2}}%
\sum_{t=p+1}^{T}\sigma _{t,p}^{2}$. Under Assumptions \ref{Assum:
non-anticipating} and \ref{Prob assign} then $\mathrm{E}^{R}\left( \hat{\tau}%
_{p}\right) =\bar{\tau}_{p}$ and $\mathrm{E}^{R}\left( \widehat{\gamma }%
_{p}\right) \geq \mathrm{Var}^{R}\left( \hat{\tau}_{p}\right) $, conditional
on $Y_{1:T}(\bullet )$.
\end{theorem}

Proof. The unbiasedness of $\hat{\tau}_{p}$ follows from Theorem \ref%
{T:p-lag}. The unbiasedness of $\widehat{\gamma }_{p}$ follows from the
randomization inducing the martingale difference property of $\left\{
u_{t,p}\right\} $ given $Y_{1:T}(\bullet )$. The last result follows
trivially.

\subsection{Stepped version and estimation}

The extension to allow $q\geq 0$ stepping is straight forward to state.
Recall that, 
\begin{equation*}
\tau _{t,p}^{(q)}(\{1\},\{0\})=\frac{1}{2^{p+q}}\sum_{\substack{ w\in
\left\{ 0,1\right\} ^{p}  \\ w^{\dag }\in \left\{ 0,1\right\} ^{q}}}\left\{
Y_{t}(w_{1:t-q-p-1}^{\text{obs}},w^{\dag },1,w)-Y_{t}(w_{1:t-q-p-1}^{\text{%
obs}},w^{\dag },0,w)\right\} .
\end{equation*}%
The Horvitz-Thompson $q$ step $p$ lag causal estimator for $t\geq p+q+1$ is 
\begin{eqnarray*}
\hat{\tau}_{t,p}^{(q)} &=&\frac{1}{2^{p+q}}\sum_{\substack{ w\in \left\{
0,1\right\} ^{p}  \\ w^{\dag }\in \left\{ 0,1\right\} ^{q}}}\left\{ \frac{%
1_{w_{t-q-p:t}^{\text{obs}}=(w^{\dag },1,w)}}{p_{t}(w^{\dag },1,w)}%
Y_{t}(w_{1:t-q-p-1}^{\text{obs}},w^{\dag },1,w)-\frac{1_{w_{t-q-p:t}^{\text{%
obs}}=(w^{\dag },0,w)}}{p_{t}(w^{\dag },0,w)}Y_{t}(w_{1:t-q-p-1}^{\text{obs}%
},w^{\dag },0,w)\right\} \\
&=&\frac{1}{2^{p+q}}\frac{Y_{t}(w_{1:t}^{\text{obs}})(-1)^{1-w_{t-p}^{\text{%
obs}}}}{p_{t}(w_{t-q-p:t}^{\text{obs}})}.
\end{eqnarray*}%
Notice that the value of $q$ has no impact on the numerator of $\hat{\tau}%
_{t,p}^{(q)}$, is just impacts the weight of each datapoint. \ For values of 
$t\in \lbrack p+1,p+q]$ we define $\hat{\tau}_{t,p}^{(q)}=\hat{\tau}%
_{t,p}^{(t-p-1)}$.

The properties of $\hat{\tau}_{t,p}^{(q)}$ exactly mimic $\hat{\tau}_{t,p}$
with no new ideas being needed to handle them. \ In particular the
estimation errors are again martingale differences. \ The details are given
in the web Appendix.

\subsection{Proxy outcomes and gains in precision}

By using proxy outcomes, we can reduce the variance of the $p$ causal effect
estimator. To illustrate this, recall $\tau
_{t,0}(\{1\},\{0\})=Y_{t}(w_{1:t-1}^{\text{obs}},1)-Y_{t}(w_{1:t-1}^{\text{%
obs}},0)$. For any $\widetilde{\mu }_{t|t-1}$ 
\begin{equation*}
\tau _{t,0}(\{1\},\{0\})=\left\{ Y_{t}(w_{1:t-1}^{\text{obs}},1)-\widetilde{%
\mu }_{t|t-1}\right\} -\left\{ Y_{t}(w_{1:t-1}^{\text{obs}},0)-\widetilde{%
\mu }_{t|t-1}\right\} .
\end{equation*}%
When $\widetilde{\mu }_{t|t-1}$ is only a function of $\mathcal{F}_{T,t-1}$
we call $\widetilde{\mu }_{t|t-1}$ a \textquotedblleft time series proxy
outcome,\textquotedblright e.g. $\widetilde{\mu }_{t|t-1}=Y_{t-1}(w_{1:t-1}^{%
\text{obs}})$. \ A good proxy outcome aims to reduce $\{Y_{t}(w_{1:t-1}^{%
\text{obs}},\bullet )-\widetilde{\mu }_{t|t-1}\}^{2}$. \ 

The corresponding causal estimator of this decomposition is 
\begin{eqnarray*}
\widetilde{\tau }_{t,0} &=&\frac{1_{W_{t}=1}\left\{ Y_{t}(w_{1:t-1}^{\text{%
obs}},1)-\widetilde{\mu }_{t|t-1}\right\} }{p_{t}(1)}-\frac{%
1_{W_{t}=0}\left\{ Y_{t}(w_{1:t-1}^{\text{obs}},0)-\widetilde{\mu }%
_{t|t-1}\right\} }{p_{t}(0)} \\
&=&\frac{\left\{ Y_{t}(w_{1:t-1}^{\text{obs}},1)-\widetilde{\mu }%
_{t|t-1}\right\} (-1)^{1-w_{t}^{\text{obs}}}}{p_{t}(w_{t}^{\text{obs}})}.
\end{eqnarray*}%
Again $\widetilde{u}_{t,0}=\widetilde{\tau }_{t,0}-\tau _{t,0}(\{1\},\{0\})$%
, is a martingale difference, with a conditional variance of%
\begin{equation*}
\mathrm{Var}^{R}(\widetilde{u}_{t,0}|\mathcal{F}_{T,t-1})=\frac{\left[
\left\{ Y_{t}(w_{1:t-1}^{\text{obs}},1)-\widetilde{\mu }_{t|t-1}\right\}
p_{t}(0)+\left\{ Y_{t}(w_{1:t-1}^{\text{obs}},0)-\widetilde{\mu }%
_{t|t-1}\right\} p_{t}(1)\right] ^{2}}{p_{t}(1)p_{t}(0)}.
\end{equation*}%
If $\widetilde{\mu }_{t|t-1}$ is a good predictor of future potential
outcomes then $\widetilde{\tau }_{t,0}$ can be much more efficient than $%
\widehat{\tau }_{t,0}$, e.g. when the potential outcomes are non-stationary.
The use of proxies appear in observational studies (e.g. \cite{Raz(90)}, 
\cite{Rosenbaum(02)} and \cite{HennessyDasguptaMiratrixPattanayakSarkar(15)}%
) and \textquotedblleft doubly robust\textquotedblright\ estimators (e.g. 
\cite{RobinsRotnitzkyZhao(94)} and \cite{BangRobins(05)}) literatures.

This approach extends to $p$ lag causal effects, but it is important that
the proxy outcome is a function of $\mathcal{F}_{T,t-p-1}$, e.g. $\widetilde{%
\mu }_{t|t-p-1}=Y_{t-p-1}(w_{1:t-p-1}^{\text{obs}})$, so it is separate from
the treatments that the estimand averages over. \ A similar extension to
stepped causal effects follows.

\section{Experiments and randomization inference}

\label{sec:inference}

To draw inference from the estimands discussed in the previous section, we
propose two non-parametric methods that rely on the random assignment of the
treatment path. \ We first focus on the sharp null of no treatment effect
and explain how to perform hypothesis tests. \ Then, using the martingale
sequence property for the estimation error, we detail a central limit
theorem that allows us to perform hypothesis tests and build confidence
intervals. \ Throughout the section, we focus our attention on the
contemporaneous causal effect. \ However, our exposition trivially
generalizes to the $p\geq 0$ and $q\geq 0$ case.

\subsection{Null of no temporal causal effects}

To assess whether the treatment has a statistically significant effect, we
consider the sharp null (in the non-time series see \cite%
{Fisher(25),Fisher(35)}) of no temporal causal effects:%
\begin{equation}
H_{0}:Y_{t}(w_{1:t})=Y_{t}(w_{1:t}^{\prime })\quad \text{for all }%
w_{1:t},w_{1:t}^{\prime },\quad t=1,2,...,T.  \label{Fisher null}
\end{equation}%
This will be tested against a portmanteau alternative. Invoking the sharp
null hypothesis implies that $Y_{t}(w_{1:t}^{\text{obs}})=Y_{t}(w_{1:t}^{%
\prime })$ for all $w_{1:t}^{\prime }$. Further, (\ref{Fisher null}) means
\textquotedblleft the null of no temporal causality at lag $p\geq 0$ of the
treatment on the outcome\textquotedblright\ holds: $H_{0,p}:\tau _{t,p}=0$,
for all $t=1,2,...,T$. In turn this forces $\bar{\tau}_{p}=0$.

In the $p=0$ case, the $\widehat{\tau }_{t,0}$ estimation error can be
written as 
\begin{equation*}
u_{t,0}=\left( \frac{1_{W_{t}=1}}{p_{t}(1)}-\frac{1_{W_{t}=0}}{p_{t}(0)}%
\right) Y_{t}(w_{1:t}^{\text{obs}}),
\end{equation*}%
and under the sharp null (\ref{Fisher null})%
\begin{equation*}
\mathrm{E}^{R}(u_{t,0}|\mathcal{F}_{T,t-1})=0\quad \text{and\quad }\mathrm{%
Var}^{R}(u_{t,0}|\mathcal{F}_{T,t-1})=\frac{Y_{t}(w_{1:t}^{\text{obs}})^{2}}{%
p_{t}(1)p_{t}(0)}.
\end{equation*}%
The $p$ lagged error variance $\mathrm{Var}^{R}(u_{t-p,p}|\mathcal{F}%
_{T,t-p-1})$ equals 
\begin{equation}
Y_{t}(w_{1:t}^{\text{obs}})^{2}\frac{1}{2^{2p}}\sum_{w\in \{0,1\}^{p}}\left( 
\frac{1}{p_{t}(1,w)}+\frac{1}{p_{t}(0,w)}\right) .  \label{E:null_variance_k}
\end{equation}

\begin{example}
For a Bernoulli experiment with $p_{t}(0)=p_{t}(1)=1/2$, then $%
u_{t,0}=2(1_{W_{t}=1}-1_{W_{t}=0})Y_{t}(w_{1:t}^{\text{obs}})$ and $\mathrm{%
Var}^{R}(u_{t}|\mathcal{F}_{T,t-1})=4Y_{t}(w_{1:t}^{\text{obs}})^{2}$. In
the $p$ lagged case $p_{t+p}(1,w)=1/2^{p+1}$, so $\mathrm{Var}^{R}(u_{t-p,p}|%
\mathcal{F}_{T,t-p-1})=4Y_{t}(w_{1:t}^{\text{obs}})^{2}$ which is uniform in 
$p$.
\end{example}

We can conduct exact tests by using the conditional distribution of the
treatment path $W_{1:T}|Y_{1:T}(\bullet )$ to simulate new treatment paths.
\ Under the sharp null, we know $Y_{1:T}(\bullet )$ and can therefore
compute the exact distribution of any causal estimand for any treatment
path. In Appendix \ref{SSS:rand_test}, we provide a simple algorithm for
conducting hypothesis tests using the $\widehat{\bar{\tau}}_{0}$ estimator
to illustrate this.


\subsection{Null of no average temporal causal effects}

\label{SS:Neyman}

The sharp null of no temporal causal effect can be relaxed to
\textquotedblleft the null of no average temporal causality at lag $p\geq 0$
of the treatment on the outcome.\textquotedblright\ \ This is written as 
\begin{equation}
H_{0}:\bar{\tau}_{p}=0,\   \label{Neyman null}
\end{equation}%
(for non-time series this type of hypothesis is often called the Neyman
null).\ Here we test this null using a central limit theorem (CLT). \ 

With Assumptions \ref{Assum: non-anticipating} and \ref{Prob assign} the
estimation error collapses to zero at rate $\sqrt{T}$. Using martingale
array CLT of Theorem 3.2 in \cite{HallHeyde(80)}, the scaled error will be
asymptotically Gaussian so long as $p$ is finite, obeys a regularity
assumption and $T\rightarrow \infty $. \ Here we detail the CLT for the $%
\bar{\tau}_{0}$ case, the extension to $p\geq 0$ and $q\geq 0$ involves no
new ideas and are given in the appendix. \ 

\begin{theorem}
\label{T:CLT} Under Assumptions \ref{Assum: non-anticipating} and \ref{Prob
assign}, $\mathrm{Var}^{R}(u_{t,0})$ is bounded, and so 
\begin{equation*}
\mathrm{Var}^{R}\left( \sqrt{T}\frac{1}{T}\sum_{t=1}^{T}u_{t,0}\right) =%
\frac{1}{T}\sum_{t=1}^{T}\mathrm{Var}^{R}(u_{t,0}).
\end{equation*}%
Consequently, as $T\rightarrow \infty $, $\sqrt{T}\left( \hat{\bar{\tau}}-%
\bar{\tau}\right) =O_{p}(1)$. Conditioning on $Y_{1:T}(\bullet )$ and
assuming that $\widehat{\eta }_{T}^{2}=\frac{1}{T}\sum_{t=1}^{T}\mathrm{Var}%
^{R}(u_{t,0}|\mathcal{F}_{T,t-1})\overset{p}{\rightarrow }\eta ^{2}$, then 
\begin{equation}
\sqrt{T}\frac{1}{T}\sum_{t=1}^{T}u_{t,0}\overset{d}{\rightarrow }N(0,\eta
^{2}).  \label{E:MD_CLT}
\end{equation}
\end{theorem}

Proof. Follows from array martingale difference property of $u_{t,0}$, the
nested filtration $\mathcal{F}_{T,t}$ and the boundedness of $Y(\bullet )$
which means the Lindeberg condition holds. \ \ 

Under $H_{0}$ (\ref{E:MD_CLT}) can be rewritten as, 
\begin{equation*}
Z=\frac{\sqrt{T}\frac{1}{T}\sum_{t=1}^{T}\hat{\bar{\tau}}_{t,0}}{\sqrt{%
\widehat{\eta }_{T}^{2}}}.
\end{equation*}%
The variance term in the denominator depends on the unobserved potential
outcomes, which can be bounded from above using the result from Theorem \ref%
{T:average_p-step}. Therefore, we can define 
\begin{equation}
\widetilde{Z}=\frac{\sqrt{T}\frac{1}{T}\sum_{t=1}^{T}\hat{\bar{\tau}}_{t,1}}{%
\sqrt{\frac{1}{T}\sum_{t=1}^{T}\hat{\sigma}_{t}^{2}}}=Z\widetilde{\gamma }%
_{T},\quad \widetilde{\gamma }_{T}=\frac{\sqrt{\widehat{\eta }_{T}^{2}}}{%
\sqrt{\frac{1}{T}\sum_{t=1}^{T}\hat{\sigma}_{t}^{2}}},
\label{E:stat_neyman_null}
\end{equation}%
where $\hat{\sigma}_{t}^{2}$ is defined in (\ref{E:var_bound}). \ Then under
the null $Z\overset{d}{\rightarrow }N(0,1)$, while $\widetilde{\gamma }_{T}$
will be below 1. \ Hence $\widetilde{Z}$ can then be used to conservatively
test the no average temporal causal effects null hypothesis (\ref{Neyman
null}). \ In practice we have noticed that this test has high power.


\section{Connection to other work}

\label{S:otherwork}

Here we discuss how our formulation of time series causal inference is
connected to other works in the literature. Particular focus will be paid to
papers which are closer to our ideas. \ 

\subsection{Robins, Murphy et al\label{sect:other works}}

Since \cite{robins1986new}, scholars have been working on longitudinal
studies where the treatment can vary over time (e.g. \cite%
{robins1999association, RobinsGreenlandHu(99),robins2000marginal}). These
papers primarily focus on estimating the total causal effect, defined in
Example \ref{ex:total cause}, at one point in time, usually at the end of
the study. Their causal estimands are expectations over a super population,
and therefore randomization is used to ensure conditional independence
between the treatment assignment and potential outcomes, rather than as an
inferential tool for hypothesis testing and building confidence intervals %
\citep{Fisher(35)}. \ As mentioned above, the Robins sequential
randomization assumption is very close to our non-anticipating treatments
assumption. \ Section 7 of \cite{RobinsGreenlandHu(99)} is particularly
interesting to us as it discusses the case where there is a single unit
being treated over quite a long period of time. \ 

\cite{BlackwellGlynn(16)} and \cite{BoruvkaAlmirallWitkiwitzMurphy(17)}
moved away from the total causal effect by defining more general super
population estimands. See also the very flexible \cite{Robins(99graph)} and 
\cite{LuoSmallLiRosenbaum(12)}. \ The causal effects they consider are a
special case of our $p$ lag causal effect. \cite{BlackwellGlynn(16)} defined
the \textquotedblleft blip\textquotedblright\ effect as a special case of
the $p>0$ lag causal effect, and their \textquotedblleft
contemporaneous\textquotedblright\ effect as a special case of our
contemporaneous causal effect when the weights equal the reciprocal adapted
propensity. \ \cite{BoruvkaAlmirallWitkiwitzMurphy(17)} defined the
\textquotedblleft lagged effect\textquotedblright\ as our $p$ lag causal
effect with the weights equal to the reciprocal adapted propensity.

Inference in Robins work rely on combining marginal structural models (MSM)
with inverse probability weighting or an application of the g-formula %
\citep{robins1986new} which leverages the entire observed joint distribution
to estimate causal effects \citep{RobinsGreenlandHu(99)}. When viewed from a
finite population perspective, using MSM imposes assumptions on the
underlying potential outcomes. For example, equation 12 of \cite%
{robins2000marginal} asserts that the marginal outcome at time $T$ is a
function of the number of times a unit was assigned to treatment, $%
\sum_{t=1}^{T}W_{t}$, and this implies that the number of potential outcomes
at time $T$ is only $T+1$ rather than $2(2^{T}-1)$. Making the MSM more
complicated does increase the number of potential outcomes, but limits the
ability to easily estimate it.

\subsection{Sims impulse response function}

\cite{Sims(80)} measures \textquotedblleft causal effects\textquotedblright\
using an impulse response function (IRF) (see also the related \cite%
{Ramey(16)}, \cite{PlagborgMoller(16)}, \cite{StockWatson(17)}), a device
which has been very influential in macroeconomics. \ In our structure the
IRF measures: 
\begin{equation*}
IRF_{t,s}=\mathrm{E}\left\{
Y_{t+s}(w_{1:t+s})|W_{1:t+s}=w_{1:t+s},Y_{1:t-1}\right\} -\mathrm{E}\left\{
Y_{t+s}(w_{1:t+s}^{\prime })|W_{1:t+s}=w_{1:t+s}^{\prime },Y_{1:t-1}\right\}
,
\end{equation*}%
where $w_{1:t-1}=w_{1:t-1}^{\prime }=0$, $w_{t+1:t+s}=w_{t+1:t+s}^{\prime
}=0 $, $w_{t}=1$, $w_{t}^{\prime }=0$ and the expectation is with respect to
a model for $Y_{t+s}(u)|\left( W_{1:t+s}=u,Y_{1:t-1}\right) $ for each path $%
u$. \cite{Sims(80)} views treatments as impulses added to the innovations of
a time series model --- see Example \ref{defn potential autoregression}%
\footnote{%
Let $\kappa $ be a versor and $w=(\underset{1\times (t-1)}{0},\kappa
^{\prime },\underset{1\times s}{0})$ and $w^{\prime }=0$. Assume $%
Y_{1:T}(\bullet )$ follows a impulse moving average from Example \ref{defn
potential autoregression} and that $\varepsilon _{t}$ is a martingale
difference sequence. Then $IRF_{t,0}=\sigma \kappa $ and $IRF_{t,1}=\theta
\sigma \kappa $. In the potential autoregression case $IRF_{t,s}=\phi
^{s}\sigma \kappa $.}. He studies how these impulses spread over the
economy. \ Many of the predictive models used to implement these IRFs have
been linear and stationary, although some recent work has seen non-linearity
introduced through regime switching, or stochastic volatility. \ 

The Sims IRF connects with our $p$ lag causal effects. \ It differs as the
IRFs are defined as conditional expectations where the expectation is with
respect to a model. \ Implicitly the lagged causal effect weights are
determined by the time series model (e.g. \cite{KoopPeseranPotter(96)}, \cite%
{GallantRossiTauchen(93)}). \ Thus it is, in turn, related to \cite%
{BlackwellGlynn(16)} and so has some links to \cite{robins1986new}.

\subsubsection{Angrist, Kuersteiner, et al}

\cite{angrist2011causal} and \cite{angrist2016semiparametric} apply the
potential outcomes framework to testing the lagged causal effect of monetary
shocks using time series observational data. \ Let $d\in \left\{ 0,1\right\} 
$ denote two possible treatments at time $t$ and, in this exposition,
suppress regressors. \ 

Starting at time $t$, \cite{angrist2011causal} generate two possible
treatment paths through the dynamic $W_{t+k|t}(d)=D_{t,t+k}\{Y_{1:t}^{\text{%
obs}}$, $W_{1:t-1}^{\text{obs}}$, $W_{t}=d$, $Y_{t+1:t+k-1|t}(d)$, $%
W_{t+1:t+k-1|t}(d)$, $\varepsilon _{t+k|t}\}$, where $k=1,2,3,...$ and $%
D_{t+k|t}$ is some (given information at time $t$) non-stochastic function.
\ Crucially, $\varepsilon _{t+k|t}$ does not vary with $d$. Angrist and
Kuersteiner play out two \textquotedblleft potential
outcomes\textquotedblright\ paths through the dynamic, for $k=0,1,2,3,...$, $%
Y_{t+k|t}(d)=G_{t+k|t}\{Y_{1:t-1}^{\text{obs}}$, $W_{1:t-1}^{\text{obs}}$, $%
W_{t}=d$, $Y_{t:t+k-1|t}(d)$, $W_{t+1:t+k-1|t}(d)$, $\eta _{t+k|t}\}$ is
some (given information at time $t$) non-stochastic function (at no point do
we need to know $G_{t+k|t}$). \ Throughout $\varepsilon _{t+k|t}$ and $\eta
_{t+k|t}$ are mutually and temporally independent. For Angrist and
Kuersteiner the causal effect at time $t$ is $Y_{t+k|t}(1)-Y_{t+k|t}(0)$.

It looks like a $p$ lag $q$ step causal effect, but it is different. \ They
keep $\varepsilon _{t+1:t+k|t}$ invariant across the two treatment paths but
this does not guarantee (as we do) that the actual treatments $W_{t+1:t+k|t}$
are the same across $Y_{t+k|t}(1)$ and $Y_{t+k|t}(0)$ for $k=1,2,...$. Hence
in principle the spirit of their approach and ours are related, but the
causal effects are different. Instead, their causal effects are a deepening
of the idea of an IRF. \ Angrist and Kuersteiner explicitly make this strong
link to IRFs in their paper. \ \cite{angrist2016semiparametric} use the same
potential outcomes formulation from \cite{angrist2011causal} to estimate the
average lagged treatment effect using an inverse propensity score weighting
similar in spirit to the work of \cite{RobinsGreenlandHu(99)}.\ \ 

\subsection{Other core approaches}

\subsubsection{Granger causality}

A less strong connection can be made to \cite{Granger(69)} causality, in the 
\cite{Chamberlain(83)} sense. \ Granger causality is used in Assumption \ref%
{Assum:non-anticipating gen}, but there is no direct connection between our
causal effects and Granger causality. \ Comparisons between potential
outcome and predictive approaches to causality are made in \cite{Lechner(11)}%
.

\subsubsection{Highly structured models}

A more remote connection is the literature on inferring causality via highly
structured models, where some equilibrium mechanism is imposed a priori. \
Examples include stochastic general equilibrium models (e.g. \cite%
{HerbstSchorfheide(15)}), behavioral game theory models (e.g. \cite%
{ToulisParkes(16)}) and reinforcement learning (e.g. \cite{Gershman(17)}). 
\cite{HarveyDurbin(86)}, \cite{Harvey(96)} and\ \cite%
{BondersenGallusserKoehlerRemyScott(15)} use state space models to study
interventions. \ Other work on interventions includes \cite{BoxTiao(75)}. \
\ 

\subsubsection{Natural\ experiments}

In macroeconomics there is a literature on identifying causal effects of
treatments through \textquotedblleft natural\textquotedblright\ experiments.
\ This was pioneered by \cite{RomerRomer(89)}, other examples include \cite%
{CochranePiazzesi(02)} and \cite{BernankeKuttner(05)}. \ The econometrics is
discussed by \cite{StockWatson(17)}. \ In our structure their basic causal
focus is on $\Theta _{t,s}=\mathrm{E}\left\{
Y_{t+s}(w_{1:t+s})|W_{1:t+s}=w_{1:t+s}\right\} -\mathrm{E}\left\{
Y_{t+s}(w_{1:t+s}^{\prime })|W_{1:t+s}=w_{1:t+s}^{\prime }\right\} $, where $%
w_{1:t-1}=w_{1:t-1}^{\prime }=0$, $w_{t+1:t+s}=w_{t+1:t+s}^{\prime }=0$, $%
w_{t}=1$, $w_{t}^{\prime }=0$ and the expectation (which is assumed to
exist) is with respect to $Y_{t+s}(u)|\left( W_{1:t+s}=u\right) $ for each
path $u$. Then the authors usually assume conditional expectations are
linear and temporally invariant in treatments, so $\mathrm{E}\left\{
Y_{t+s}(w)|W_{1:t+s}=w_{1:t+s}\right\} =\alpha _{s}+\beta _{s}w_{t}$, which
means that $\Theta _{t,s}=\beta _{s}W_{t}$. They then estimate $\beta _{s}$
by regressing $Y_{t+s}^{\text{obs}}$ on $W_{t}^{\text{obs}}$ and then view $%
\beta _{s}$ as \textquotedblleft causal\textquotedblright . \ \cite%
{StockWatson(17)} discuss many extensions to this basic framework.


\section{Multiple units\label{Muliple units}}

\subsection{Basic structure}

Our focus has been on experimenting on one unit over time. We now propose
three methods for generalizing our experimental and inference approaches to
multiple units who are receiving the same class of treatment over time. \
The literature on longitudinal studies is very large, and important work
related to our own includes, for example, \cite{robins1986new}, \cite%
{RobinsGreenlandHu(99)}, \cite{AbbringBerg(03)}, \cite{Lok(08)}, \cite%
{Lechner(09)}, \cite{BoruvkaAlmirallWitkiwitzMurphy(17)} and \cite%
{RicciardiMatteiMealli(16)}. \ Our approach differs as it is distinctly
model free. \ 

Let time vary over the interval $1\leq t\leq T$, while the different units
are indexed as $i=1,2,...,n$. Let $N_{t,i}$ count the number of observations
for the $i$-th unit up to time $t$. \ This notation allows units to start
and stop their experiments at different times. For the $i$-th unit, the time
of $j$-th treatment and value of treatment are written as $t_{j,i}\in
\lbrack 1,T]$, $W_{t_{j,i},i}\in \left\{ 0,1\right\} $, $j=1,2,...,N_{T,i}$.
Throughout we will think of the times $\left\{ t_{j,i}\right\} $ as
non-stochastic (or that we can make inference conditional on them) and will
write the collection of all times in our sample as $\mathcal{T}_{T,n}$. \ We
collect treatment path up to time $t$ as $%
W_{1:t,i}=(W_{t_{1,i},i},...,W_{t_{N_{t,i}},i})$, $W_{1:t}=\left\{
W_{1:t,1},...,W_{1:t,n}\right\} $. Below we use the stochastic process
notation, for an arbitrary series $\left\{ X_{s},s\in \lbrack 1,T]\right\} $%
, that $X_{t-}=\lim_{\varepsilon \downarrow 0}X_{t-\varepsilon }$. \ Now we
state two types of assumptions.

\begin{assumption}[Temporal stable unit treatment value assumption]
\ For all $j=1$, ...,$N_{T_{n}}$, and $i=1$, ...,$n$, $Y_{j,i}(\bullet
)=\left\{ Y_{j,i}(w_{1:j}):w_{1:j}\in \left\{ 0,1\right\} ^{j}\right\} $.
\end{assumption}

This setup sits on the shoulders of \cite{Cox(58book)} and \cite{Rubin(80)}.
\ It says that the $j,i$-th potential outcome only functionally depends upon
the $i$-th individual's treatment path. \ 

We collect terms as $Y_{1:t,i}(\bullet )$=$\{Y_{1,i}(\bullet
),...,Y_{N_{t},i}(\bullet )\}$, $Y_{1:t}(\bullet )=\left\{ Y_{1:t,1}(\bullet
),...,Y_{1:t,n}(\bullet )\right\} $.

\begin{assumption}
For all $t\in \mathcal{T}_{T,n}$ and $w_{1:t}$, $\Pr
(W_{1:t}=w_{1:t}|W_{1:t-}=w_{1:t-},Y_{1:T}(\bullet ))=\Pr
(W_{1:t}=w_{1:t}|W_{1:t-}=w_{1:t-},Y_{1:t-}(w_{1:t-}))$.
\end{assumption}

We assume this probability is bounded away from 0 and 1. \ This
non-anticipating structure allows the treatment or outcome of one series to
potentially change the chance another series is treated in the future. \
There is no need to assume that $Y_{1:t,i}(\bullet )$ and $Y_{1:t,k}(\bullet
)$ are independent for $i\neq k$. Instead this structure allows the use of
randomization based inference, conditioning on all the potential outcomes,
but this time in the context of multiple units.

\subsection{Aggregating}

In time series experiments $N_{T,i}$ is often larger than $n$. \ Here we
combine information across units to gain more accurate estimates of the
effect of treatment, without adding assumptions. \ 

For unit $i$, let $\bar{\tau}_{p,i}$ be the average $p$ lagged effect of
treatment. Then the weighted population averaged $p$ lagged effect of
treatment is given by, $\overline{\tau }_{p}=\sum_{i=1}^{n}c_{p,i}\bar{\tau}%
_{p,i}$, $c_{p,i}>0$, $\sum_{i=1}^{n}c_{p,i}=1$, where $\left\{
c_{p,i}\right\} $ are non-stochastic. \ We will focus on the equally
weighted case $\overline{\tau }_{p}=\frac{1}{n}\sum_{i=1}^{n}\bar{\tau}%
_{p,i} $. \ This can be interpreted as the average effect of the
intervention across time and units. \ The $q>0$ stepped version is defined
in the analogous way.

\subsubsection{Randomization}

An almost identical procedure to the one described in Section \ref%
{SSS:rand_test}, for conduct a randomization-based hypothesis test for the
sharp null of no unit level treatment effect, can be applied to the multiple
unit scenario. \ The difference is that at each step of the procedure we
sample a new treatment path for each of the units and can compute a pooled
statistic, e.g. 
\begin{equation}
\hat{\bar{\tau}}_{p}=\frac{1}{\gamma _{p}^{2}}\sum_{i=1}^{n}\frac{\hat{\bar{%
\tau}}_{p,i}}{\gamma _{p,i}^{2}/(T_{i}-p)},
\label{E:pooled_estimator_fisher}
\end{equation}%
where $\gamma _{p,i}^{2}$ is the known variance, given in (\ref%
{E:null_variance_k}), and $\gamma _{p}^{2}=\sum_{i=1}^{n}(T_{i}-p)/\gamma
_{p,i}^{2}$. Again $\hat{\bar{\tau}}_{p}$ is a conditionally unbiased
estimator of $\overline{\tau }_{p}$, whatever the dependence across units. \
\ 

\subsubsection{Conservative test}

\label{sub:conservative_test} 

The no average temporal causal effects null hypothesis (\ref{Neyman null})
style conservative test, described in Section \ref{SS:Neyman}, also
generalizes to the multiple unit setting. The null hypothesis now assumes
that each unit level average $p$ lagged effect of treatment is equal to
zero. \ Under this hypothesis the exact variance is not assumed known,
instead we replace all of the variance terms in (\ref%
{E:pooled_estimator_fisher}) by their estimates, as given in Lemma \ref%
{L:var_bound_p}. \ The pooled estimator is then%
\begin{equation}
\hat{\bar{\tau}}_{p}=\frac{1}{\hat{\gamma}_{p}^{2}}\sum_{i=1}^{n}\frac{\hat{%
\bar{\tau}}_{p,i}}{\hat{\gamma}_{p,i}^{2}/(T_{i}-p)},
\label{E:pooled_estimator}
\end{equation}%
where $\hat{\gamma}_{p,i}^{2}$ is an estimate of the variance, given in
Theorem \ref{T:p-lag}, and $\hat{\gamma}_{p}^{2}=\sum_{i=1}^{n}(T_{i}-p)/%
\hat{\gamma}_{p,i}^{2}$. A simple reference distribution of this estimator
can be calculated if treatment paths are independent over units. \ This is
formalized below. \ \ 

\begin{assumption}
\label{Assump: indepent treatment}For all $t\in \mathcal{T}_{T,n}$, then,
for all $w_{0:t}$, 
\begin{equation*}
\Pr \left( W_{0:t,i}=w_{0:t,i}|W_{0:t-}=w_{0:t-},Y_{0:T}(\bullet )\right)
=\Pr \left(
W_{0:t,i}=w_{0:t,i}|W_{0:t-,i}=w_{0:t-,i},Y_{0:t-,i}(w_{0:t-,i})\right) .
\end{equation*}
\end{assumption}

Then (\ref{E:pooled_estimator}) can be compared to $N\left( 0,\hat{\gamma}%
_p^{-2}\right) $. \ 

\subsubsection{Fisher's method}

\label{sub:fisher_s_method}

The unit level hypothesis tests described in Section \ref{SSS:rand_test} and %
\ref{SS:Neyman} yield $n$ independent $p$-values under Assumption \ref%
{Assump: indepent treatment}. \ We can combine them using Fisher's method
for combining independent $p$-values testing the same null hypothesis. \
Under the null, let $X^{2}=-2\sum_{i=1}^{n}\log (p_{i})\sim \chi _{2n}^{2}$,
where $p_{i}$ is the $i$-th $p$-value obtained from the $i$-th unit. \ The
reference distribution of this statistic, under the null, is asymptotically $%
\chi _{2n}^{2}$. \ When the variance of each of the estimates is different,
then this will not be the most powerful test. \ Alternatives exist, such as
the one proposed in \cite{Jordan(95)}.


\section{Simulation study\label{sect:simulation study}}

Our aim for this small simulation study is threefold. \ First, we want to
show that the CLT provides a reasonable guide for relatively low sample
sizes. \ Second, we want to show that our conservative test procedure has
approximately the correct type I error rates. \ Third, we want to show that
our proposed tests have good power.

\subsection{Study design}

The general causal effect for a potential autoregression given in Example %
\ref{defn potential autoregression} is 
\begin{equation*}
\tau _{t}(w_{1:t},w_{1:t}^{\prime })=Y_{t}(w_{1:t})-Y_{t}(w_{1:t}^{\prime
})=\left( \mu _{w_{1:t}}-\mu _{w_{1:t}^{\prime }}\right) +\phi \tau
_{t-1}(w_{1:t-1},w_{1:t-1}^{\prime })+\left( \sigma _{w_{1:t}}-\sigma
_{w_{1:t}^{\prime }}\right) \varepsilon _{t}.
\end{equation*}%
When $\mu _{w_{1:t}}=\mu _{w_{t}}$ and $\sigma _{w_{1:t}}=\sigma _{w_{t}}$,
then the lagged causal effect, given in Definition \ref{defn:lag cause
effect}, is $\tau _{t,p}(\{1\},\{0\})=\phi ^{p}\left\{ \left( \mu _{1}-\mu
_{0}\right) +\left( \sigma _{1}-\sigma _{0}\right) \varepsilon
_{t-p}\right\} $, for $p=0,1,2,...$, for any choice of weights.

\begin{example}
Consider a univariate impulse potential autoregression with $T=100$, $\mu
_{1}=0.5$, $\mu _{0}=0$, $\phi =0.5$, $\sigma =1$, $\varepsilon _{t}\overset{%
iid}{\sim }N(0,1)$. \ Figure \ref{fig:sim_study_sample} (left), shows $%
Y_{t}^{\text{obs}}$ together with a symbol for $W_{t}^{\text{obs}}$ which is
either 0 (control) or 1 (treatment). \ The orange dotted line shows $%
\overline{\tau }_{0}$, which is around $0.8$. \ \ \ \ 
\begin{figure}[h]
\centering%
\begin{minipage}[t]{0.48\textwidth}
\includegraphics[width=\textwidth]{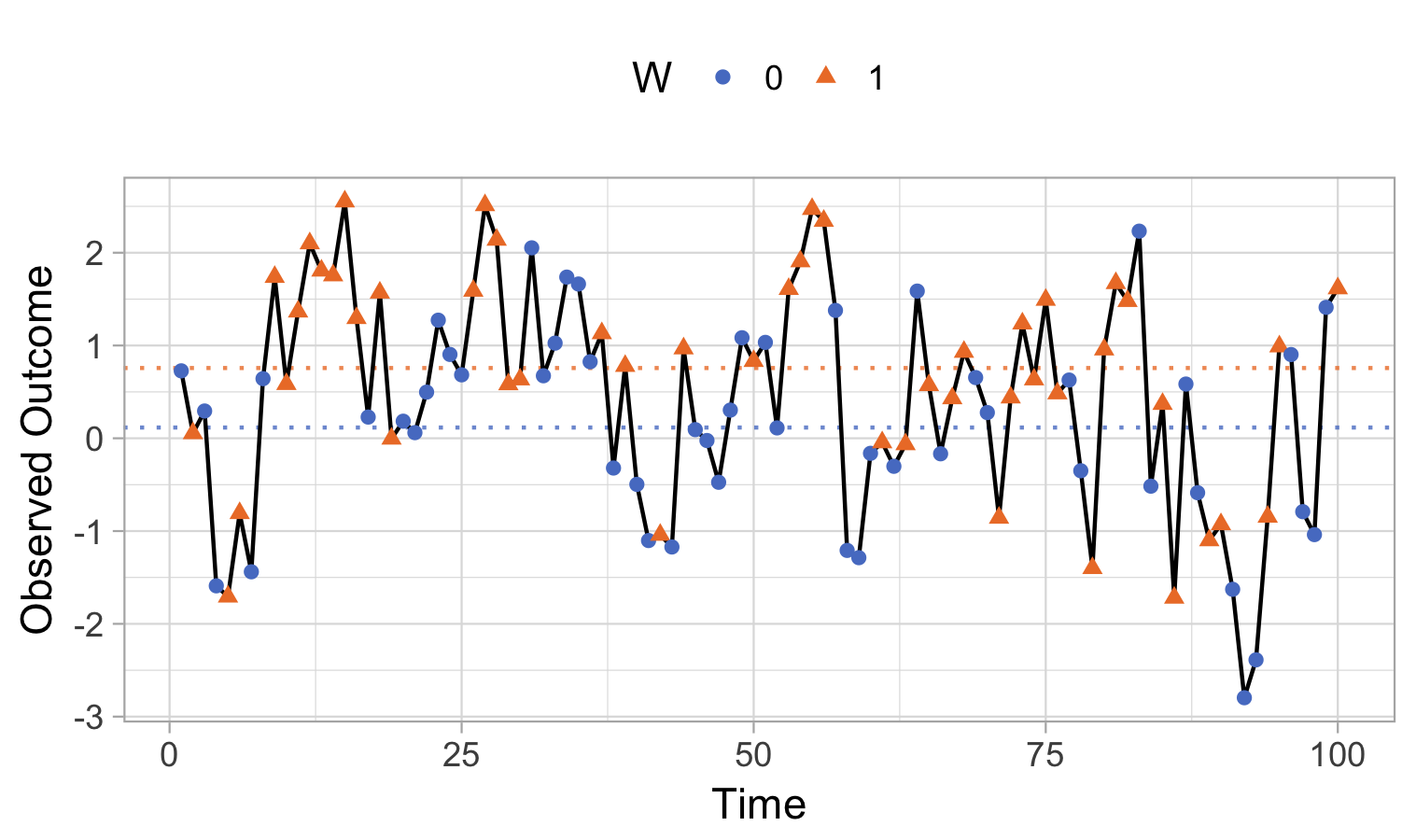}
Observed outcomes: $Y_t^{\text{obs}}$
\end{minipage}%
\begin{minipage}[t]{0.48\textwidth}
\includegraphics[width=\textwidth]{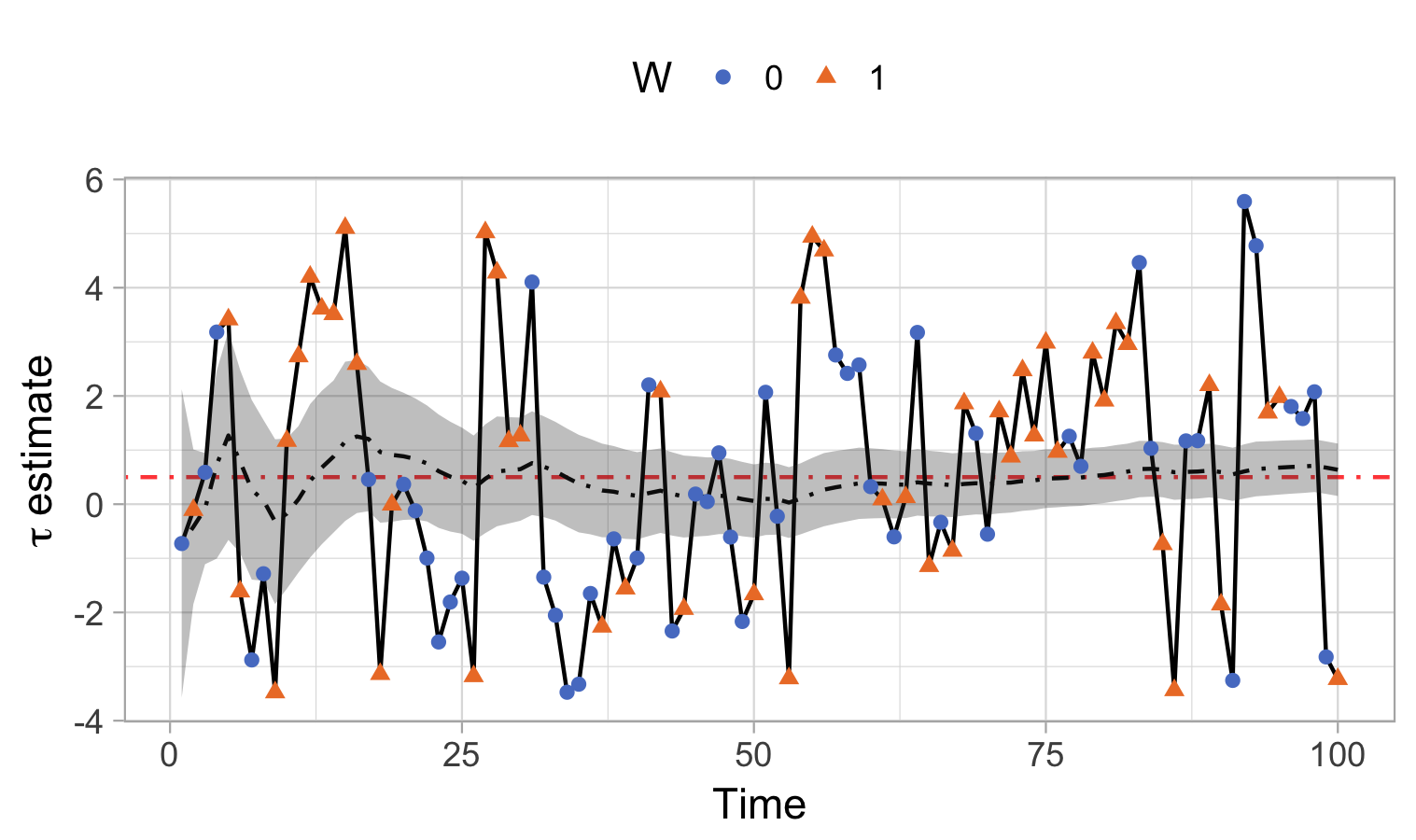}
Estimates $\hat{\tau}_{t,0}$ and $\hat{\overline{\tau}}_t$ and estimand ${\overline{\tau}}$
\end{minipage}\newline
\caption{Left: $Y_{1:T}^{\text{obs}}$ from a potential autoregression with $%
\protect\mu _{1}=0.5$, $\protect\mu _{0}=0$, $\protect\phi =0.5$, $\protect%
\sigma _{1}=\protect\sigma _{0}=1$, the color indicate the received
treatment and the dotted lines are the observed average. Right: the estimate
of $\hat{\protect\tau}_{t,0}$ at each point, the black dotted line is the
running average of the estimates, the gray polygon is the corresponding 95\%
confidence interval and the red dotted line is the true value of $\bar{%
\protect\tau}_{0}$.}
\label{fig:sim_study_sample}
\end{figure}
Figure \ref{fig:sim_study_sample} (right), shows $\widehat{\tau }_{t,0}$
plotted against time. \ Also plotted is the sequential average of these
differences and a 95\% confidence interval. \ The plotted orange line, which
again indicates $\overline{\tau }_{0}$, is close to $\widehat{\overline{\tau 
}}_{0}$.
\end{example}

\subsection{Simulation results}

We study the sampling performance of our testing procedures using a
potential autoregression with $\mu _{w_{1:t}}=\mu _{w_{t}}$, $\sigma
_{w_{1:t}}=\sigma _{w_{t}}$, $\phi =0.5$, $\sigma _{1}=\sigma _{0}=1$ and%
\begin{equation}
\text{Null}\text{: }\mu _{1}=\mu _{0}=0,\quad \text{Alternative}\text{: }\mu
_{1}=0.2,\quad \mu _{0}=0.  \label{potential AR}
\end{equation}%
We look at where $\varepsilon _{t}\overset{iid}{\sim }N(0,1)$ and $%
\varepsilon _{t}\overset{iid}{\sim }$ Cauchy. \ The latter produces heavy
tailed data.

\subsubsection{Fixing the potential outcomes}

\begin{figure}[h]
\centering%
\begin{minipage}[t]{0.31\textwidth}
$\protect\widehat{\bar{\protect\tau}}_{0}$
\includegraphics[width=\textwidth]{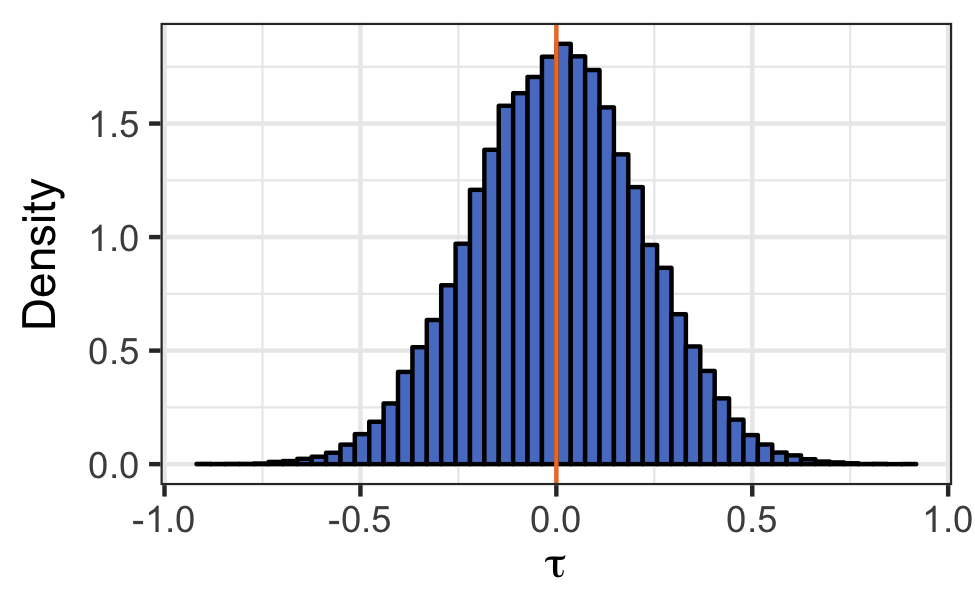}
\end{minipage}%
\begin{minipage}[t]{0.31\textwidth}
$\protect\widehat{\bar{\protect\tau}}_{0}$
\includegraphics[width=\textwidth]{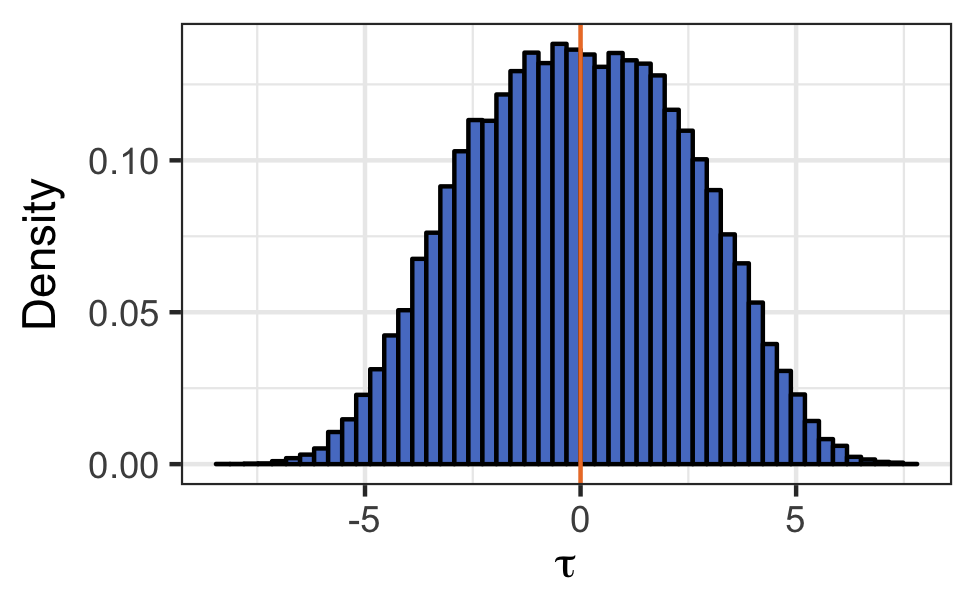}
\end{minipage}%
\begin{minipage}[t]{0.31\textwidth}
$\protect\widehat{\bar{\protect\tau}}_{0}$
\includegraphics[width=\textwidth]{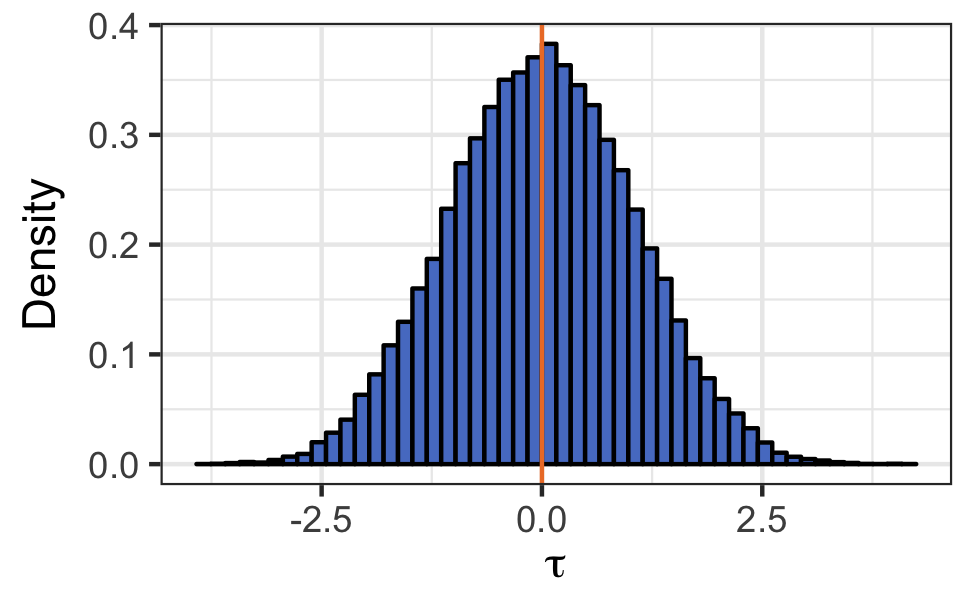}
\end{minipage}\newline
\begin{minipage}[t]{0.31\textwidth}
\includegraphics[width=\textwidth]{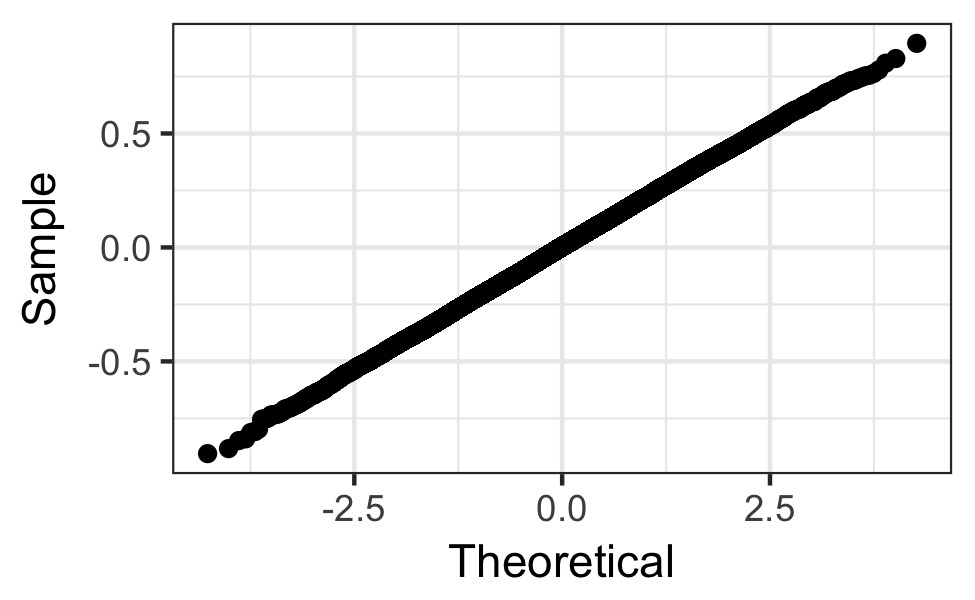}
$N(0,1), T=100$
\end{minipage}%
\begin{minipage}[t]{0.31\textwidth}
\includegraphics[width=\textwidth]{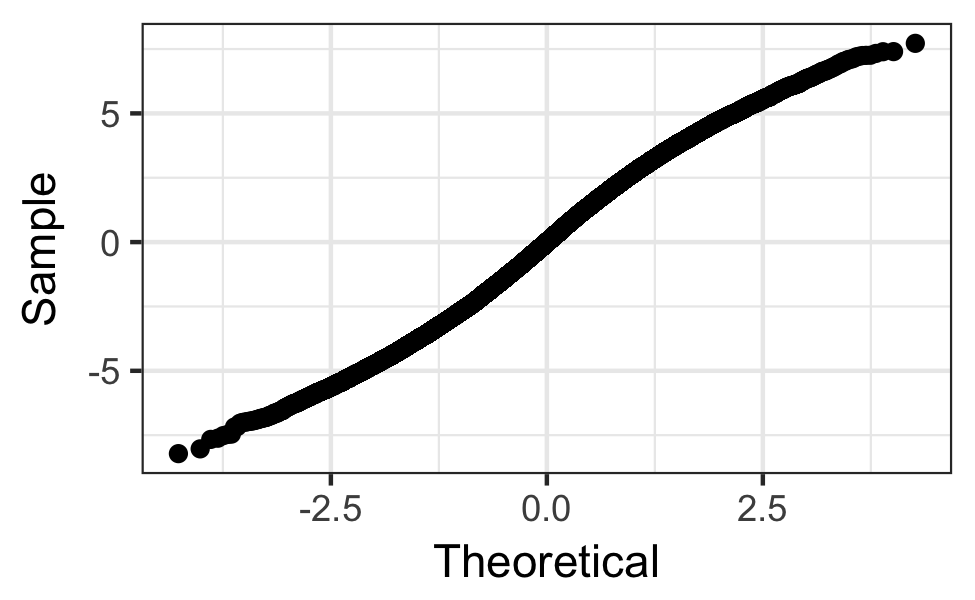}
Cauchy, $T=1,000$ 
\end{minipage}%
\begin{minipage}[t]{0.31\textwidth}
\includegraphics[width=\textwidth]{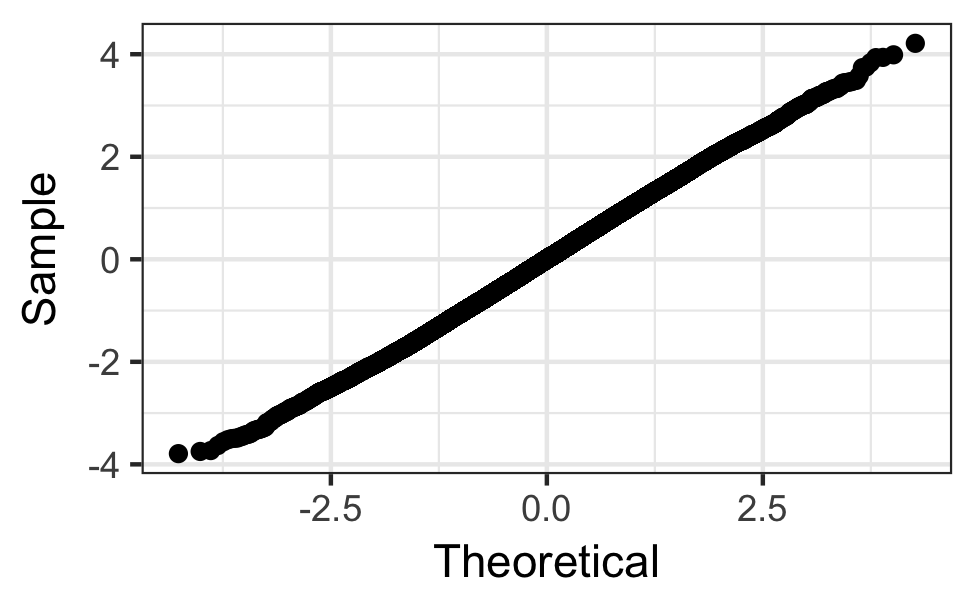}
Cauchy, $T=10,000$ 
\end{minipage}
\caption{Sampling behavor of estimators. \ Top: histograms of $\protect%
\widehat{\bar{\protect\tau}}_{0}$ over $W_{1:T}|Y_{1:T}(\bullet )$ based on
conditioning on a single draw from $Y_{1:T}(\bullet )$, i.e. fixing the
potential outcomes. Bottom: Q--Q normal plot. For heavy tailed distributions
a larger sample is need to achieve asymptotic normality. The vertical orange
line represents the true value of the estimand. }
\label{fig:CLT1}
\end{figure}

To illustrate the central limit approximation we first generate the
potential outcomes and then, fixing $Y_{1:T}(\bullet )$, simulate over
different $W_{1:T}$. Figure \ref{fig:CLT1} shows our results for the
randomization distribution of $\widehat{\overline{\tau }}_{0}$ given in (\ref%
{E:1-step_estimator}). When $\varepsilon _{t}\overset{iid}{\sim }N(0,1)$ the
estimates obtained from different treatment paths quickly converge to a
normal distribution. When $\varepsilon _{t}\overset{iid}{\sim }$ Cauchy a
longer experimental time is require to reach approximate normality. This is
due to the heavy tails of the noise. Figure \ref{fig:CLT_dif_est}, in the
web appendix, shows examples of $\widehat{\overline{\tau }}_{1}$, $\widehat{%
\overline{\tau }}_{0}^{(1)}$ and $\widehat{\overline{\tau }}_{1}^{(1)}$ also
appear roughly Gaussian for $T=100$, for different values of $\mu _{t}$. 

\subsubsection{Replicating over potential outcomes\label{sect:power simul}}

We also need to evaluate the frequentist properties over the sampling
distribution -- explicitly averaging over the potential outcomes using the
potential autoregression. To this end, we generated 50,000 independent
copies of $Y_{1:T}$, under the null assumption of no treatment effect, and
computed the randomization based $p$-value distribution for $\widehat{%
\overline{\tau }}_{0}$. \ For each replication the tests are exact, and the $%
p$-values follows a discrete uniform distribution. \ Of more interest is
that we also applied the conservative test which relies on the approximate
normality of $\widehat{\overline{\tau }}_{0}$.

\begin{figure}[h]
\centering%
\begin{minipage}[t]{0.31\textwidth}
\includegraphics[width=\textwidth]{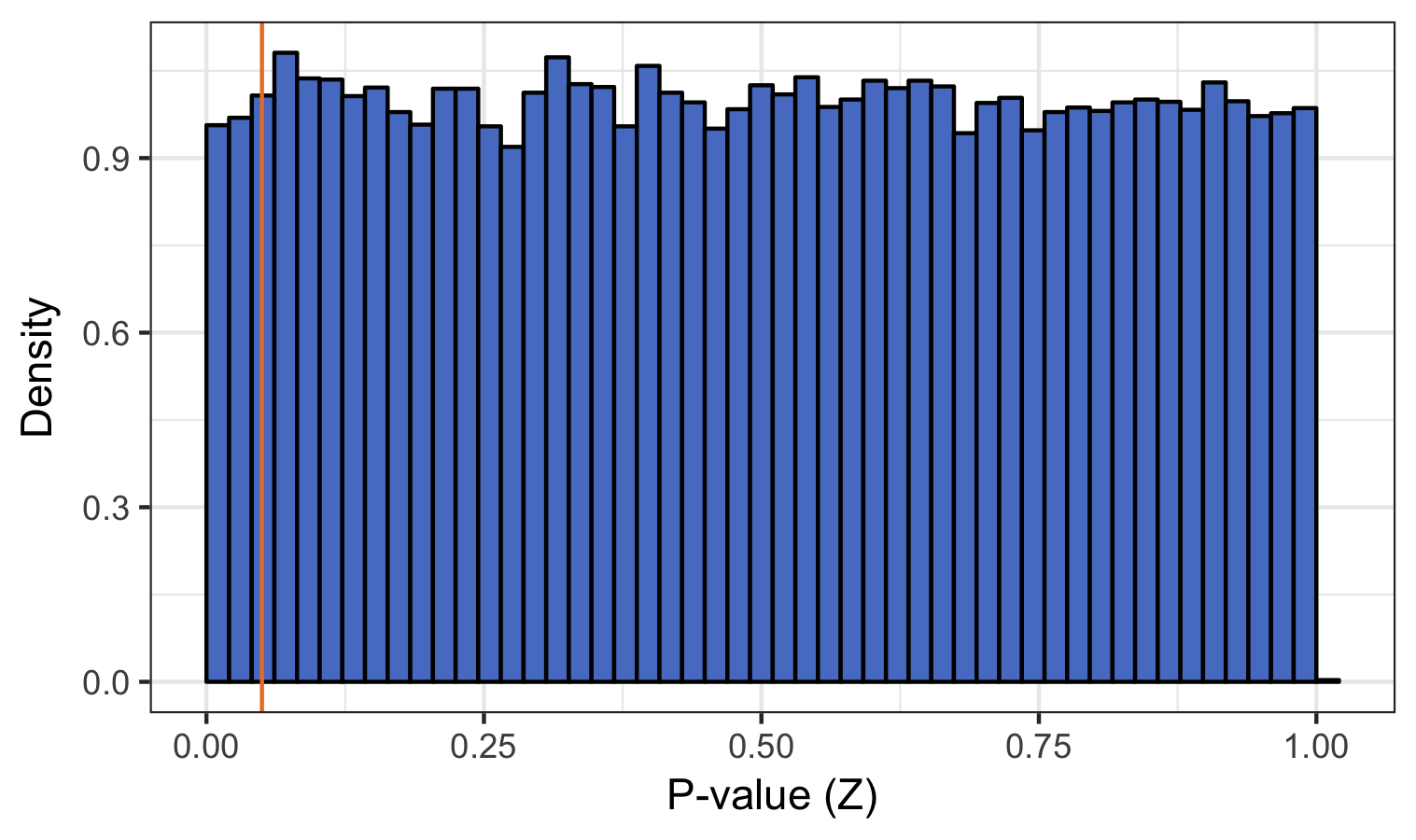}
Conservative test: $\widehat{\bar{\protect\tau}}_{0}$ 
\end{minipage}%
\begin{minipage}[t]{0.31\textwidth}
\includegraphics[width=\textwidth]{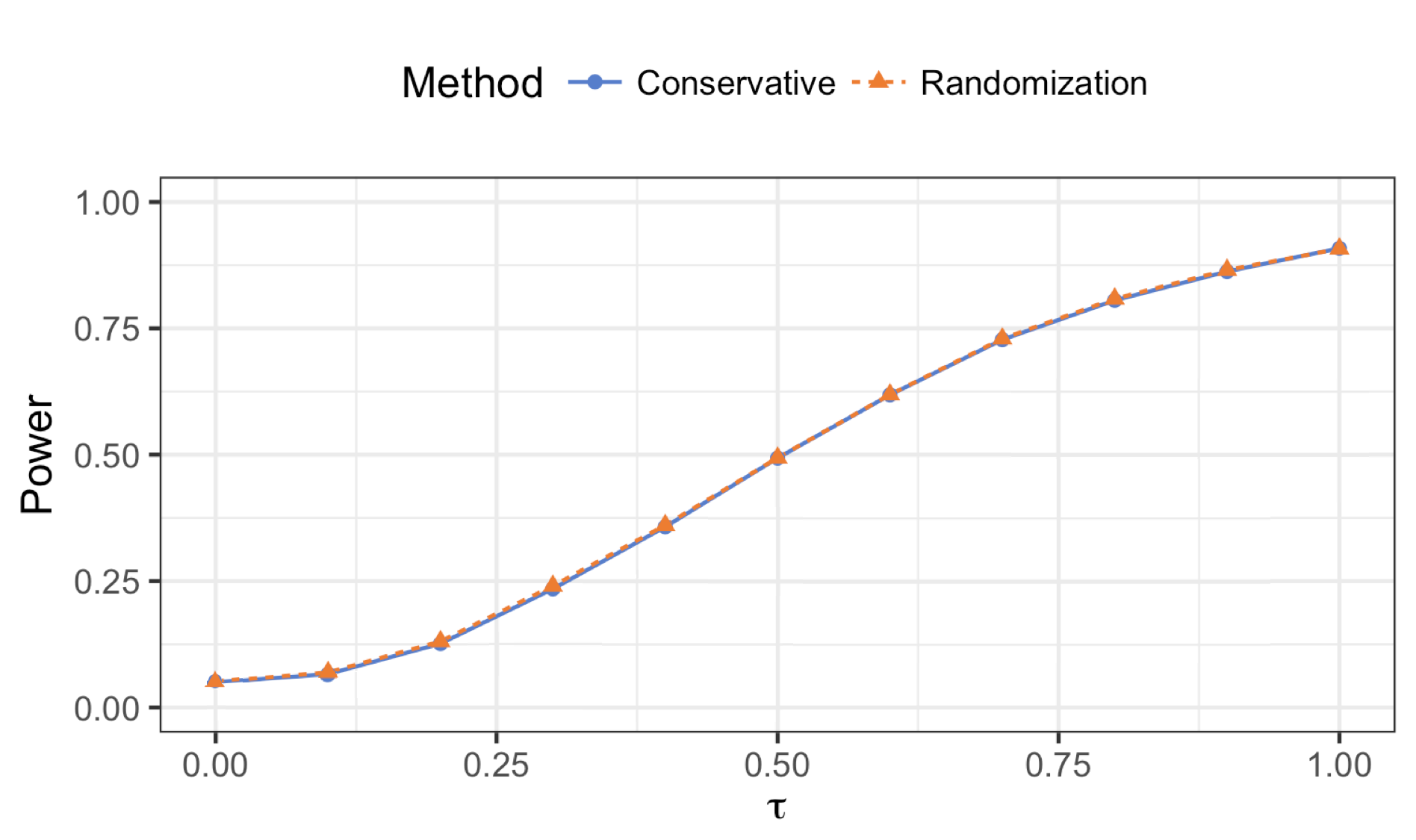}
For $\widehat{\bar{\protect\tau}}_{0}$ as $\bar{\protect\tau}_{0}$ varies 
\end{minipage}%
\begin{minipage}[t]{0.31\textwidth}
\includegraphics[width=\textwidth]{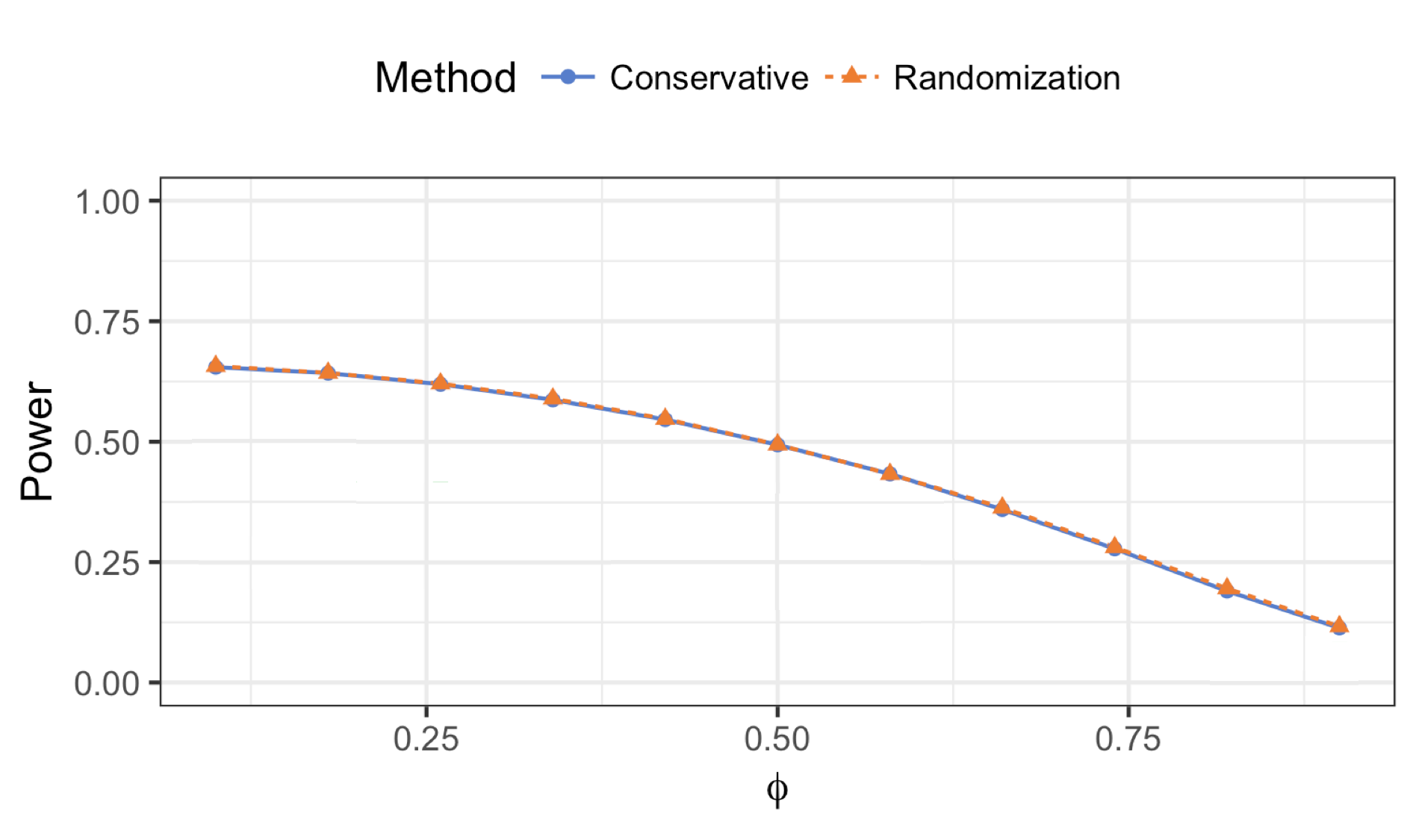}
For $\widehat{\bar{\protect\tau}}_{1}$ as $\phi$ varies
\end{minipage}\newline
\caption{Left: distributions of $p$-values over $50,000$ simulations for $%
T=100$ when there is no treatment effect, using the conservative test using
the normal approximation. The orange line represents the $0.05$ cut off.
Middle and right: The relative power for the tests as $\protect\tau $
increases for fixed $\protect\phi =0.5$ (middle) and as the $\protect\phi $
parameter increases for fixed $\protect\tau =0.5$ (right). The conservative
test performs only slightly worse that the exact randomization test. The
noise distribution is $N(0,1)$.}
\label{fig:power}
\end{figure}

The left hand side of Figure \ref{fig:power} shows the $p$-value
distributions for $T=100$ in the conservative test cases. \ We omit the
results for the exact unstandardized test. \ The $p$-values obtained from
the conservative tests and the randomization test show similar patterns, but
the conservative test has slight lower $\alpha $ level due to the
overestimation of the variance. Figure \ref{fig:power} also shows the
relative power of the two methods as the treatment effect increases for $%
\widehat{\overline{\tau }}_{0}$, holding all other factors fixed. \ Notice
how the conservative test is only slightly less powerful than the
randomization based test. The second figure shows the relative power of the
two methods as a function of the $\phi $ parameter for the 1 lag case.

\subsection{Pooled estimation: averaging over multiple units}

To investigated the behavior of the pooled estimator, we generated two
independent experiments for $T_{1}=T_{2}=100$ and combined them, as
described in (\ref{E:pooled_estimator_fisher}). Figure \ref{fig:CLT_pool},
in the web appendix, shows how the distribution of the pooled estimator is
well approximated using the CLT. \ The variance of the pooled estimator is
lower than that obtained from each of the individual experiments.

\section{Empirical example from finance\label{sect:empirical example}}

\subsection{Trading futures contracts}

Here we analyze experiments carried out by AHL Partners, a quantitative
trading group that mainly trades in financial futures. Within each futures
market their desired positions are decided solely by algorithms and
currently available financial data. Once their target position changes, they
need to carry out \textquotedblleft an order\textquotedblright\ within a
prescribe time period e.g. buy \$20 million of gold futures over the next
trading day. These orders are often large, and they make such orders
frequently, so trading well is vital to their performance as a group.

The firm has two ways of executing an order, either using a human trader or
an algorithm. Both methods typically avoid executing the whole order in one
go as this could deliver a poor execution price. Instead, they break up the
order into smaller pieces and make a sequence of trades to \textquotedblleft
fill\textquotedblright\ the order. As they trade the price will often move
against them, i.e. rise if they are executing a series of buys, or fall if
they are carrying out a series of sells. \ However, these rises and falls
are somewhat masked by the general volatility in the market as filling the
order takes time.

This group allocates the subset of orders which have order size between a
fixed lower and upper bound (these bounds are time invariant in the
experiments covered by our data, but differ over the markets. \ For
confidentiality reasons we are unable to reveal these bounds), to humans
traders and algorithmic traders randomly, to experimentally work out which
is more effective at trading their orders. The treatments were always i.i.d.
Bernoulli, ignoring past data. \ Hence this setup obeys our non-anticipating
treatments assumption.\ 
\begin{table}[t]
\centering
\par
{\footnotesize \label{t:summary} 
\begin{tabular}{r|rr|rr||rr|rr}
& \multicolumn{4}{c||}{} & \multicolumn{4}{c}{Median per Order} \\ \hline
& \multicolumn{2}{c|}{Prob of Treatment} & \multicolumn{2}{c||}{Numbers of
orders} & \multicolumn{2}{c|}{Number of trades} & \multicolumn{2}{c}{
Execution time} \\ 
& $<$July 12 & $>=$ July 12 &  &  &  &  & \multicolumn{2}{c}{(in minutes)}
\\ 
Market &  &  & A & B & A & B & A & B \\ \hline
1 & 0.50 & 0.25 & 147 & 105 & 62 & 132 & 29.7 & 6.6 \\ 
2 & 0.50 & 0.50 & 85 & 64 & 123 & 118 & 34.6 & 13.3 \\ 
3 & 0.50 & 0.25 & 281 & 95 & 109 & 108 & 44.1 & 14.3 \\ 
4 & 0.50 & 0.50 & 36 & 42 & 102 & 71 & 21.9 & 8.0 \\ 
5 & 0.25 & 0.25 & 81 & 22 & 118 & 103 & 40.9 & 10.9 \\ 
6 & 0.50 & 0.50 & 39 & 41 & 19 & 5 & 34.6 & 4.4 \\ 
7 & 0.50 & 0.50 & 118 & 129 & 82 & 71 & 28.9 & 5.0 \\ 
8 & 0.50 & 0.50 & 71 & 72 & 38 & 28 & 24.2 & 7.7 \\ 
9 & 0.50 & 0.50 & 178 & 239 & 26 & 15 & 19.6 & 0.6 \\ 
10 & 0.50 & 0.25 & 272 & 154 & 62 & 27 & 44.7 & 4.4%
\end{tabular}%
}
\caption{Summary information for the ten different markets in 2016, with
method A being control ($W_{t}=0$) and B being treatment ($W_{t}=1$). The
number of orders is the number of experiments conducted in 2016 in each
market. In three markets the probability of treatment was changed on
midnight 12 July 2016. \ }
\label{tab:sample sizes}
\end{table}

AHL Partners provided us with 10 sets of data from 2016; all are from
trading index equity futures markets, where the underlying indices are from
the US, Europe, and Asia. We will regard each of these 10 markets as
separate units. \ 

Table \ref{tab:sample sizes} provides the number of trades in each market
during 2016, the median number of trades per order and the median number of
minutes it takes to execute the order. \ In both cases the numbers are
broken down into the two trading methods. \ To protect their confidential
information they did not tell us which of methods \textquotedblleft
A\textquotedblright\ and \textquotedblleft B\textquotedblright\ correspond
to a human trader or algorithmic trader nor the identity of the individual
markets. \ Throughout we label method \textquotedblleft A\textquotedblright\
as Control ($W_{t}=0$) and method \textquotedblleft B\textquotedblright\ as
Treatment ($W_{t}=1$).

Table \ref{tab:sample sizes} shows that method \textquotedblleft
A\textquotedblright\ typically trades more often when filling an order, but
this varies over the market. \textquotedblleft A\textquotedblright\
generally fills the order roughly three times slower than method
\textquotedblleft B\textquotedblright . \ The group changed on midnight 12th
July 2016 the probability of allocating to the treatment, method
\textquotedblleft B\textquotedblright , in three of the markets. \ This is
detailed in the Table. \ 

\subsection{Definition of financial slippage}

The quality of each order's execution is measured by \textquotedblleft
slippage,\textquotedblright\ which is a simple function of the trade prices
and volumes at which the order was executed. \ Slippage will be the
\textquotedblleft outcome\textquotedblright\ $Y_{t}^{\text{obs}}$ from these
financial experiments. \ The traders would like $Y_{t}^{\text{obs}}$ to be
as low as possible (minimizing trading costs). \ It will be measured using a
signed volume weighted average price (VWAP) minus the mid-price scaled by
mid-price (e.g. \cite{BerkowitzLogueNoser(88)} and \cite%
{CalvoriCipolliniGallo(13)}). \ The details are in the Web Appendix \ref%
{sect:slip}. \ \ 
\begin{figure}[t]
\centering%
\begin{minipage}[t]{0.48\textwidth}
\includegraphics[width=\textwidth]{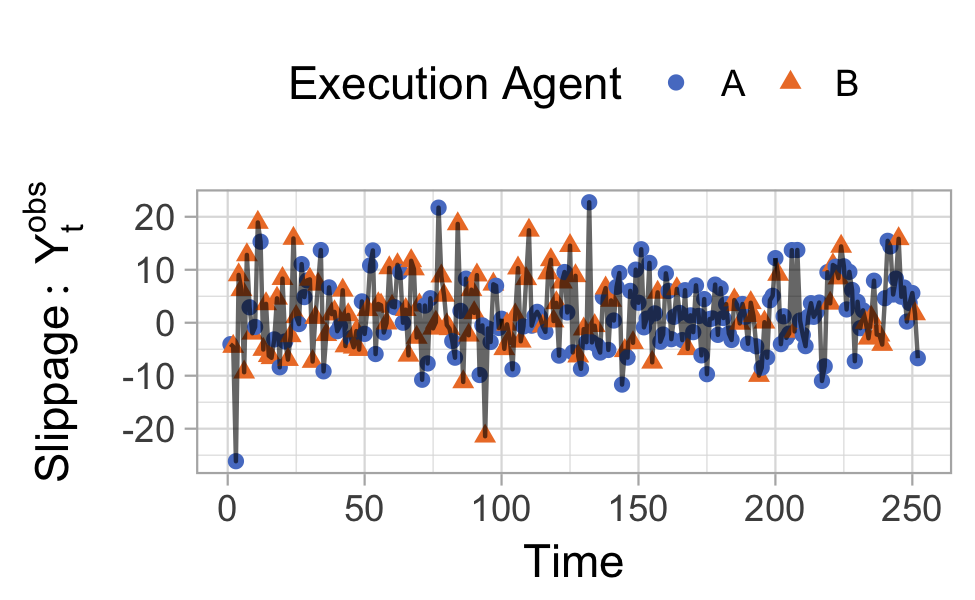}
\end{minipage}%
\begin{minipage}[t]{0.48\textwidth}
\includegraphics[width=\textwidth]{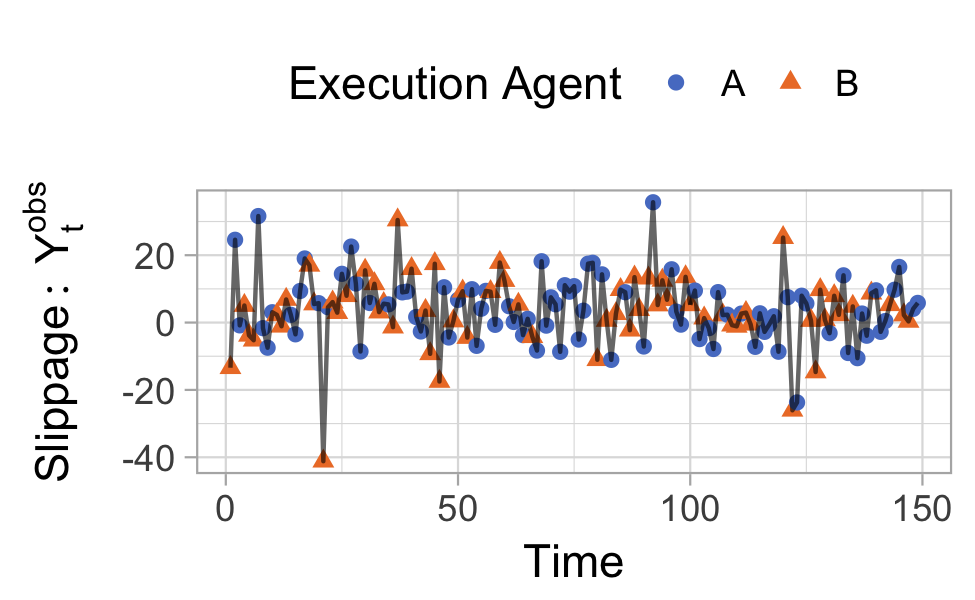}
\end{minipage}\newline
\caption{Left: The total slippage $Y_{t}^{\text{obs}}$ per order for Market
1. Orange indicates method \textquotedblleft A\textquotedblright\ and blue
\textquotedblleft B\textquotedblright . Right: The total slippage $Y_{t}^{%
\text{obs}}$ per order for Market 2.}
\label{fig:slippage_example}
\end{figure}

From these experiments we will use the time series of slippages $Y_{t}^{%
\text{obs}}$ as our primary outcome of interest, and the treatment $%
w_{t}^{obs}$ which will be 0 (the control) if method \textquotedblleft
A\textquotedblright\ is used to trade and a 1 (the treatment) if method
\textquotedblleft B\textquotedblright\ is used. The right plot of Figure \ref%
{fig:slippage_example} shows an example of the primary outcome for Markets 1
and 2, the remaining eight are given in the web appendix.

Taking a step back, modelling the dynamics of asset price returns is
challenging as returns are very thick tailed and exhibit long-range
volatility clustering (e.g. \cite{Taylor(05)}). \ Our approach, which just
uses the randomization of the experiment, does not need to make a stand on
modelling the dynamics of returns and hence is entirely objective. \ \ 

\subsection{Inference on $\protect\widehat{\bar{\protect\tau}}_{p}$}

Table \ref{tab:average slippage in practice} shows the average slippage of A
and B as well as the estimated $\widehat{\bar{\tau}}_{0}$, $\widehat{\bar{%
\tau}}_{1}$ and $\widehat{\bar{\tau}}_{2}$ and the corresponding $p$-values.
\ 
\begin{table}[t]
\centering
\par
{\footnotesize \label{t:rand_infer_res} 
\begin{tabular}{r|rr|rr|rr|rr}
\multirow{2}{*}{Market} & \multicolumn{2}{c|}{Average Slippage} & $\widehat{%
\overline{\tau }}_{0}$ & p.val & $\widehat{\overline{\tau }}_{1}$ & p.val & $%
\widehat{\overline{\tau }}_{2}$ & p.val \\ 
& A & B &  &  &  &  &  &  \\ \hline\hline
1 & 1.48 & 2.29 & 0.52 & 0.628 & -0.78 & 0.382 & -0.04 & 0.963 \\ 
2 & 4.11 & 3.35 & -1.80 & 0.315 & 0.53 & 0.771 & -1.86 & 0.300 \\ 
3 & -0.05 & 1.07 & 0.54 & 0.634 & -0.26 & 0.813 & 0.30 & 0.762 \\ 
4 & 3.38 & 3.23 & 0.36 & 0.900 & 2.10 & 0.450 & 3.11 & 0.283 \\ 
5 & 0.57 & 0.63 & -0.42 & 0.743 & -0.76 & 0.369 & -0.04 & 0.448 \\ 
6 & -1.48 & 3.72 & 5.26 & 0.008 & -1.33 & 0.507 & 0.54 & 0.791 \\ 
7 & 1.99 & 1.64 & -0.19 & 0.881 & -2.50 & 0.043 & -3.34 & 0.009 \\ 
8 & -0.08 & -0.07 & 0.01 & 0.996 & -0.79 & 0.732 & -3.71 & 0.112 \\ 
9 & -2.19 & 0.64 & 2.60 & 0.000 & 0.19 & 0.803 & -0.44 & 0.567 \\ 
10 & 0.80 & 2.10 & 0.57 & 0.603 & 0.58 & 0.514 & 0.24 & 0.770 \\ \hline
Overall & 0.55 & 1.60 & 1.11 & 0.010 & -0.33 & 0.396 & -0.49 & 0.161%
\end{tabular}%
}
\caption{Randomization based inference results for $\protect\widehat{\bar{%
\protect\tau}}_{0}$ and $\protect\widehat{\bar{\protect\tau}}_{1}$,
\textquotedblleft B\textquotedblright\ is considered treatment and
\textquotedblleft A\textquotedblright\ is considered control. The $p$-values
(p.val) were obtained using the randomization method. The overall statistics
are the pooled statistics. }
\label{tab:average slippage in practice}
\end{table}
To calibrate $\widehat{\bar{\tau}}_{0}$ it may be helpful to recall that the
current annual expenses for holding the Vanguard 500 Index Fund Admiral
Shares (VFIAX) is five basis points. \ From Table \ref{t:rand_infer_res} the
firm is suffering an average slippage rate in equity futures of very roughly
0 to 2 basis points. \ Thus if the firm trades in and out of the market
with, say, \$10M once during the year it would be likely to pay less in
transaction charges than it would in expenses for holding \$10M in the
Vanguard fund for the year. \ However, if it trades more often its trading
costs could exceed the Vanguard level of costs. \ Of course, there may be
compensating advantages of more frequent trading, but we do not study that
topic here. \ \ \ From the results for $\widehat{\bar{\tau}}_{0}$, we can
see that in markets 6 and 9 \textquotedblleft A\textquotedblright\ performed
significantly better than \textquotedblleft B\textquotedblright . There are
no markets where there is evidence of out performance by method
\textquotedblleft B\textquotedblright . \ The lagged versions, $\widehat{%
\bar{\tau}}_{1}$ and $\widehat{\bar{\tau}}_{2}$, suggest little lagged
casual dependence, with only 1 out of 20 statistics being statistically
significant, that at lagged 2 for market 7. \ 

\begin{figure}[t]
\centering%
\begin{minipage}[t]{0.48\textwidth}
\includegraphics[width=\textwidth]{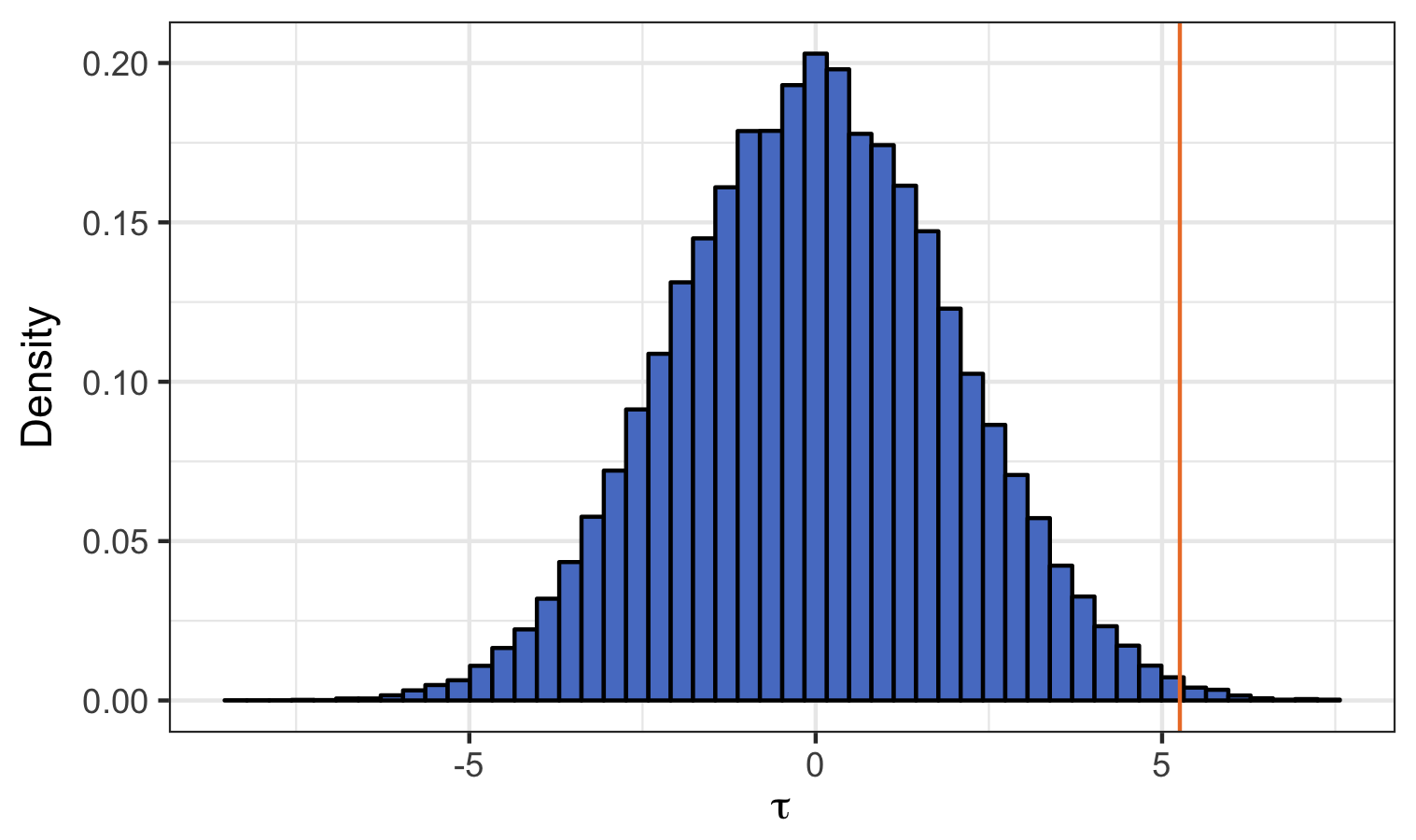}
Market 6: $\widehat{\bar{\protect\tau}}_{0}$
\end{minipage}%
\begin{minipage}[t]{0.48\textwidth}
\includegraphics[width=\textwidth]{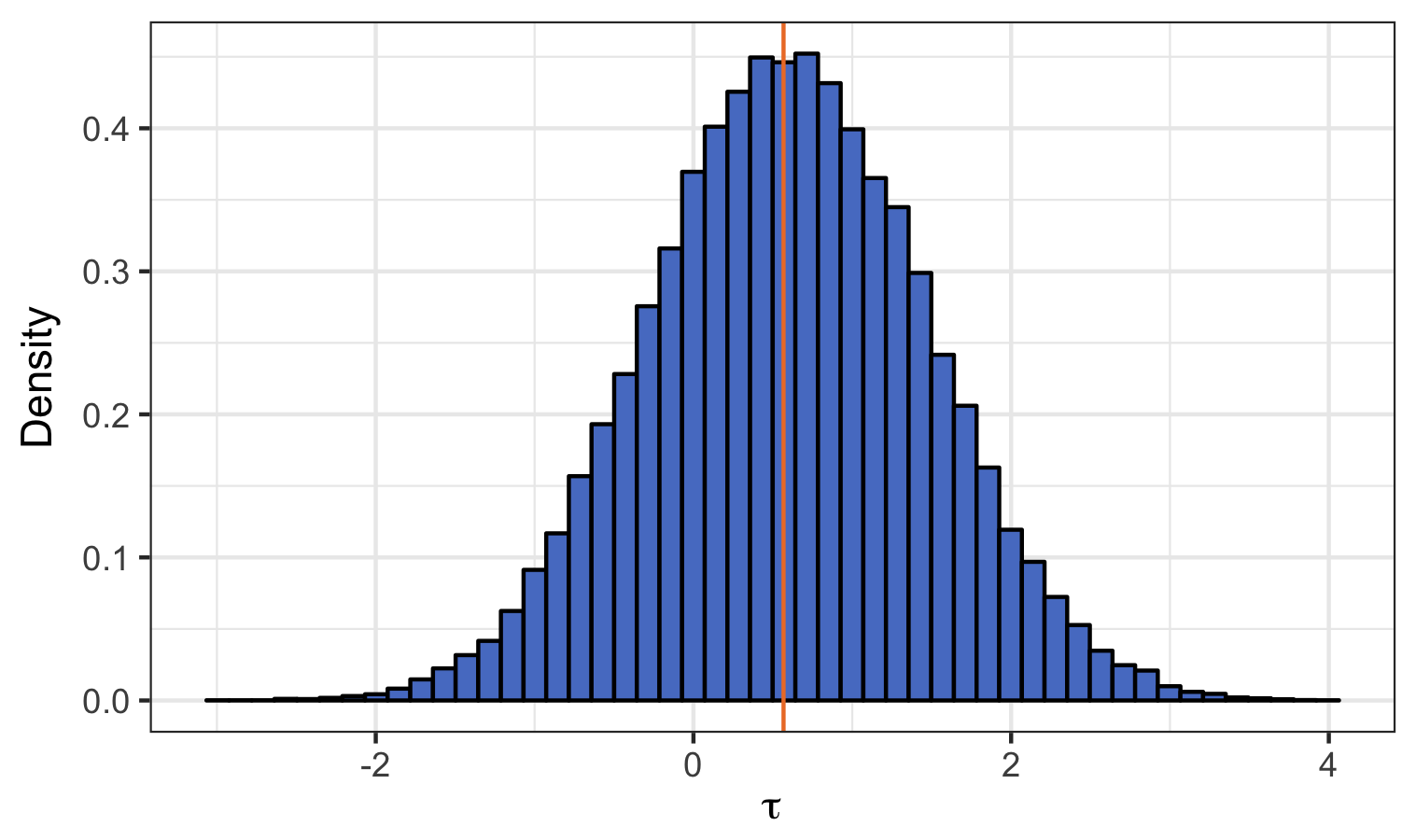}
Market 10: $\widehat{\bar{\protect\tau}}_{0}$
\end{minipage}
\caption{Randomization distribution for $\protect\widehat{\bar{\protect\tau}}%
_{0}$ Market 7 and 10. The vertical orange line indicated the observed value
of the statistic $\protect\widehat{\bar{\protect\tau}}_{0}$. }
\label{fig:rand_dist}
\end{figure}

Figure \ref{fig:rand_dist} shows the randomization distribution of the
statistic $\widehat{\bar{\tau}}_{0}$ for markets 6 and 10. The value of the
observed statistic is shown by the vertical orange line. \ The
unstandardized statistic's randomization distribution looks symmetric and
smooth around 0. \ Figure \ref{fig:rand_dist1}, in the web appendix, shows
the results for the other markets. \ 

\subsection{Pooled estimation and lagged estimation}

\label{sub:pooled_estimation}

In Section \ref{Muliple units}, we explained how to jointly analyze the
results for multiple units. \ The treatments are i.i.d. Bernoulli, so
continue to be independent of potential outcomes.

The most powerful procedure we proposed is the randomization based test. \
When we applied it to all 10 markets we obtain a highly significant result,
indicating that the slippage for method \textquotedblleft
A\textquotedblright\ is likely to be lower than that for method
\textquotedblleft B\textquotedblright .

\begin{figure}[t]
\centering%
\begin{minipage}[t]{0.24\textwidth}
\includegraphics[width=\textwidth]{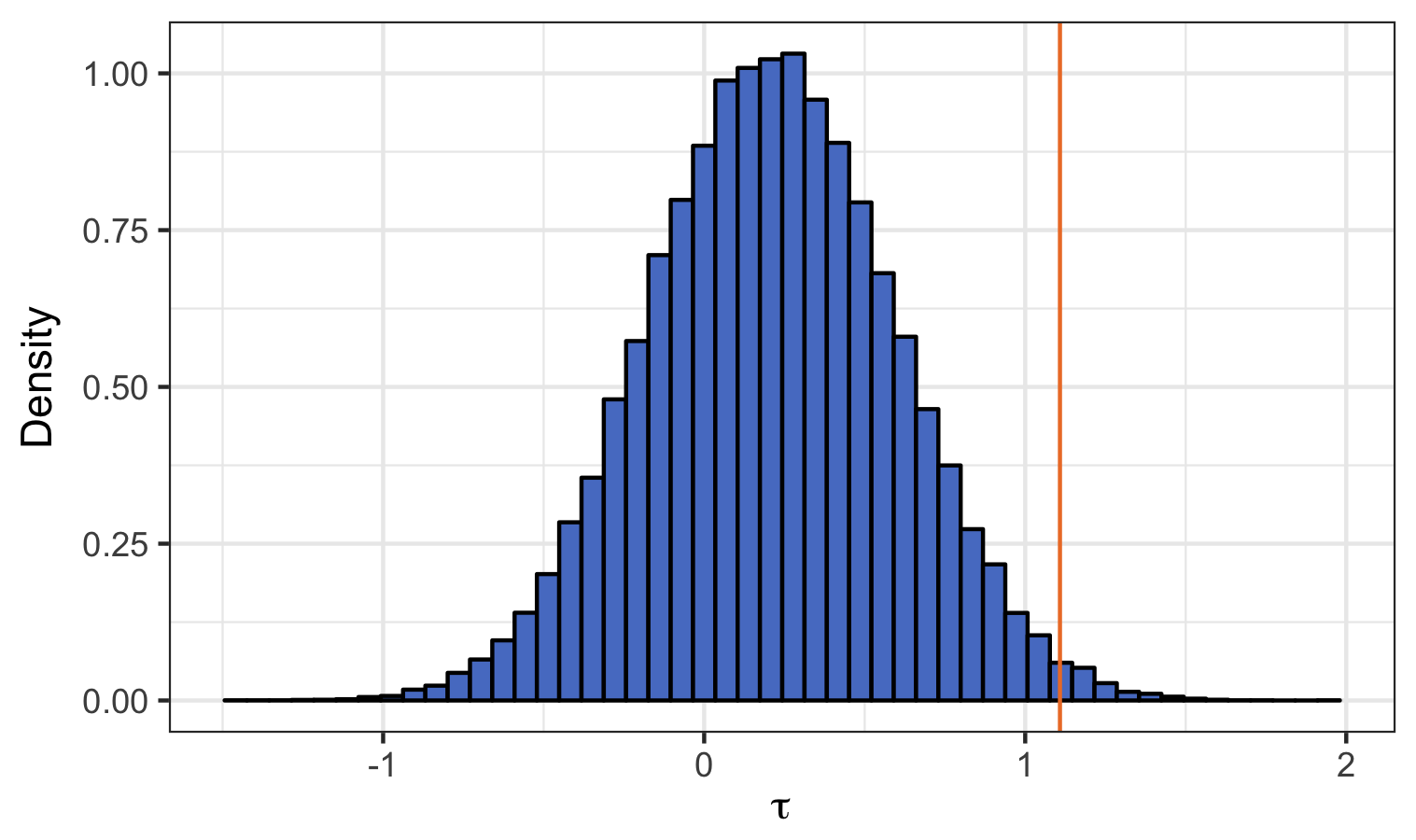}
Pooled $\overline{\tau }_{0}$
\end{minipage}%
\begin{minipage}[t]{0.24\textwidth}
\includegraphics[width=\textwidth]{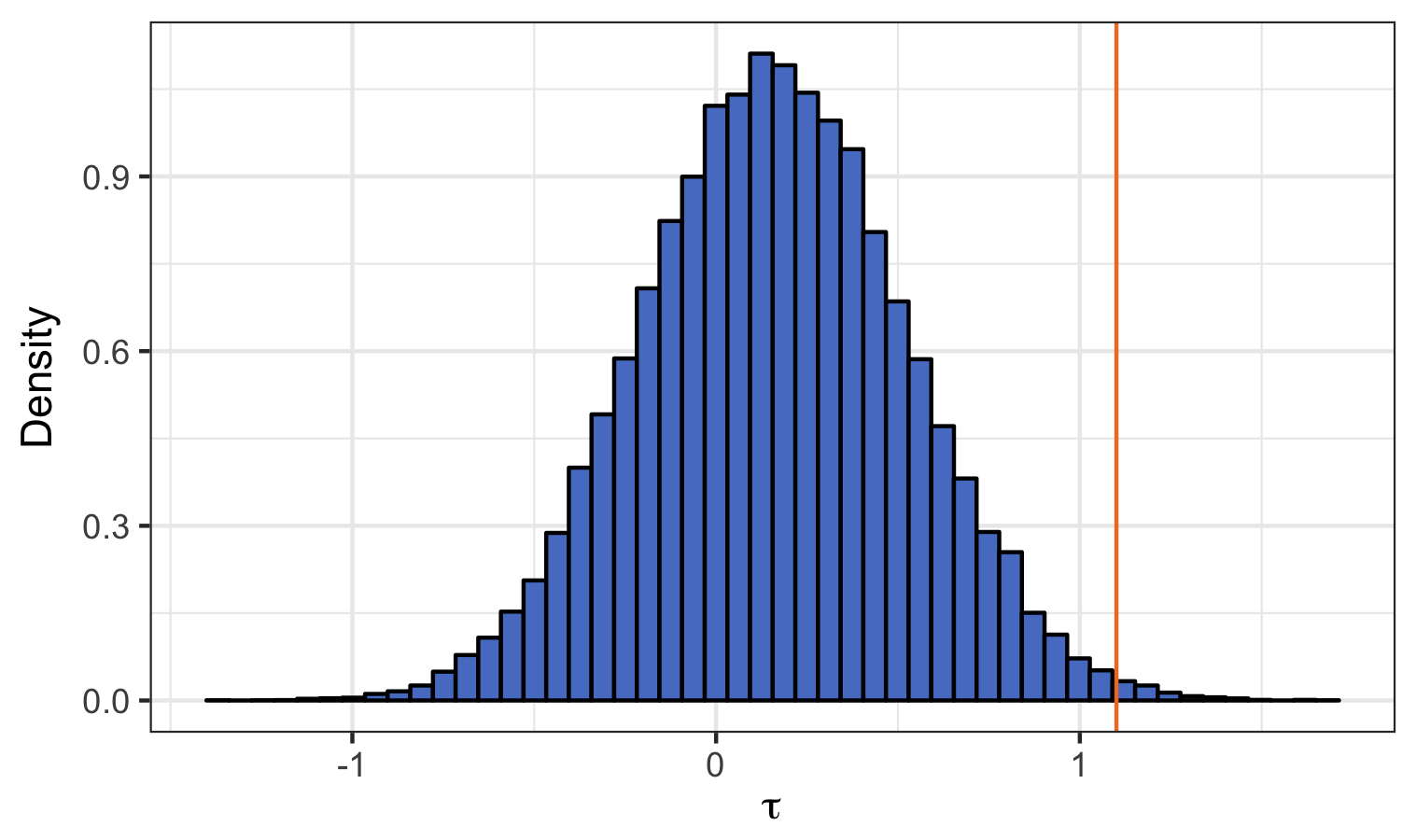}
Pooled $\overline{\tau }_{0}^{(1)}$
\end{minipage}%
\begin{minipage}[t]{0.24\textwidth}
\includegraphics[width=\textwidth]{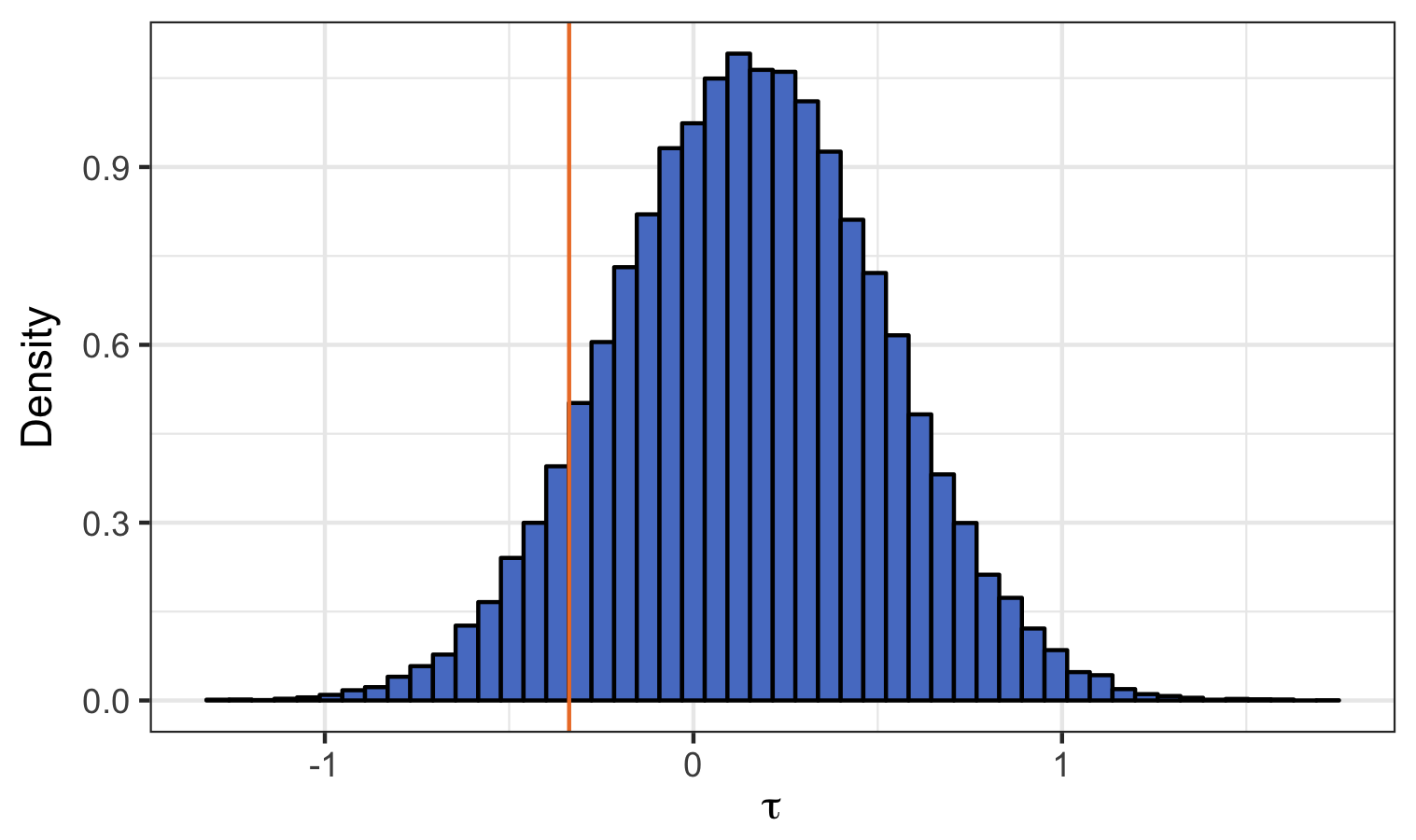}
Pooled $\overline{\tau }_{1}$
\end{minipage}%
\begin{minipage}[t]{0.24\textwidth}
\includegraphics[width=\textwidth]{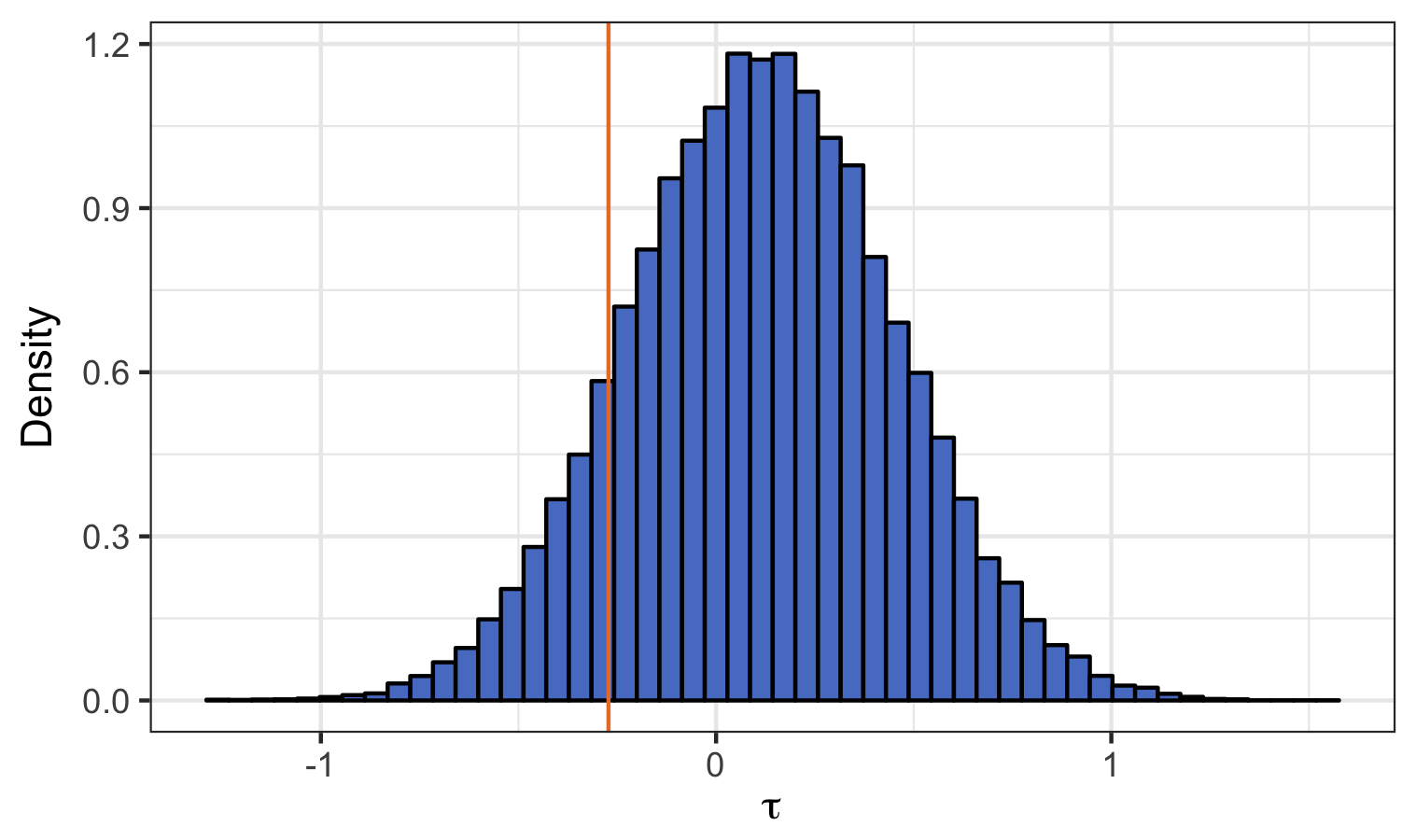}
Pooled $\overline{\tau }_{1}^{(1)}$
\end{minipage}
\caption{Randomization distribution for the pooled statistics, using the $%
\overline{\protect\tau }_{0}$ and $\overline{\protect\tau }_{1}$ measures. \
Both are also shown with 1 step. The vertical orange line indicated the
observed value of the statistic. Notice that there is strong evidence
indicated that there is a contemporaneous effect whereas there is little
evidence indicated that there is a lagged effect.}
\label{fig:rand_dist_pool}
\end{figure}

The randomization distribution under the no temporal causal effects null (%
\ref{Fisher null}) is shown in Figure \ref{fig:rand_dist_pool}, plotting the
pooled $\widehat{\overline{\tau }}_{0}$ and $\widehat{\overline{\tau }}_{1}$
and step cases $\widehat{\overline{\tau }}_{0}^{(1)}$ and $\widehat{%
\overline{\tau }}_{1}^{(1)}$. \ As usual the observed value is shown by the
orange vertical line. \ The observed value is in the extreme right hand tail
of the distribution. \ The contemporaneous results are strongly in favor of
method A out performing method B in a causal sense across the 10 markets,
with a $p$-value close to zero. \ There seems to be almost no lagged effect.
\ For financial data this is not surprising as there is a modest amount of
serial correlation in financial returns through time. \ These results do not
change for higher order stepped and lagged step causal effects. \ More
results along these lines are shown in Table \ref{tab:pool table results}. \ 

\begin{table}[tbp]
\centering
\par
{\footnotesize 
\begin{tabular}{r|rrrr|rrrr}
& \multicolumn{4}{c|}{No temporal causal effects?} & \multicolumn{4}{c}{No
average temporal causal effects?} \\ \hline
$p$ & $\widehat{\bar{\tau}}_{p}$ & p.val & $\widehat{\bar{\tau}}_{p}^{(1)}$
& p.val & $\widehat{\bar{\tau}}_{p}$ & p.val & $\widehat{\bar{\tau}}%
_{p}^{(1)}$ & p.val \\ \hline\hline
0 & 1.108 & 0.010 & 1.101 & 0.005 & 0.941 & 0 & 0.936 & 0 \\ 
1 & -0.337 & 0.396 & -0.272 & 0.455 & -0.327 & 0.087 & -0.264 & 0.189 \\ 
2 & -0.494 & 0.161 & -0.414 & 0.212 & -0.412 & 0.041 & -0.368 & 0.053 \\ 
3 & -0.029 & 0.933 & -0.018 & 0.953 & -0.061 & 0.750 & -0.032 & 0.853 \\ 
4 & -0.046 & 0.882 & -0.068 & 0.801 & -0.123 & 0.478 & -0.093 & 0.539%
\end{tabular}%
}
\caption{Results from pooled hypothesis tests for the 10 Markets for the $%
p\geq 0$ lagged causal effects. The no temporal causal effects refers to the
Fisher style sharp null test. \ The no average temporal causal effects
refers to the Neyman style null implemented using the CLT.}
\label{tab:pool table results}
\end{table}

\section{Conclusion\label{sect:conclusion}}

In this paper, we use a potential outcome and treatment path framework to
define causal estimands for experiments carried out on a single unit over
time. \ We define a broad class of estimands and proposed how to estimate
them without any assumptions on the underlying potential outcomes. \
Instead, we require that the treatments be non-anticipating and
probabilistic. \ We further propose two inferential strategies that utilize
the probabilistic assignment of the treatment path. \ Finally, we provide
three strategies for generalizing our framework to multiple units.

Our first inferential procedure tests the sharp null of no temporal causal
effect using the underlying randomization distribution. \ These
randomization based tests are exact and can be conducted using any of our
proposed causal estimands. \ Our second inferential procedure weakens the no
temporal causal effect null hypothesis to the no average temporal causal
effects null hypothesis. \ Inference in this setting is conducted using a
CLT and estimating an upper bound of the variance.

We apply our new methods on a large database of experiments carried out by a
quantitative hedge fund, who decide to execute orders either using human
traders or computers. \ We show that one of these trading methods has a
lower slippage rate than the alternative. \ 

\baselineskip=17pt

\bibliographystyle{chicago}
\bibliography{reference,neil}

\baselineskip=20pt

\appendix

\section{Appendix}

\subsection{Proof of Theorem \protect\ref{T:p-lag}}

\label{ss:proof_thrm_kstep}

For fixed $t$, we will further condense the adapted propensity score and use
the notation $p_{1,w}=p_{t+p}(1,w)$, $p_{0,w}=p_{t+p}(0,w)$. For simplicity
of exposition we use uniform weights, the extension to the non-uniform case
is straightforward. \ Then 
\begin{eqnarray}
\tau _{t+p,p}({1},{0}) &=&\frac{1}{2^{p}}\sum_{w\in
\{0,1\}^{p}}Y_{t+p}(w_{1:t-1}^{\text{obs}},1,w)-Y_{t+p}(w_{1:t-1}^{\text{obs}%
},0,w) \\
\hat{\tau}_{t+p,p} &=&\frac{1}{2^{p}}\sum_{w\in \{0,1\}^{p}}\left\{ \frac{%
1_{W_{t:t+p}=\left( 1,w\right) }}{p_{1,w}}Y_{t+p}(w_{1:t-1}^{\text{obs}%
},1,w)-\frac{1_{W_{t:t+p}=\left( 0,w\right) }}{p_{0,w}}Y_{t+p}(w_{1:t-1}^{%
\text{obs}},0,w)\right\} .  \notag
\end{eqnarray}%
Hence $u_{t,p}=\hat{\tau}_{t+p,p}-\tau _{t+p,p}({1},{0})$ equals 
\begin{equation*}
\frac{1}{2^{p}}\sum_{w\in \{0,1\}^{p}}\left[ \left\{ \frac{%
1_{W_{t:t+p}=\left( 1,w\right) }}{p_{1,w}}-1\right\} Y_{t+p}(w_{1:t-1}^{%
\text{obs}},1,w)-\left\{ \frac{1_{W_{t:t+p}=\left( 0,w\right) }}{p_{0,w}}%
-1\right\} Y_{t+p}(w_{1:t-1}^{\text{obs}},0,w)\right]
\end{equation*}

Now $\mathrm{E}(u_{t,p}|\mathcal{F}_{T,t-1})=0$, and $\mathrm{E}(\left\vert
u_{t,p}\right\vert )<\infty $. Hence these errors are a martingale
difference sequence and so uncorrelated through time. \ The remaining issue
is the variance. \ Let, 
\begin{equation*}
R=\sum_{w\in \{0,1\}^{p}}\left\{ c_{1,w}\left( 1_{W=\left( 1,w\right)
}-p_{1,w}\right) -c_{0,w}\left( 1_{W=\left( 0,w\right) }-p_{0,w}\right)
\right\} ,
\end{equation*}%
where $c_{1,w}=Y_{t+p}(w_{1:t-1}^{\text{obs}},1,w)/p_{1,w}$, $%
c_{0,w}=Y_{t+p}(w_{1:t-1}^{\text{obs}},0,w)/p_{0,w}\ $are known at $t-1$. Now%
\begin{eqnarray*}
\mathrm{E}\left\{ \left( 1_{W=\left( 1,w\right) }-p_{1,w}\right) \left(
1_{W=\left( 0,w^{\prime }\right) }-p_{0,w^{\prime }}\right) |\mathcal{F}%
_{T,t-1}\right\} &=&\mathrm{E}\left( 1_{W=\left( 1,w\right) }1_{W=\left(
0,w^{\prime }\right) }|\mathcal{F}_{T,t-1}\right) -p_{1,w}p_{0,w^{\prime }}
\\
&=&-p_{1,w}p_{0,w^{\prime }}, \\
\mathrm{E}\left\{ \left( 1_{W=\left( j,w\right) }-p_{j,w}\right) \left(
1_{W=\left( j,w^{\prime }\right) }-p_{j,w^{\prime }}\right) |\mathcal{F}%
_{T,t-1}\right\} &=&1_{w=w^{\prime }}p_{j,w}-p_{j,w}p_{j,w^{\prime }}\quad 
\text{for }j=0,1.
\end{eqnarray*}%
So $\mathrm{Var}^{R}(R|\mathcal{F}_{T,t-1})$ equals 
\begin{eqnarray*}
&&\sum_{w,w^{\prime }\in \{0,1\}^{p}}c_{1,w}c_{1,w^{\prime }}\left(
1_{w=w^{\prime }}p_{1,w}-p_{1,w}p_{1,w^{\prime }}\right)
+c_{0,w}c_{0,w^{\prime }}\left( 1_{w=w^{\prime
}}p_{0,w}-p_{0,w}p_{0,w^{\prime }}\right) \\
&&+\left( c_{1,w}c_{0,w^{\prime }}p_{1,w}p_{0,w^{\prime }}+c_{1,w^{\prime
}}c_{w,0}p_{1,w^{\prime }}p_{0,w}\right) \\
&=&\sum_{w\in
\{0,1\}^{p}}c_{1,w}^{2}p_{1,w}+c_{0,w}^{2}p_{0,w}+\sum_{w,w^{\prime }\in
\{0,1\}^{p}}c_{1,w}c_{0,w^{\prime }}p_{1,w}p_{0,w^{\prime }}+c_{1,w^{\prime
}}c_{0,w}p_{1,w^{\prime }}p_{0,w} \\
&&-c_{1,w}c_{1,w^{\prime }}p_{1,w}p_{1,w^{\prime }}-c_{0,w}c_{0,w^{\prime
}}p_{0,w}p_{0,w^{\prime }} \\
&=&\sum_{w\in \{0,1\}^{p}}\left( \frac{Y_{t+p}(w_{1:t-1}^{\text{obs}%
},1,w)^{2}}{p_{1,w}}+\frac{Y_{t+p}(w_{1:t-1}^{\text{obs}},0,w)^{2}}{p_{0,w}}%
\right) \\
&&+\sum_{w,w^{\prime }\in \{0,1\}^{p}}Y_{t+p}(w_{1:t-1}^{\text{obs}%
},1,w)Y_{t+p}(w_{1:t-1}^{\text{obs}},0,w^{\prime })-Y_{t+p}(w_{1:t-1}^{\text{%
obs}},0,w)Y_{t+p}(w_{1:t-1}^{\text{obs}},0,w^{\prime }) \\
&&+Y_{t+p}(w_{1:t-1}^{\text{obs}},1,w^{\prime })Y_{t+p}(w_{1:t-1}^{\text{obs}%
},0,w)-Y_{t+p}(w_{1:t-1}^{\text{obs}},1,w^{\prime })Y_{t+p}(w_{1:t-1}^{\text{%
obs}},1,w) \\
&=&\sum_{w\in \{0,1\}^{p}}\left( \frac{Y_{t+p}(w_{1:t-1}^{\text{obs}%
},1,w)^{2}}{p_{1,w}}+\frac{Y_{t+p}(w_{1:t-1}^{\text{obs}},0,w)^{2}}{p_{0,w}}%
\right) \\
&&-\sum_{w,w^{\prime }\in \{0,1\}^{p}}\left\{ Y_{t+p}(w_{1:t-1}^{\text{obs}%
},1,w^{\prime })-Y_{t+p}(w_{1:t-1}^{\text{obs}},0,w^{\prime })\right\}
\left\{ Y_{t+p}(w_{1:t-1}^{\text{obs}},1,w)-Y_{t+p}(w_{1:t-1}^{\text{obs}%
},0,w)\right\} .
\end{eqnarray*}

\subsection{Proof of Lemma \protect\ref{L:var_bound_p}\label{appendix:proof
of lemma}}

Note that for any $w$ and $w^{\prime }$ we have the following bound, $%
\left\{ Y_{t}(w)+Y_{t}(w^{\prime })\right\}
^{2}=Y_{t}(w)^{2}+Y_{t}(w^{\prime })^{2}+2Y_{t}(w)Y_{t}(w^{\prime })$ is
non-negative, so $Y_{t}(w)^{2}+Y_{t}(w^{\prime })^{2}\geq
2Y_{t}(w)Y_{t}(w^{\prime })$. By proof of Theorem \ref{T:p-lag} we can write 
$\mathrm{Var}^{R}(u_{t,p}|\mathcal{F}_{T,t-1})$, ignoring the scaling, in
the following way, 
\begin{align*}
& \sum_{w\in \{0,1\}^{p}}\left( \frac{Y_{t+p}(w_{1:t-1}^{\text{obs}},1,w)^{2}%
}{p_{1,w}}+\frac{Y_{t+p}(w_{1:t-1}^{\text{obs}},0,w)^{2}}{p_{0,w}}\right)
+\sum_{w,w^{\prime }\in \{0,1\}^{p}}\left\{ Y_{t+p}(w_{1:t-1}^{\text{obs}%
},1,w)Y_{t+p}(w_{1:t-1}^{\text{obs}},0,w^{\prime })\right. \\
& +Y_{t+p}(w_{1:t-1}^{\text{obs}},1,w^{\prime })Y_{t+p}(w_{1:t-1}^{\text{obs}%
},0,w)-Y_{t+p}(w_{1:t-1}^{\text{obs}},1,w^{\prime })Y_{t+p}(w_{1:t-1}^{\text{%
obs}},1,w) \\
& -Y_{t+p}(w_{1:t-1}^{\text{obs}},0,w)Y_{t+p}(w_{1:t-1}^{\text{obs}%
},0,w^{\prime }) \\
& \leq \sum_{w\in \{0,1\}^{p}}\left( \frac{Y_{t+p}(w_{1:t-1}^{\text{obs}%
},1,w)^{2}}{p_{1,w}}+\frac{Y_{t+p}(w_{1:t-1}^{\text{obs}},0,w)^{2}}{p_{0,w}}%
\right) -2\sum_{w\in \{0,1\}^{p}}\left\{ Y_{t+p}(w_{1:t-1}^{\text{obs}%
},1,w)^{2}+Y_{t+p}(w_{1:t-1}^{\text{obs}},0,w)^{2}\right\} \\
& +\frac{1}{2}\sum_{w,w^{\prime }\in \{0,1\}^{p}}\left\{ Y_{t+p}(w_{1:t-1}^{%
\text{obs}},1,w)^{2}+Y_{t+p}(w_{1:t-1}^{\text{obs}},1,w^{\prime
})^{2}+Y_{t+p}(w_{1:t-1}^{\text{obs}},0,w^{\prime })^{2}\right. \\
& +\left. Y_{t+p}(w_{1:t-1}^{\text{obs}},0,w)^{2}+Y_{t+p}(w_{1:t-1}^{\text{%
obs}},0,w^{\prime })^{2}+Y_{t+p}(w_{1:t-1}^{\text{obs}%
},1,w)^{2}+Y_{t+p}(w_{1:t-1}^{\text{obs}},1,w^{\prime
})^{2}+Y_{t+p}(w_{1:t-1}^{\text{obs}},0,w)^{2}\right\} \\
& =\sum_{w\in \{0,1\}^{p}}\left[ \frac{Y_{t+p}(w_{1:t-1}^{\text{obs}%
},1,w)^{2}}{p_{1,w}}+\frac{Y_{t+p}(w_{1:t-1}^{\text{obs}},0,w)^{2}}{p_{0,w}}%
+2(2^{p}-1)\left\{ Y_{t+p}(w_{1:t-1}^{\text{obs}%
},1,w)^{2}+Y_{t+p}(w_{1:t-1}^{\text{obs}},0,w)^{2}\right\} \right] \\
& =\sum_{w\in \{0,1\}^{p}}\left( \frac{Y_{t+p}(w_{1:t-1}^{\text{obs}%
},1,w)^{2}\left[ 1+2p_{1,w}(2^{p}-1)\right] }{p_{1,w}}+\frac{%
Y_{t+p}(w_{1:t-1}^{\text{obs}},0,w)^{2}\left[ 1+2p_{0,w}(2^{p}-1)\right] }{%
p_{0,w}}\right) .
\end{align*}

\newpage

\section{Web appendix}

\subsection{Properties of the stepped estimator}

\begin{theorem}[Properties of $q$ step $p$ lag estimators]
Define $u_{t-p-q,p}=\hat{\tau}_{t,p}^{(q)}-\tau _{t,p}^{(q)}(\{1\},\{0\})$
as the estimation error. \ Then $\mathrm{E}^{R}(u_{t,p}^{(q)}|\mathcal{F}%
_{T,t-1})=0$, and $\mathrm{Var}^{R}(u_{t,p}^{(q)}|\mathcal{F}_{T,t-1})$
equals 
\begin{align*}
& \frac{1}{2^{2p}}\sum_{\substack{ w^{\dag }\in \left\{ 0,1\right\} ^{p}  \\ %
w\in \left\{ 0,1\right\} ^{p}}}\left( \frac{Y_{t+p+q}(w_{1:t-1}^{\text{obs}%
},w^{\dag },1,w)^{2}}{p_{t+p+q}(w^{\dag },1,w)}+\frac{Y_{t+p+q}(w_{1:t-1}^{%
\text{obs}},w^{\dag },0,w)^{2}}{p_{t+p+q}(w^{\dag },0,w)}\right) \\
-& \frac{1}{2^{2p}}\sum_{\substack{ w^{\dag },w^{\prime \dag }\in \left\{
0,1\right\} ^{p}  \\ w,w^{\prime }\in \left\{ 0,1\right\} ^{p}}}\left\{
Y_{t+p+q}(w_{1:t-1}^{\text{obs}},w^{\prime \dag },1,w^{\prime
})-Y_{t+p+q}(w_{1:t-1}^{\text{obs}},w^{\prime \dag },0,w^{\prime })\right\}
\\
& \qquad \qquad \times \left\{ Y_{t+p+q}(w_{1:t-1}^{\text{obs}},w^{\dag
},1,w)-Y_{t+p+q}(w_{1:t-1}^{\text{obs}},w^{\dag },0,w)\right\} .
\end{align*}%
Averaging over the randomization (always conditioning on $Y_{1:T}(\bullet )$%
), then $\mathrm{E}^{R}(\left\vert u_{t,p}^{(q)}\right\vert )<\infty $ and 
\begin{equation*}
\mathrm{E}^{R}(u_{t,p}^{(q)})=0,\quad \mathrm{Cov}%
^{R}(u_{t,p}^{(q)},u_{s,p}^{(q)})=0,\quad s\neq t.
\end{equation*}
\end{theorem}

The proof is identical to the proof of the Theorem \ref{T:p-lag}, with a
minor change of the subscripts and the replacement of $(j,w)$ with $(w^\dag,
j, w)$.

The variance can again be bounded.

\begin{lemma}
Under non-anticipating treatments Assumption \ref{Assum: non-anticipating}
and probabilistic assignment Assumption \ref{Prob assign}, the variance of $%
u_{t,p}^{(q)}$, and in turn $\hat{\tau}^{(q)}_{t,p}$, is bounded above by 
\begin{equation*}
\mathrm{Var}^{R}(u_{t,p}^{(q)}|\mathcal{F}_{T,t-1})\leq \frac{1}{2^{2(p+q)}}
\sum_{w\in \{0,1\}^{p+q+1}}\frac{Y_{t+p+q}^{2}(w_{1:t-1}^{\text{obs}},w)%
\left[ 1+2p_{t+p+q}(w)(2^{p-1}-1)\right] }{p_{t+p+q}(w)}=(%
\sigma^{(q)}_{t+p,p})^{2}.
\end{equation*}%
Moreover, this upper bound can be estimated by, 
\begin{equation}
\widehat{\sigma^{(q)} }_{t+p,p}^{2}=\frac{1}{2^{2(p+w)}}\sum_{w\in
\{0,1\}^{p+q+1}}\frac{1_{W_{t:t+p+q}=w}Y_{t+p+q}(w_{1:t-1}^{\text{obs}%
},w)^{2}\left[ 1+2p_{t+p+q}(w)(2^{p+q-1}-1)\right] }{p_{t+p+q}^{2}(w)}
\label{E:var_bound_p}
\end{equation}%
and is conditionally unbiased, i.e. $\mathrm{E}^{R}(\widehat{\sigma^{(q)} }%
_{t+p,p}^{2}|\mathcal{F}_{T,t-1})=(\sigma^{(q)} _{t+p,p})^{2}$.
\end{lemma}

The proof follows directly from the proof of Lemma \ref{L:var_bound_p}

\subsection{$m$-period causal impact\label{sect:m dependent}}

In some applications outcomes at time $t$ only depend upon treatments which
go back $m\geq 0$ time periods. \ This is formalized in the following
assumption. \ 

\begin{assumption}
\label{Assump: m depend} ($m$-period causal impact). If, for all $%
u_{1:t-m-1},u_{1:t-m-1}^{\prime }$ and $w_{t-m:t}$, 
\begin{equation*}
Y_{t}(u_{1:t-m-1},w_{t-m:t})=Y_{t}(u_{1:t-m-1}^{\prime },w_{t-m:t})
\end{equation*}%
then the treatment path $W_{1:t}$ is said to have $m$-period causal impact
on $Y_{t}$. \ 
\end{assumption}

The $m=0$ case assumes the treatment only influences the current value of
the outcome, where as the $m=1$ case appeared in Example \ref{defn potential
autoregression} in the potential moving average.

Under Assumption \ref{Assump: m depend}, the causal effects is, for all $%
u_{1:t-m-1}$, 
\begin{eqnarray*}
\tau _{t}(w_{t-m:t},w_{t-m:t}^{\prime })
&=&Y_{t}(u_{1:t-m-1},w_{t-m:t})-Y_{t}(u_{1:t-m-1},w_{t-m:t}^{\prime }) \\
&=&Y_{t}(w_{1:t-m-1}^{\text{obs}},w_{t-m:t})-Y_{t}(w_{1:t-m-1}^{\text{obs}%
},w_{t-m:t}^{\prime }),
\end{eqnarray*}%
while the temporal average $m$-period causal effect is $\frac{1}{T}%
\sum_{t=1}^{T}\tau _{t}(w_{t-m:t},w_{t-m:t}^{\prime })$.

In the case of a comparison of time-invariant treatments $w$ to $w^{\prime }$
(which are each $\left( m+1\right) $-dimensional) $\tau _{t}(w,w^{\prime
})=Y_{t}(w_{1:t-m-1}^{\text{obs}},w)-Y_{t}(w_{1:t-m-1}^{\text{obs}%
},w^{\prime })$, then the average treatment effect simplifies to $\frac{1}{T}%
\sum_{t=1}^{T}\tau _{t}(w,w^{\prime })$. This returns us to the causal
effects we discussed above: $\widehat{\overline{\tau }}_{0}$ and $\widehat{%
\overline{\tau }}_{0}^{(k)}$ for $k=1,2,...$. Now $\tau _{t}(w,w^{\prime })$%
\ can be unbiasedly estimated by 
\begin{equation*}
\widehat{\tau }_{t}(w,w^{\prime })=\left( \frac{1_{w_{t-m:t}^{\text{obs}}=w}%
}{p_{t}(w)}-\frac{1_{w_{t-m:t}^{\text{obs}}=w^{\prime }}}{p_{t}(w^{\prime })}%
\right) Y_{t}(w_{1:t}^{\text{obs}}),\quad \text{where\quad }p_{t}(w)=\Pr
(W_{t-m:t}=w|\mathcal{F}_{T,t-m-1}).
\end{equation*}

\subsection{Randomization test}

\label{SSS:rand_test}

We can do exact hypothesis testing using a randomization (or permutation)
test for $\widehat{\overline{\tau }}_{p}^{(q)}$. \ The implementation and
analysis of this is standard (e.g. section 5.8 of \cite{ImbensRubin(15)}). \ 

First, fix $M>0$ and simulate $M$ estimates of the $\hat{\bar{\tau}}_{0}$
estimator using the algorithm:

\begin{enumerate}
\item Set $m=1$.

\item Sample a new treatment assignment path $w_{1:T}^{[m]}$ and record the
adapted propensity score path $p_{t}^{[m]}=\Pr
(W_{t}=w_{t}^{[m]}|W_{1:t-1}=w_{1:t-1}^{[m]},Y_{1:t-1}^{\text{obs}})$, $%
t=1,2,...,T$.

\item Compute 
\begin{equation*}
\hat{\tau}_{t,0}^{[m]}=\left\{ \frac{1_{w_{t}^{[m]}=1}}{p_{t}^{[m]}(1)}-%
\frac{1_{w_{t}^{[m]}=0}}{p_{t}^{[m]}(0)}\right\} Y_{t}(w_{1:t}^{\text{obs}%
}),\quad t=1,2,...,T.
\end{equation*}

\item Store $\hat{\bar{\tau}}_{0}^{[m]}=T^{-1}\sum_{t=1}^{T}\hat{\tau}%
_{t,0}^{[m]}$.

\item If $m<M$, set $m=m+1$ and go to 2. \ 
\end{enumerate}

Second, compute $\widehat{p}=M^{-1}\sum_{m=1}^{M}1_{\left\vert \hat{\bar{\tau%
}}_{0}^{[m]}\right\vert >\left\vert \hat{\bar{\tau}}_{0}\right\vert }\ $,
which compares the simulations $\left\{ \hat{\bar{\tau}}_{0}^{[m]}\right\} $
to the estimated $\hat{\bar{\tau}}_{0}$. \ 

This average $\widehat{p}$ simulate estimates the $p$-value of $\hat{\bar{%
\tau}}_{0}$ under the null. \ As M gets large this procedure becomes exact.
\ 

\subsection{Standardized measures of lagged causality}

Recall the 
\begin{equation*}
\hat{\tau}_{t,0}=\frac{1_{W_{t}=1}Y_{t}(w_{1:t-1}^{\text{obs}},1)}{p_{t}(1)}-%
\frac{1_{W_{t}=0}Y_{t}(w_{1:t-1}^{\text{obs}},0)}{p_{t}(0)}
\end{equation*}%
so the estimation error can be written as 
\begin{equation*}
u_{t,0}=\left( \frac{1_{W_{t}=1}}{p_{t}(1)}-\frac{1_{W_{t}=0}}{p_{t}(0)}%
\right) Y_{t}(w_{1:t}^{\text{obs}}),
\end{equation*}%
and under the sharp null (\ref{Fisher null})%
\begin{equation*}
\mathrm{E}^{R}(u_{t,0}|\mathcal{F}_{T,t-1})=0\quad \text{and\quad }\mathrm{%
Var}^{R}(u_{t,0}|\mathcal{F}_{T,t-1})=\frac{Y_{t}(w_{1:t}^{\text{obs}})^{2}}{%
p_{t}(1)p_{t}(0)}.
\end{equation*}

Since the variance is known we can define the $\widehat{\tau }_{t,0}$
standardized estimator as, 
\begin{equation*}
v_{t,0}=\frac{\widehat{\tau }_{t,0}}{\sqrt{\mathrm{Var}^{R}(u_{t,0}|\mathcal{%
F}_{T,t-1})}}=\left( \sqrt{\frac{p_{t}(0)}{p_{t}(1)}}1_{W_{t}=1}-\sqrt{\frac{%
p_{t}(1)}{p_{t}(0)}}1_{W_{t}=0}\right) sign\left\{ Y_{t}(w_{1:t}^{\text{obs}%
})\right\} .
\end{equation*}%
Under the sharp null, $\mathrm{E}^{R}(v_{t,0}|\mathcal{F}_{T,t-1})=0$\ and\ $%
\mathrm{Var}^{R}(v_{t,0}|\mathcal{F}_{T,t-1})=1$, and $v_{t}$ is bounded (as 
$p_{t}(1)\in (0,1)$ by Assumption \ref{Prob assign}) and is always a
martingale difference sequence. In particular, universally, as $T\rightarrow
\infty $ so $\sqrt{T}\frac{1}{T}\sum_{t=1}^{T}v_{t,0}\underset{H_{0}}{%
\overset{d}{\rightarrow }}N(0,1)$. \ 

In the Tables below we will write $\widehat{\overline{v}}_{p}=\frac{1}{T-p}%
\sum_{t=p+1}^{T}v_{t,p}$, where $v_{t,p}=\widehat{\tau }_{t,p}/\sqrt{\mathrm{%
Var}^{R}(u_{t,p}|\mathcal{F}_{T,t-p-1})}$.

\subsubsection{Simulation evidence}

\begin{figure}[h]
\centering%
\begin{minipage}[t]{0.48\textwidth}
\includegraphics[width=\textwidth]{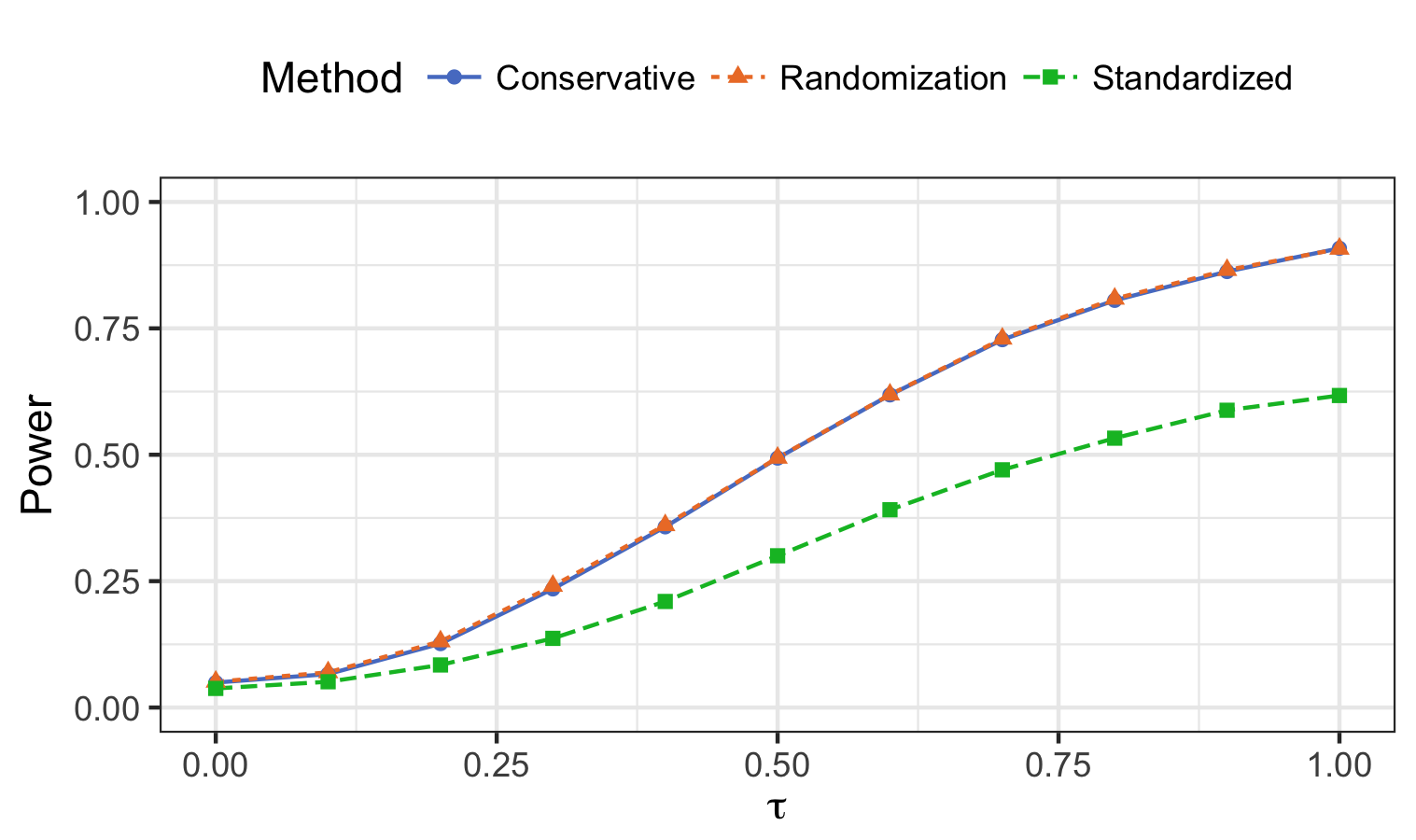}
For $\widehat{\bar{\protect\tau}}_{0}$ and $\widehat{\bar{ v}}_{0}$ as $\bar{\protect\tau}_{0}$ varies 
\end{minipage}%
\begin{minipage}[t]{0.48\textwidth}
\includegraphics[width=\textwidth]{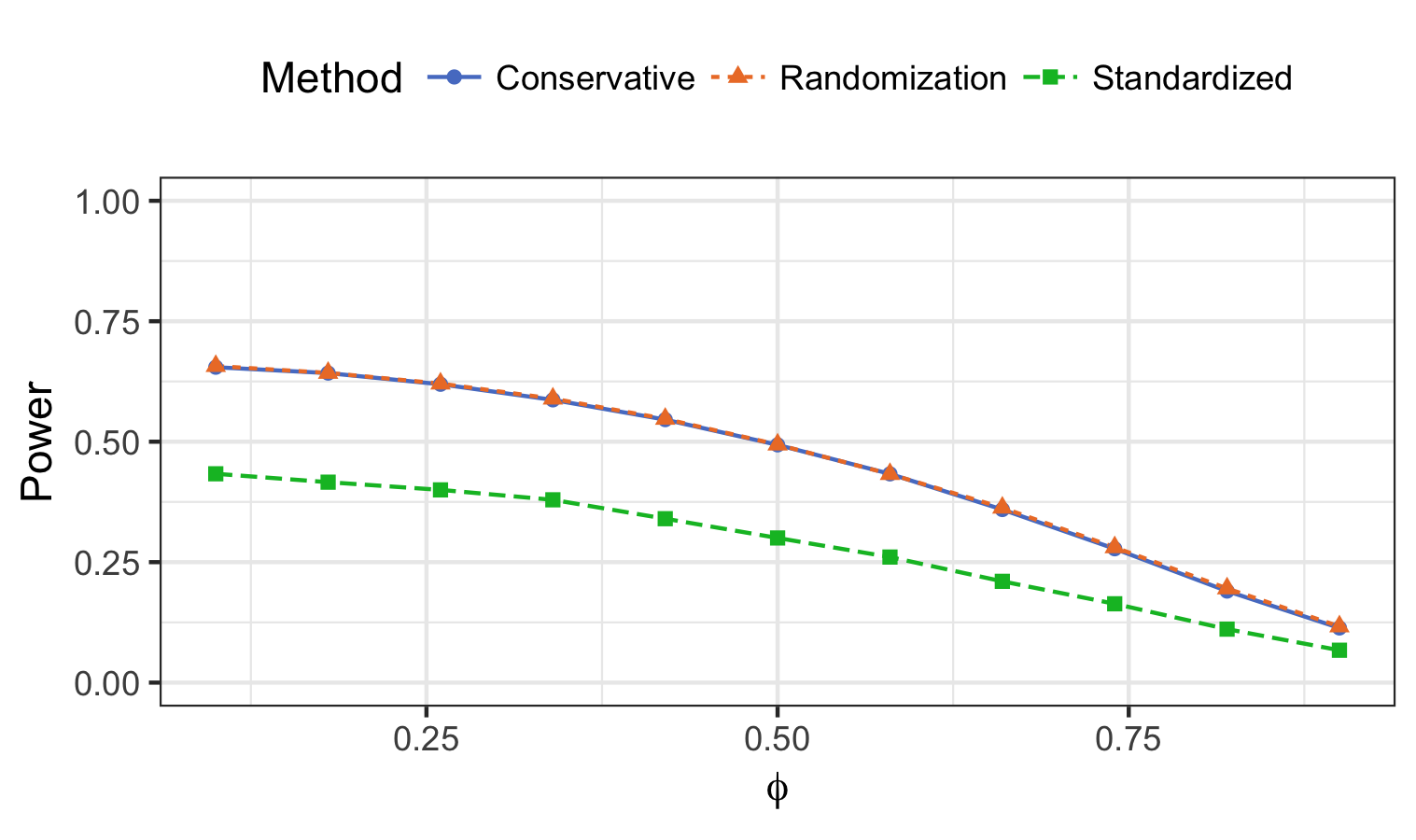}
For $\widehat{\bar{\protect\tau}}_{1}$ and $\widehat{\bar{v}}_{1}$ as $\phi$ varies
\end{minipage}\newline
\caption{The two plots shows the relative power for the three tests as the
treatment effect increases for fixed $\protect\phi =0.5$ (left) and as the $%
\protect\phi $ parameter increases for fixed $\protect\tau =0.5$ (right).
The conservative test performs only slightly worse that the exact
randomization test. The noise distribution is Gaussian with variance 1.}
\label{fig:standard power1}
\end{figure}

We now return to the simulation experiments discussed in Section \ref%
{sect:power simul}. The results are given in Figure \ref{fig:standard power1}%
. \ 
It shows that the standardized test has worse power than the exact and the
conservative tests discussed above. \ 

\subsubsection{Empirical results\ }

We have repeated our empirical analysis using the standardized statistics. 
\begin{table}[t]
\centering
\par
\label{t:rand_infer_res2} 
\begin{tabular}{r|rr|rr|rr}
\multirow{2}{*}{Market} & \multicolumn{2}{c}{Average Slippage} & %
\multirow{2}{*}{$\hat{\bar\tau}_0$} & \multirow{2}{*}{p.val} & %
\multirow{2}{*}{$\hat{\bar v}_0$} & \multirow{2}{*}{p.val} \\ 
& A & B &  &  &  &  \\ \hline\hline
1 & 1.48 & 2.29 & 0.52 & 0.628 & 0.03 & 0.678 \\ 
2 & 4.11 & 3.35 & -1.80 & 0.315 & 0.03 & 0.741 \\ 
3 & -0.05 & 1.07 & 0.54 & 0.634 & 0.01 & 0.893 \\ 
4 & 3.38 & 3.23 & 0.36 & 0.900 & -0.03 & 0.912 \\ 
5 & 0.57 & 0.63 & -0.42 & 0.743 & -0.03 & 0.671 \\ 
6 & -1.48 & 3.72 & 5.26 & 0.008 & 0.25 & 0.034 \\ 
7 & 1.99 & 1.64 & -0.19 & 0.881 & 0.06 & 0.374 \\ 
8 & -0.08 & -0.07 & 0.01 & 0.996 & 0.06 & 0.557 \\ 
9 & -2.19 & 0.64 & 2.60 & 0.000 & 0.14 & 0.005 \\ 
10 & 0.80 & 2.10 & 0.57 & 0.603 & 0.06 & 0.143 \\ \hline
Overall & 0.55 & 1.60 & 1.11 & 0.010 & 0.060 & 0.003%
\end{tabular}%
\caption{{\protect\small Randomization based inference results,
\textquotedblleft B\textquotedblright\ is considered treatment and
\textquotedblleft A\textquotedblright\ is considered control. The overall $p$%
-values (p.val) were obtained using the randomization method.}}
\label{tab:standard test1}
\end{table}
\ The basic results are given in Table \ref{tab:standard test1}. \ The
pooled results are given in Table \ref{tab:pooled standard1}. \ Overall, the
standardized statistics are broadly in line with the unstandardized ones. \ 

\begin{table}[tbp]
\centering
\par
\begin{tabular}{r|rr|rr|}
& \multicolumn{2}{c|}{Unstandardized} & \multicolumn{2}{c|}{Standardized} \\ 
\hline
$p$ & $\widehat{\bar{\tau}}_{p}$ & p.val & $\widehat{\bar{v}}_{p}$ & p.val
\\ \hline\hline
0 & 1.108 & 0.010 & 0.060 & 0.003 \\ 
1 & -0.337 & 0.396 & 0.001 & 0.953 \\ 
2 & -0.494 & 0.161 & -0.044 & 0.018 \\ 
3 & -0.029 & 0.933 & -0.025 & 0.167 \\ 
4 & -0.046 & 0.882 & -0.020 & 0.278%
\end{tabular}%
\caption{{\protect\small The results from the pooled hypothesis test for the
10 Markets for $\protect\widehat{\overline{\protect\tau }}_{p}$ and $\protect%
\widehat{\overline{v}}_{p}$. }}
\label{tab:pooled standard1}
\end{table}
\ 

\subsection{Definition of financial slippage\label{sect:slip}}

We write the time the $t$-th randomization is carried out as $\zeta _{0,t}$
and at precisely that time the mid-price of the asset (the average of the
best advertised bid (buying price) and offer (selling price)) is recorded as 
$P_{\zeta _{0,t}}^{mid}$. \ The trading performance will be compared to this
mid-price. \ Let $b_{t}=1$ if this is a sell order and $b_{t}=-1$ if this is
a buy order. \ 

Suppose the trades are made at times $\zeta _{j,t}$ where $j=1,2,...,J_{t}$
and the fraction of the fill of $t$-th order achieved on the $\zeta _{j,t}$%
-th trade is $v_{j,t}>0$. \ All trades are filled so $%
\sum_{j=1}^{J_{t}}v_{j,t}=1$. Then the \textquotedblleft
slippage\textquotedblright\ rate, in terms of basis points (one basis point
is 0.01\%), will be $Y_{t}^{\text{obs}}=b_{t}r_{t}$ where writing $P_{\zeta
_{j,t}}$ as the price of the trade made by the company (not the mid-price)
at time $\zeta _{j,t}$,%
\begin{eqnarray*}
r_{t} &=&10000\frac{1}{P_{\zeta _{0,t}}^{mid}}\left(
\sum_{j=1}^{J_{t}}v_{j,t}P_{\zeta _{j,t}}-P_{\zeta _{0,t}}^{mid}\right)
=10000\sum_{j=1}^{J_{t}}v_{j,t}\frac{P_{\zeta _{j,t}}-P_{\zeta _{0,t}}^{mid}%
}{P_{\zeta _{0,t}}^{mid}} \\
&=&\sum_{j=1}^{J_{t}}v_{\zeta _{j,t}}r_{\zeta _{j,t}},\quad r_{\zeta
_{j,t}}=10000\frac{P_{\zeta {_{j,t}}}-P_{\zeta _{0,t}}^{mid}}{P_{\zeta
_{0,t}}^{mid}}.
\end{eqnarray*}%
Thus $r_{t}$ is a volume weighted average price (VWAP) minus the mid-price
scaled by mid-price (e.g. \cite{BerkowitzLogueNoser(88)} and \cite%
{CalvoriCipolliniGallo(13)}). \ \ 

\subsection{Simulation study extra figures}

\begin{figure}[h]
\centering
\begin{minipage}[t]{0.23\textwidth}
$\widehat{\bar{\protect\tau}}_{0}$
\includegraphics[width=\textwidth]{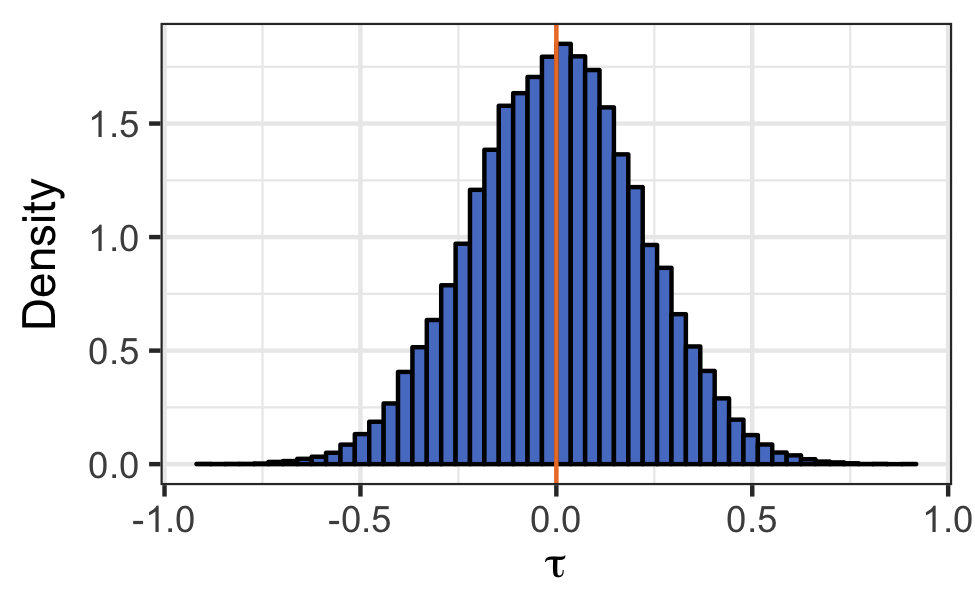}
$\mu_1=0.0$, $\mu_0=0.0$
\end{minipage}
\begin{minipage}[t]{0.23\textwidth}
$\widehat{\bar{\protect\tau}}_{0}^{(1)}$
\includegraphics[width=\textwidth]{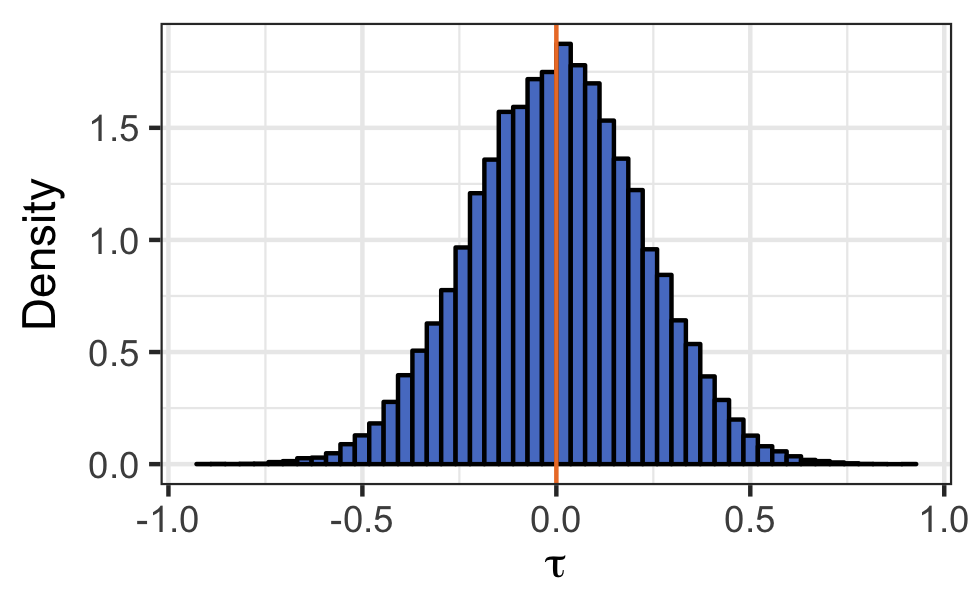}
\end{minipage}
\begin{minipage}[t]{0.23\textwidth}
$\widehat{\bar{\protect\tau}}_{1}$
\includegraphics[width=\textwidth]{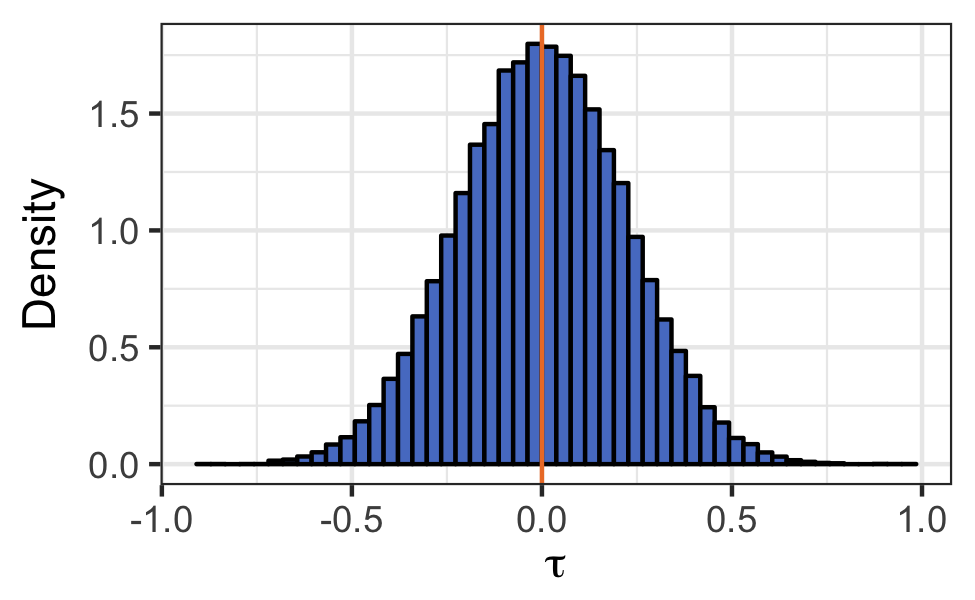}
\end{minipage}
\begin{minipage}[t]{0.22\textwidth}
$\widehat{\bar{\protect\tau}}_{2}$
\includegraphics[width=\textwidth]{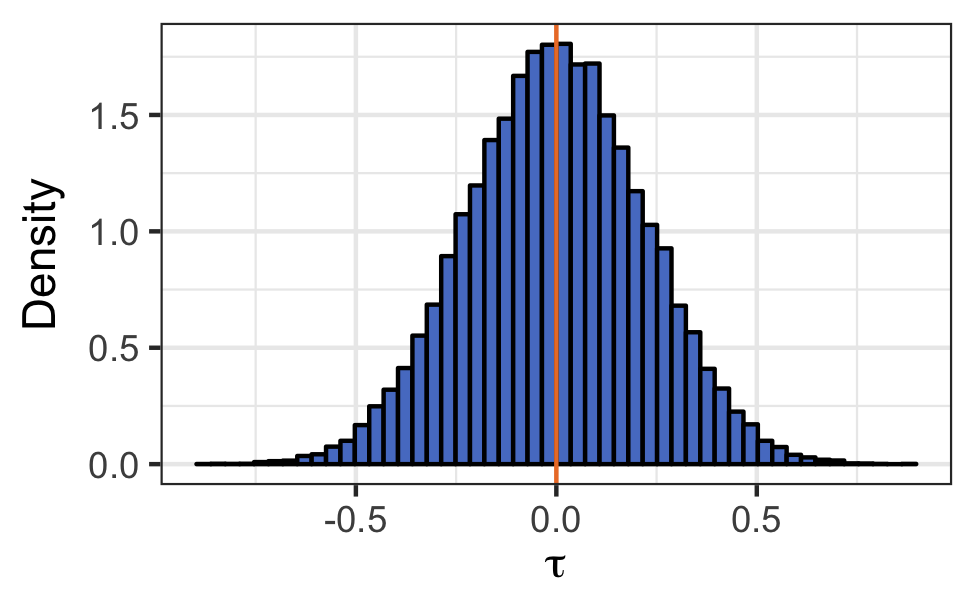}
\end{minipage}
\newline
\begin{minipage}[t]{0.23\textwidth}
\includegraphics[width=\textwidth]{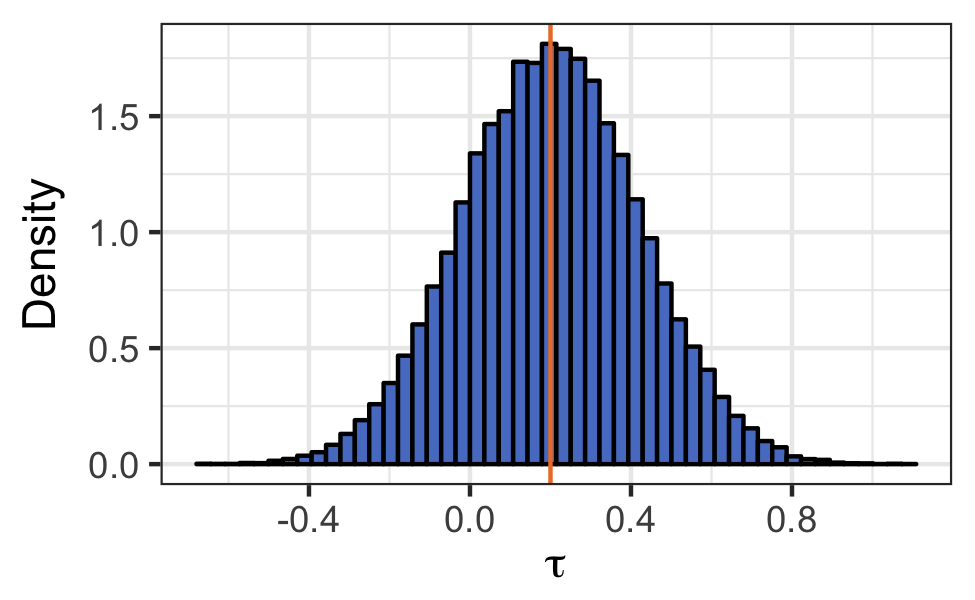}
$\mu_1=0.2$, $\mu_0=0.0$
\end{minipage}
\begin{minipage}[t]{0.23\textwidth}
\includegraphics[width=\textwidth]{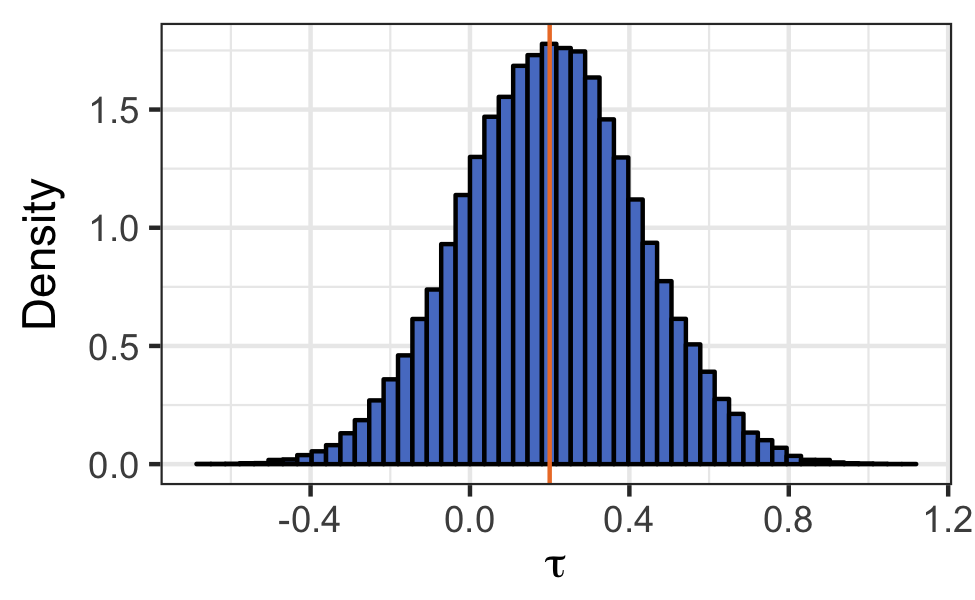}
\end{minipage}
\begin{minipage}[t]{0.23\textwidth}
\includegraphics[width=\textwidth]{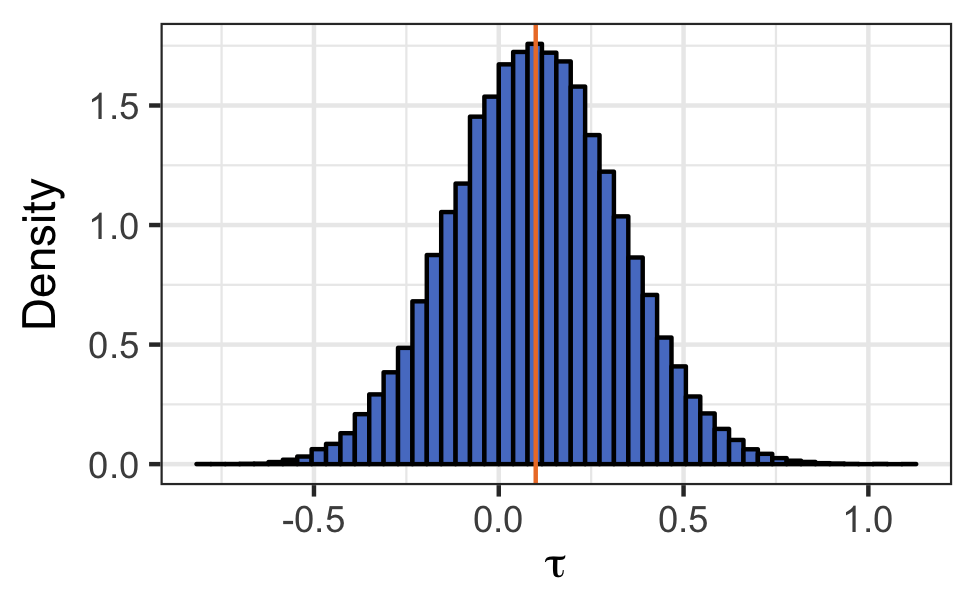} 
\end{minipage}
\begin{minipage}[t]{0.23\textwidth}
\includegraphics[width=\textwidth]{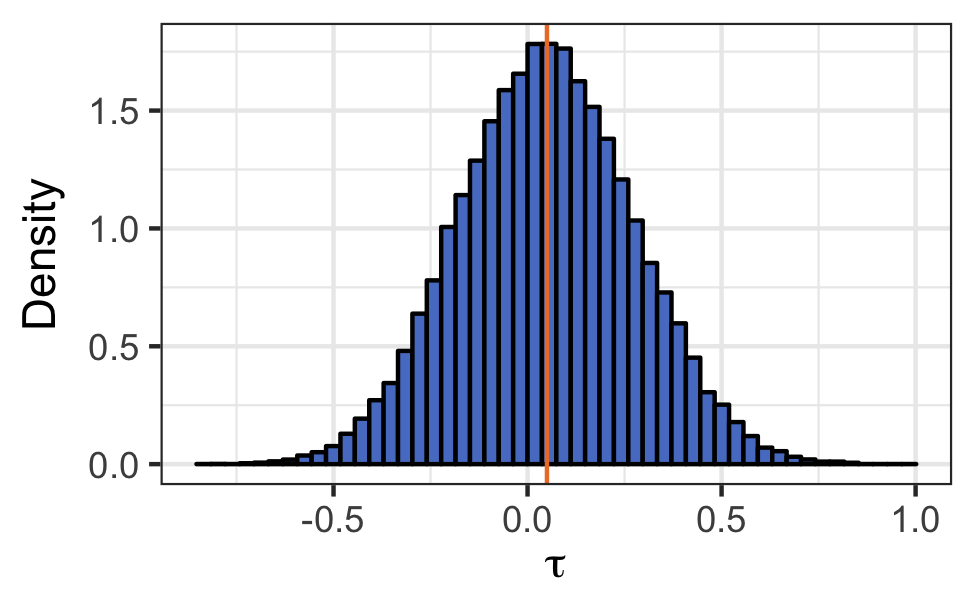} 
\end{minipage}
\newline
\begin{minipage}[t]{0.23\textwidth}
\end{minipage}
\begin{minipage}[t]{0.23\textwidth}
\end{minipage}
\begin{minipage}[t]{0.23\textwidth} 
\end{minipage}
\begin{minipage}[t]{0.23\textwidth} 
\end{minipage}
\caption{ Histograms of different estimates obtain from 50,000 treatment
paths for the same $Y_{1:T}$ with $T=100$ \& $\protect\phi =0.5$ where the
treatment effect is either $\protect\mu _{1}=0$ (top) or $\protect\mu %
_{1}=0.2$ and the stocks are Gaussian. The 1st column is from the $\protect%
\widehat{\bar{\protect\tau}}_{0}$ estimator, the 2nd column is the $\protect%
\widehat{\bar{\protect\tau}}_{0}^{(1)}$ estimator, the 3rd column is the $%
\protect\widehat{\bar{\protect\tau}}_{1}$ estimator \& the 4th column is the 
$\protect\widehat{\bar{\protect\tau}}_{2}$ estimator. All of the estimators
have similar variance, \& are centered at the true data generating value.}
\label{fig:CLT_dif_est}
\end{figure}

\begin{figure}[h]
\centering
\begin{minipage}[t]{0.48\textwidth}
\includegraphics[width=\textwidth]{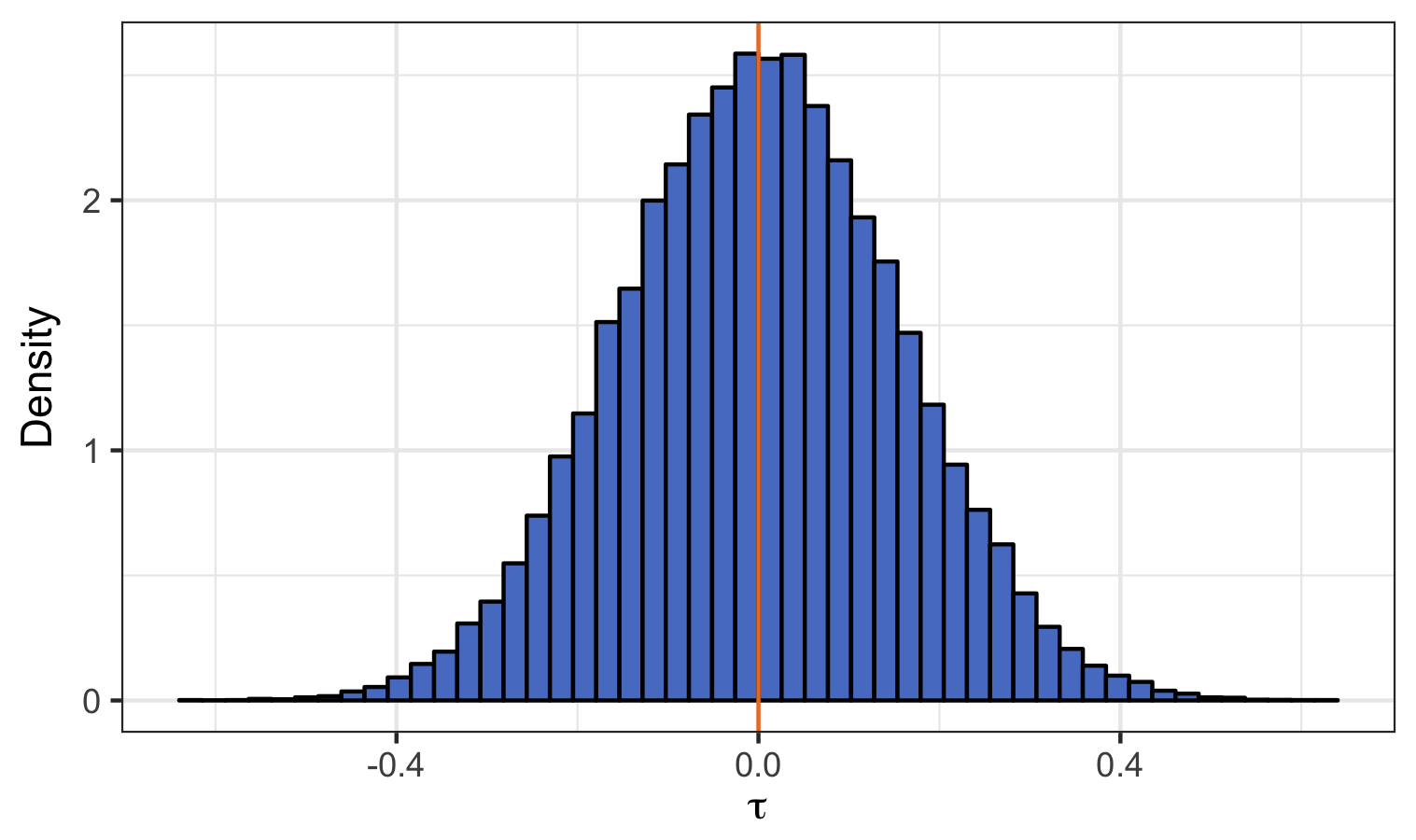}
\end{minipage}
\begin{minipage}[t]{0.48\textwidth}
\includegraphics[width=\textwidth]{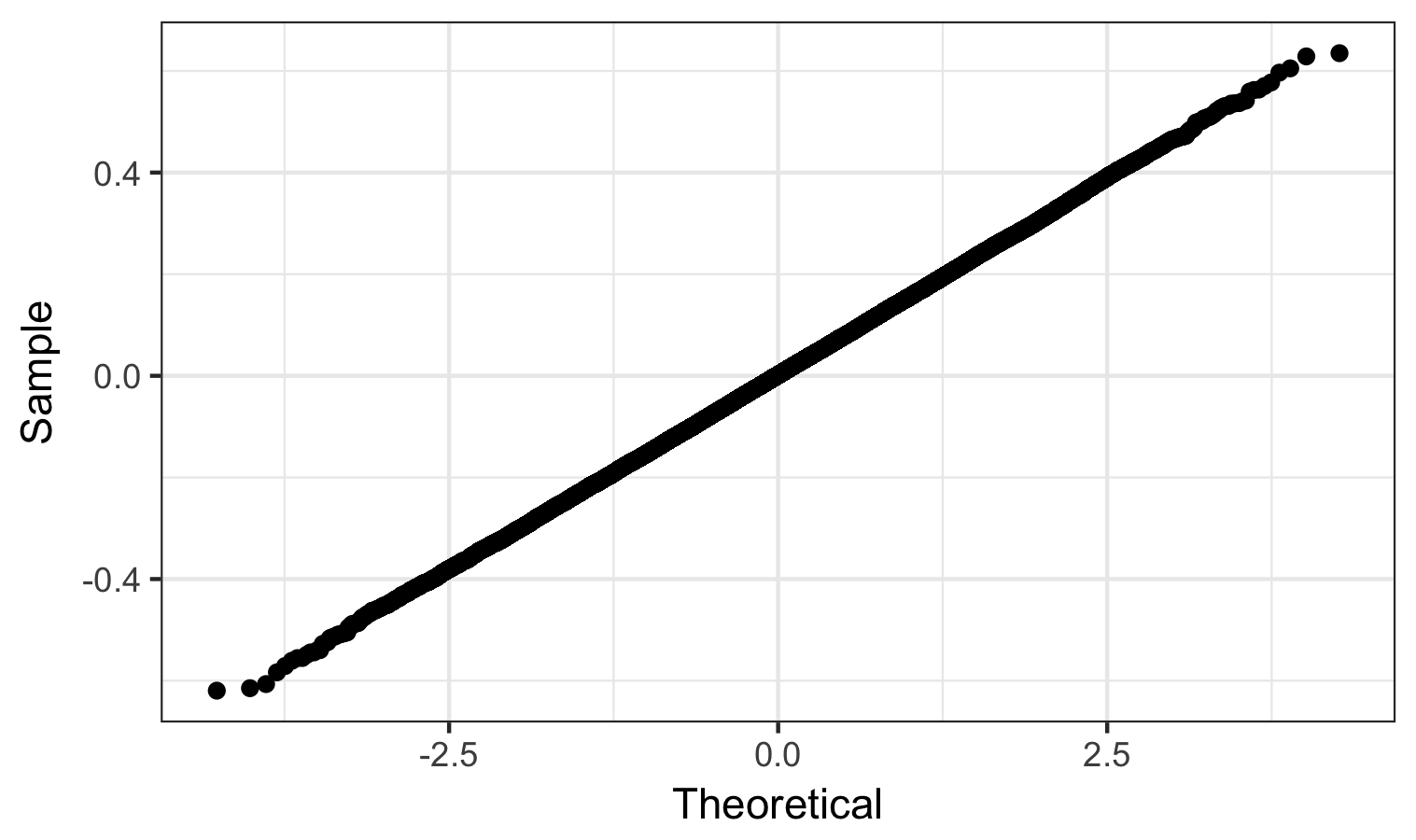}
\end{minipage}
\newline
\caption{The distribution of the pooled estimator of $\bar{\protect\tau}$
over 50,000 simulations, for $n=2$ experiments and $T=100$. Notice how the
variance is smaller than the unpooled version in Figure \protect\ref%
{fig:CLT_dif_est}. }
\label{fig:CLT_pool}
\end{figure}
\newpage

\subsection{Extra figures from the empirical example}

\begin{figure}[h]
\centering
\begin{minipage}[t]{0.23\textwidth}
\includegraphics[width=\textwidth]{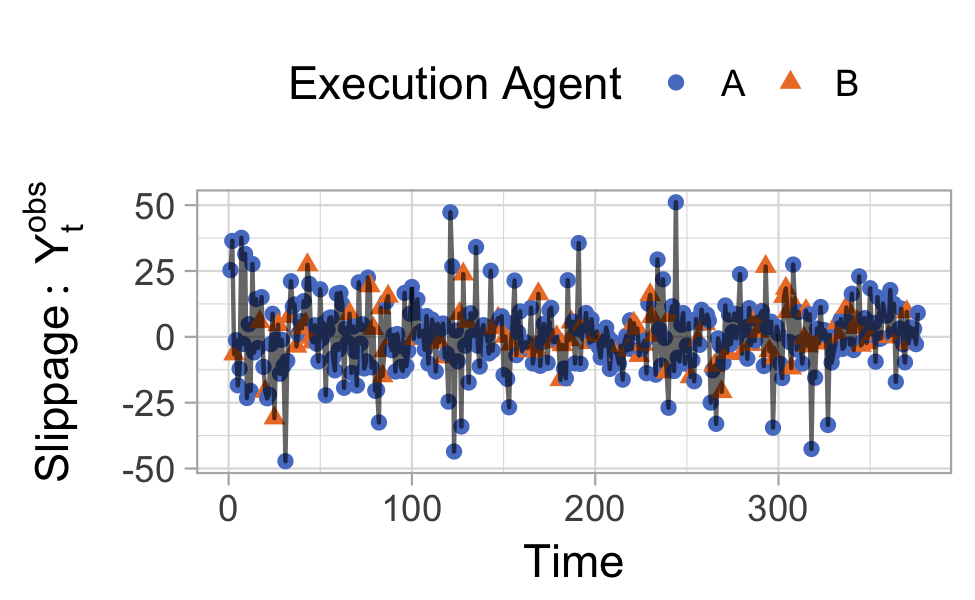}
Market 3
\end{minipage}
\begin{minipage}[t]{0.23\textwidth}
\includegraphics[width=\textwidth]{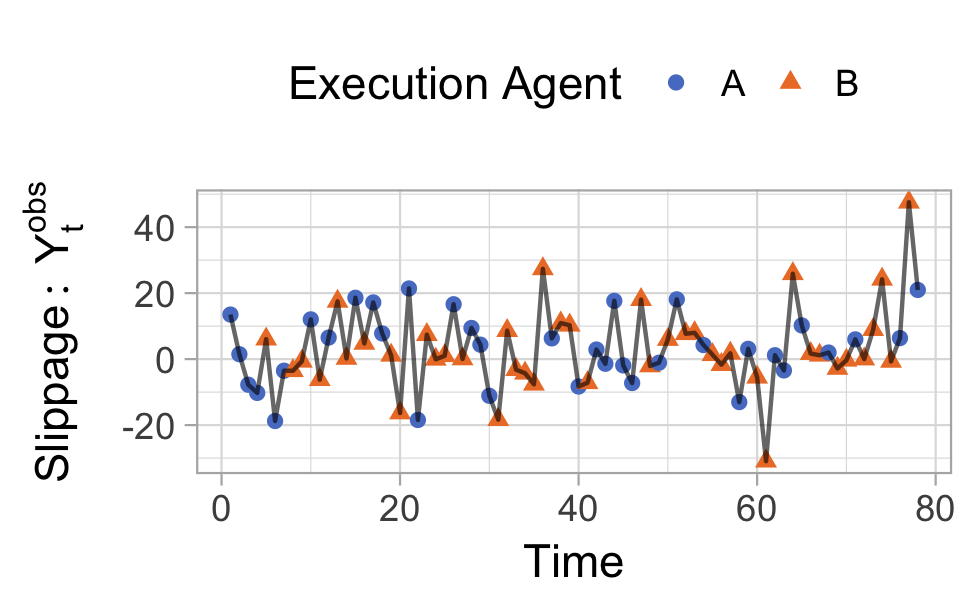}
Market 4
\end{minipage}
\begin{minipage}[t]{0.23\textwidth}
\includegraphics[width=\textwidth]{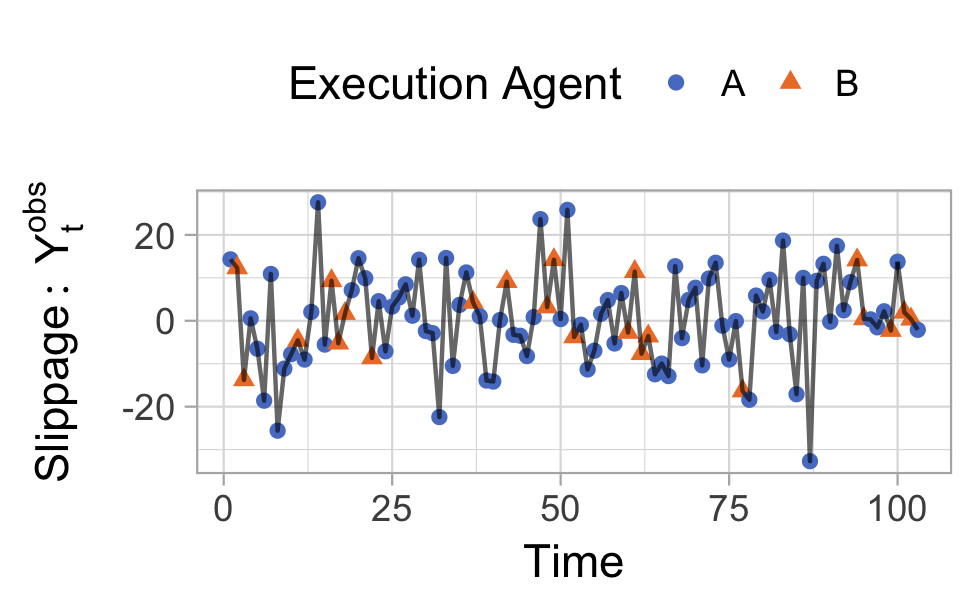}
Market 5
\end{minipage}
\begin{minipage}[t]{0.23\textwidth}
\includegraphics[width=\textwidth]{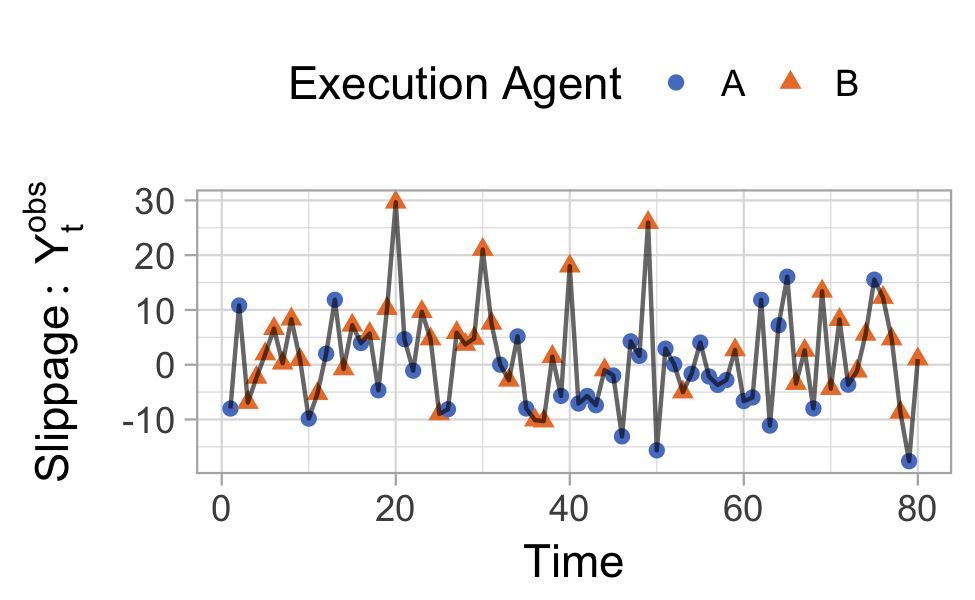}
Market 6
\end{minipage}
\begin{minipage}[t]{0.23\textwidth}
\includegraphics[width=\textwidth]{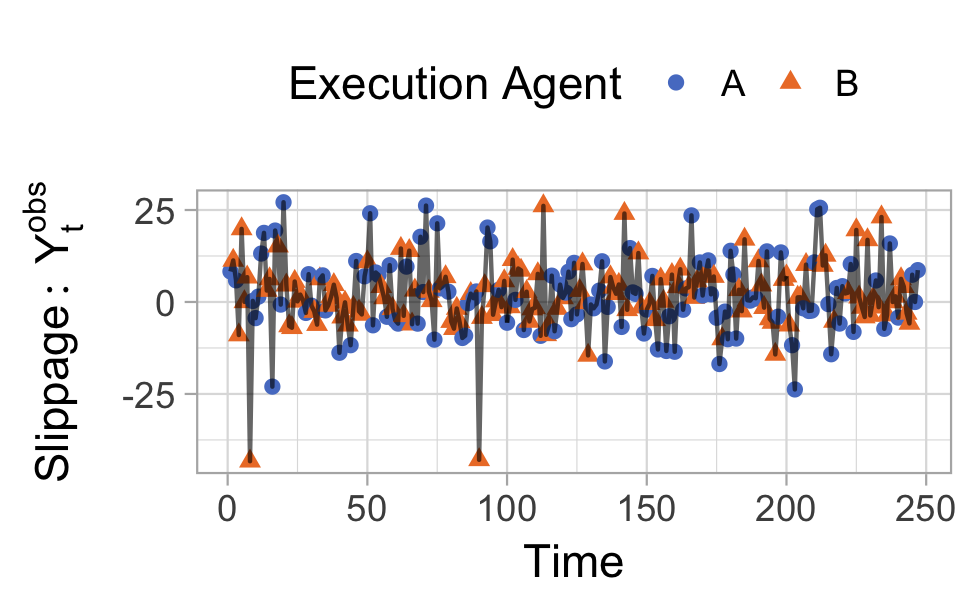}
Market 7
\end{minipage}
\begin{minipage}[t]{0.23\textwidth}
\includegraphics[width=\textwidth]{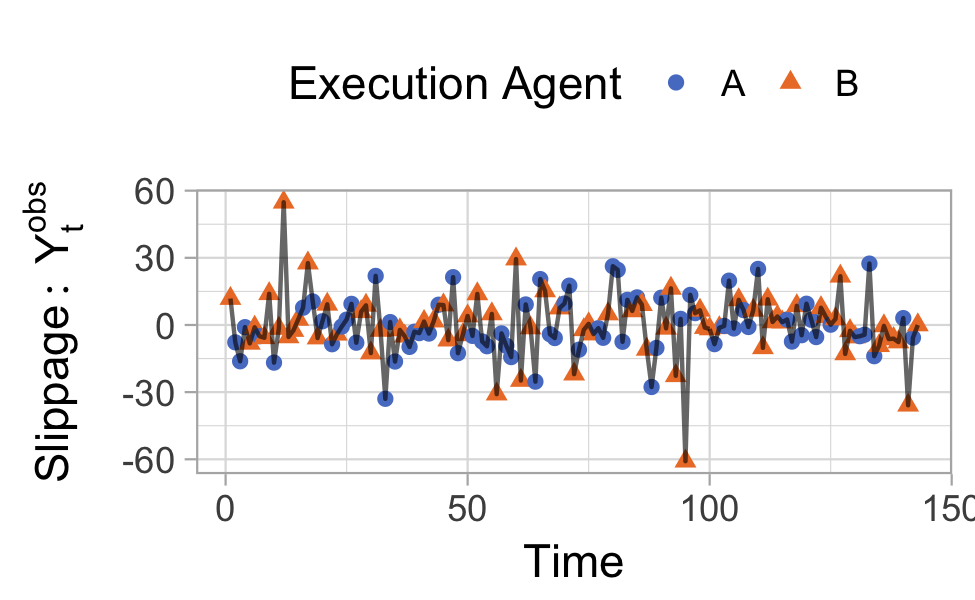}
Market 8
\end{minipage}
\begin{minipage}[t]{0.23\textwidth}
\includegraphics[width=\textwidth]{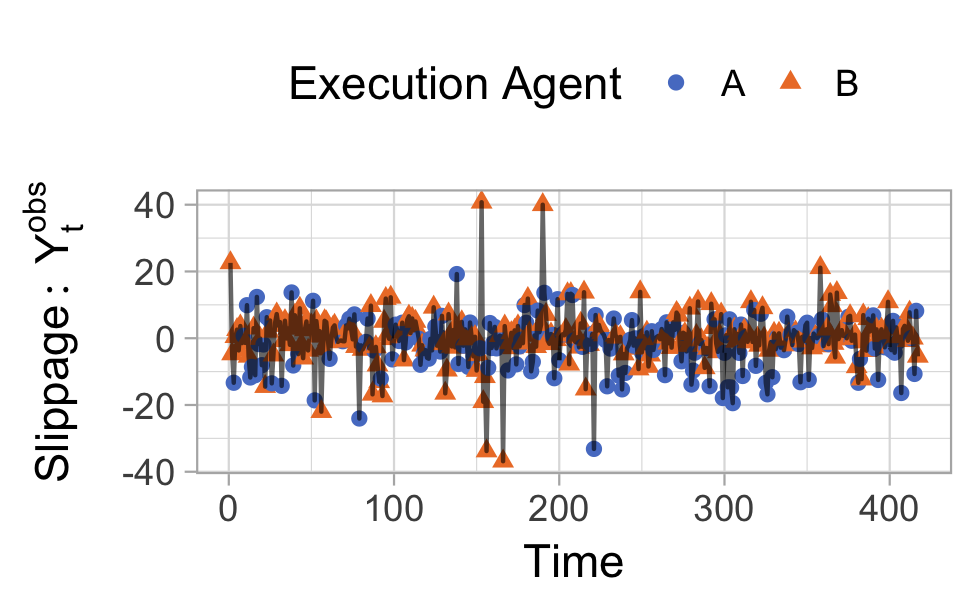}
Market 9
\end{minipage}
\begin{minipage}[t]{0.23\textwidth}
\includegraphics[width=\textwidth]{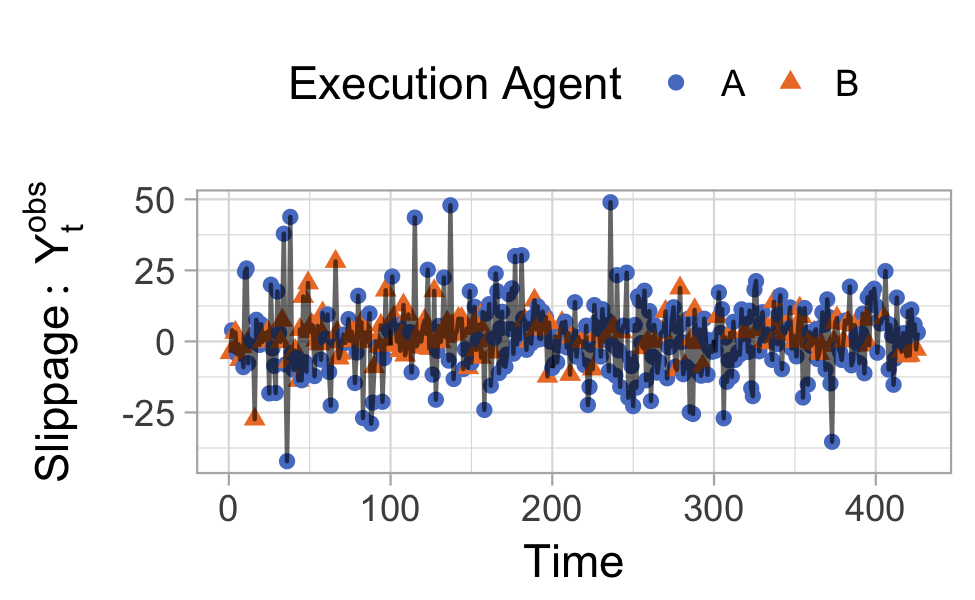}
Market 10
\end{minipage}
\newline
\caption{Slippage $Y_{t}^{\text{obs}}$ as a function of time for eight
markets, red indicates agent A and blue indicated agent B.}
\label{fig:slippage_total_web}
\end{figure}

\begin{figure}[h]
\centering
\begin{minipage}[t]{0.23\textwidth}
\includegraphics[width=\textwidth]{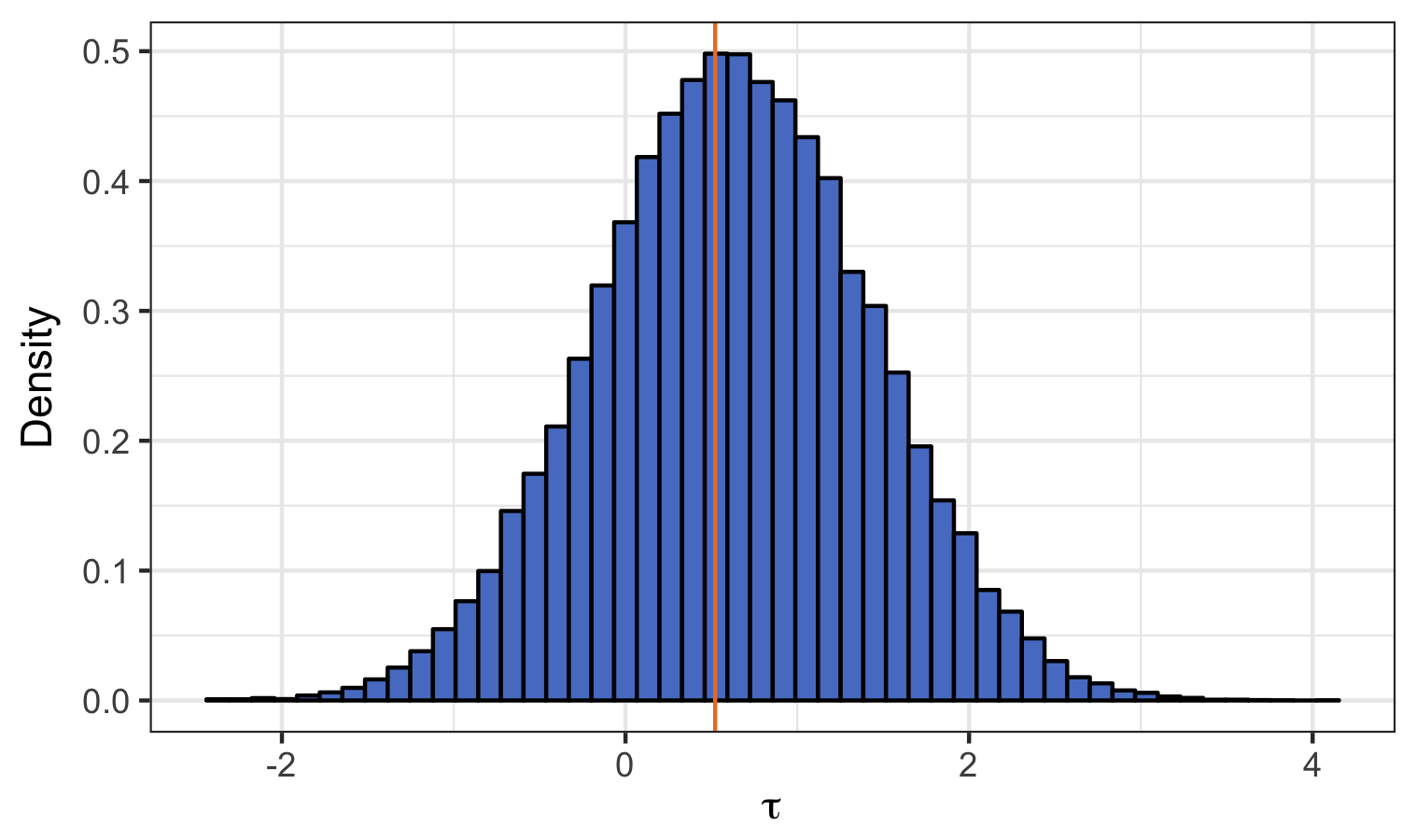}
Market 1
\end{minipage}
\begin{minipage}[t]{0.23\textwidth}
\includegraphics[width=\textwidth]{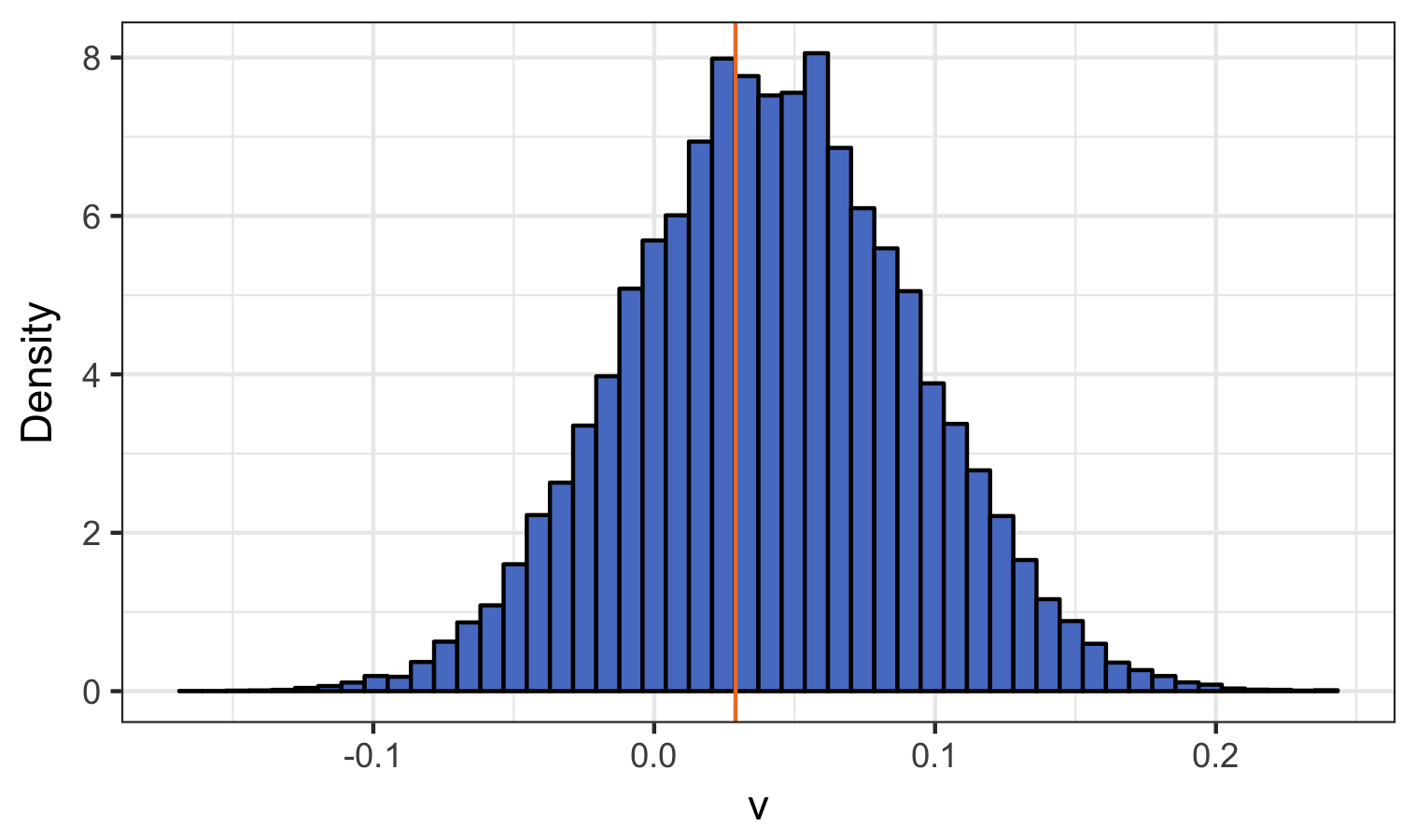}
\end{minipage}
\begin{minipage}[t]{0.23\textwidth}
\includegraphics[width=\textwidth]{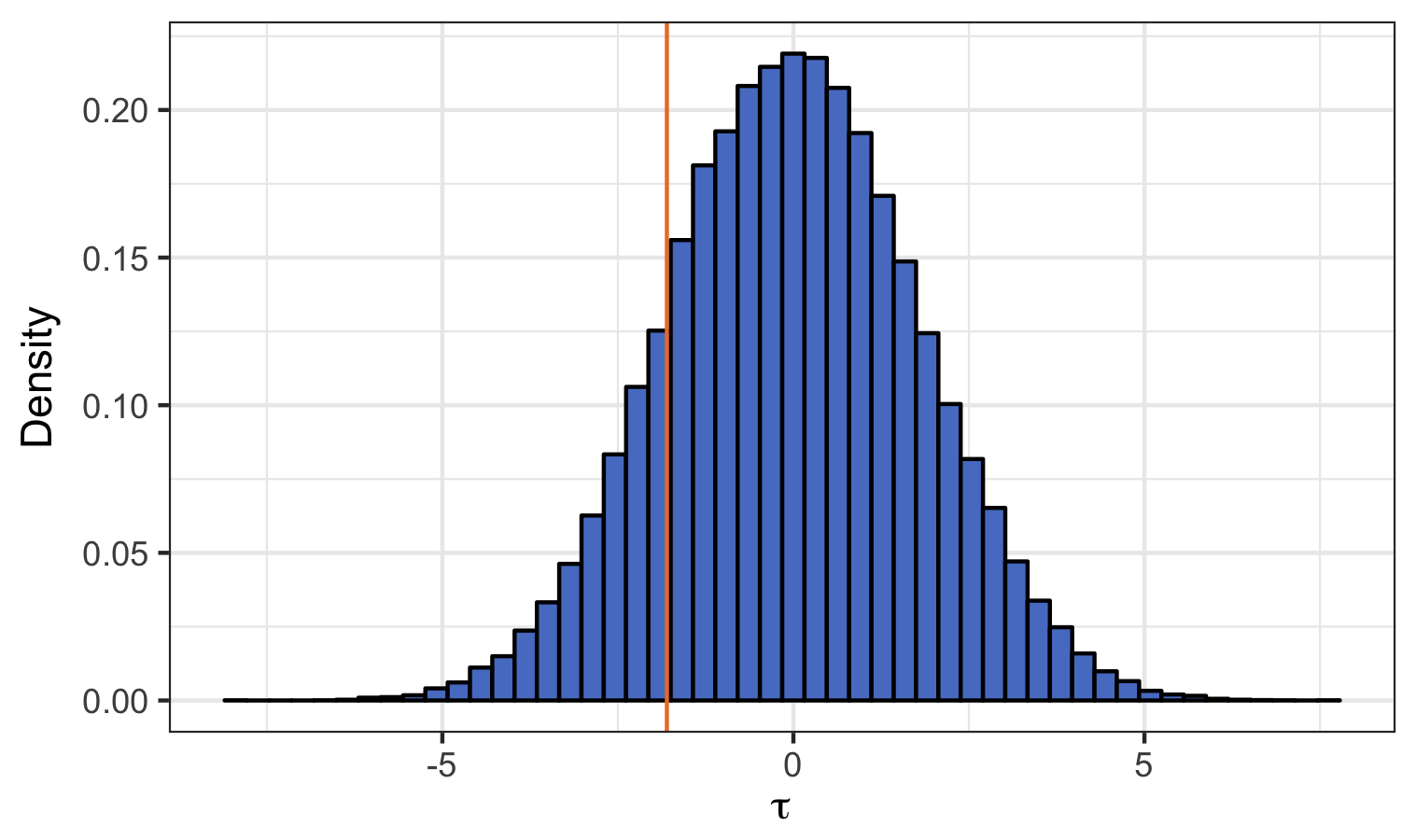}
Market 2
\end{minipage}
\begin{minipage}[t]{0.23\textwidth}
\includegraphics[width=\textwidth]{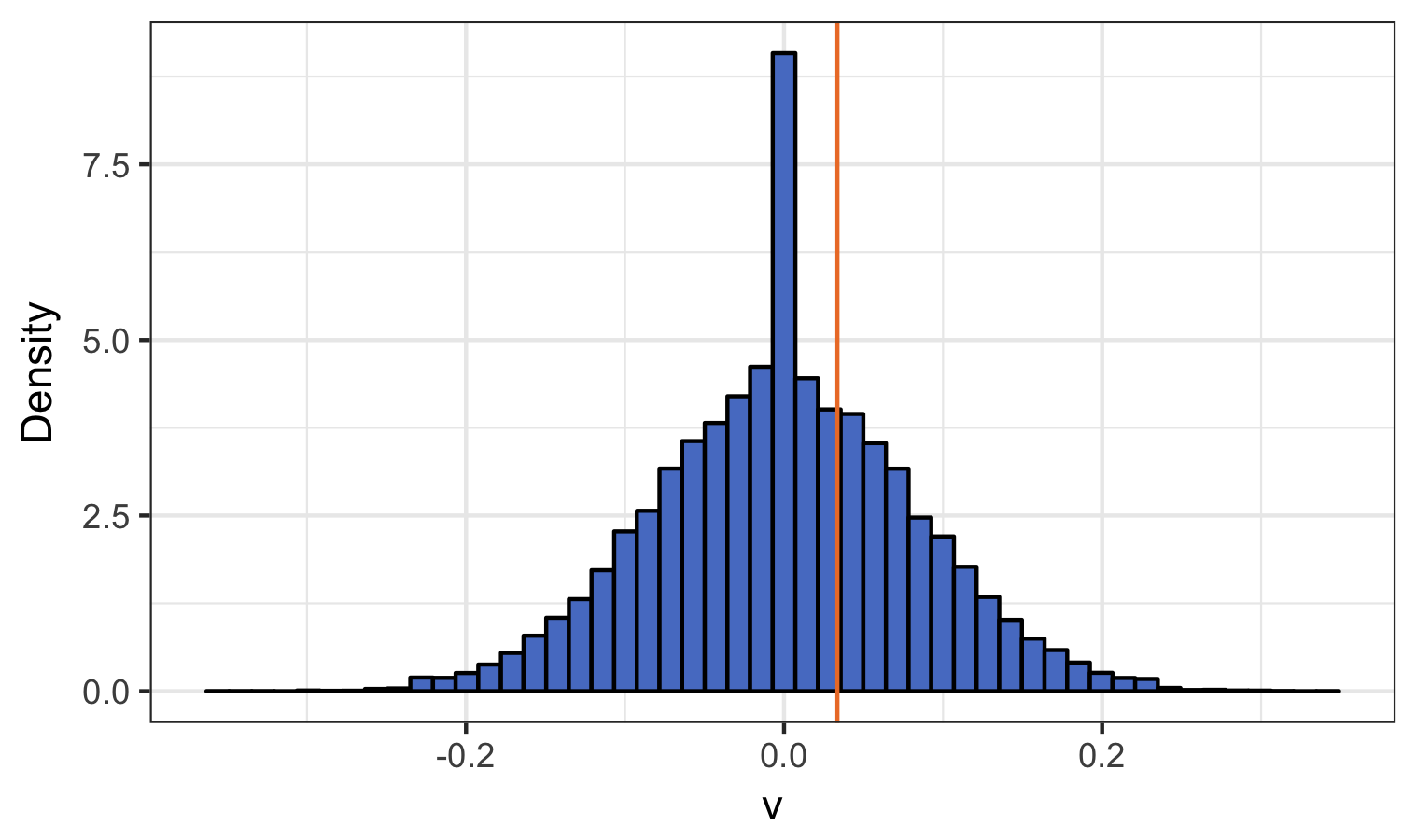}
\end{minipage}
\begin{minipage}[t]{0.23\textwidth}
\includegraphics[width=\textwidth]{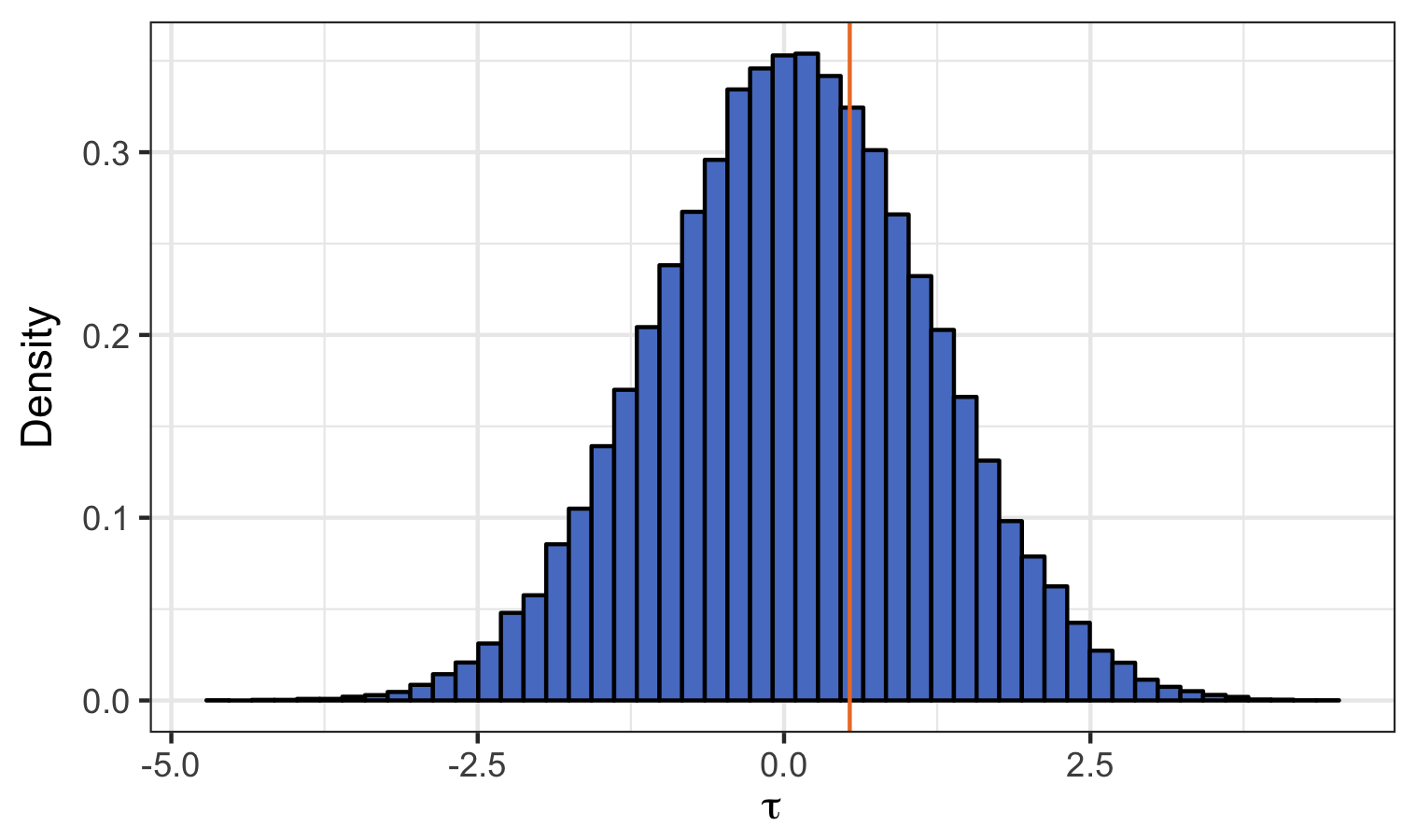}
Market 3
\end{minipage}
\begin{minipage}[t]{0.23\textwidth}
\includegraphics[width=\textwidth]{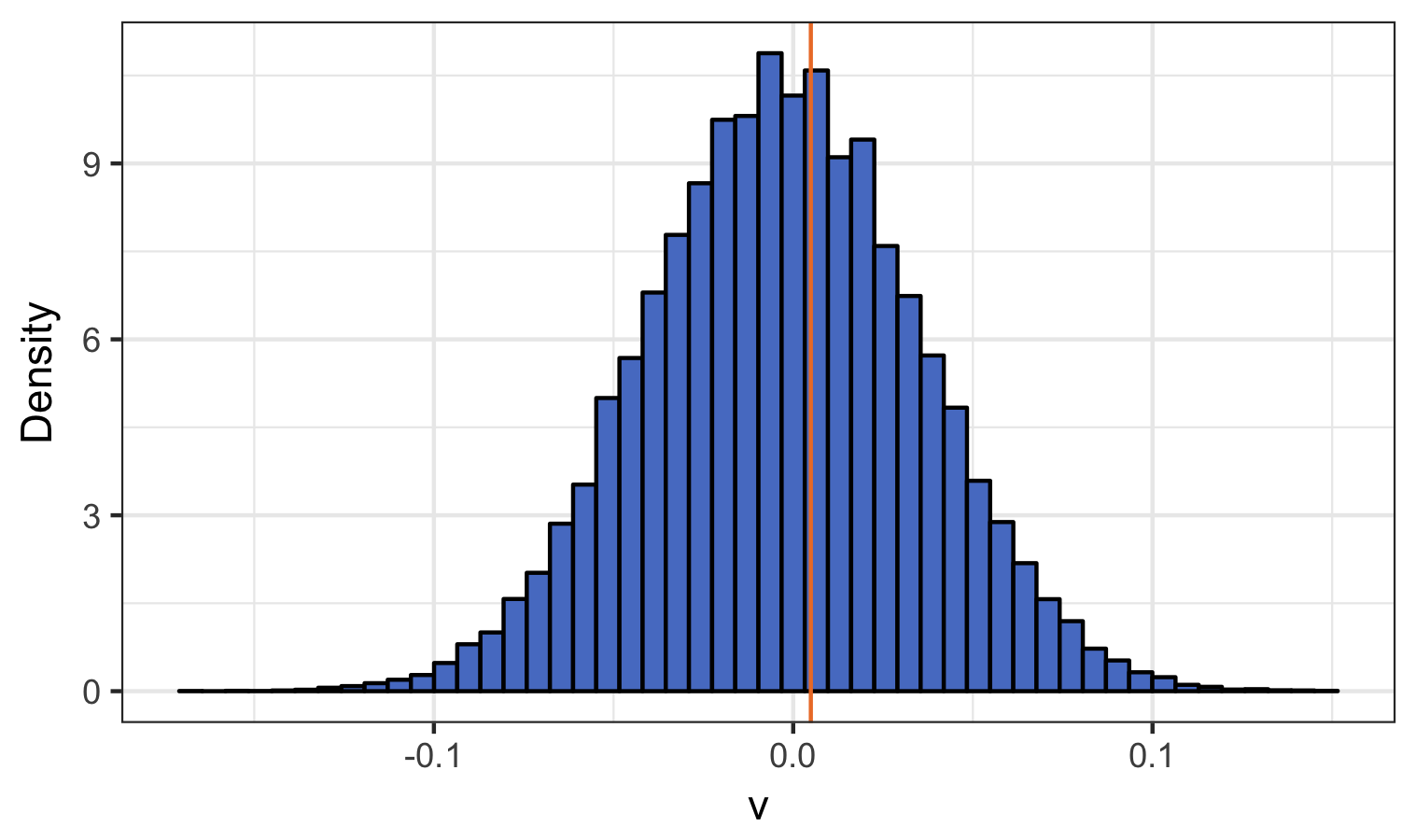}
\end{minipage}
\begin{minipage}[t]{0.23\textwidth}
\includegraphics[width=\textwidth]{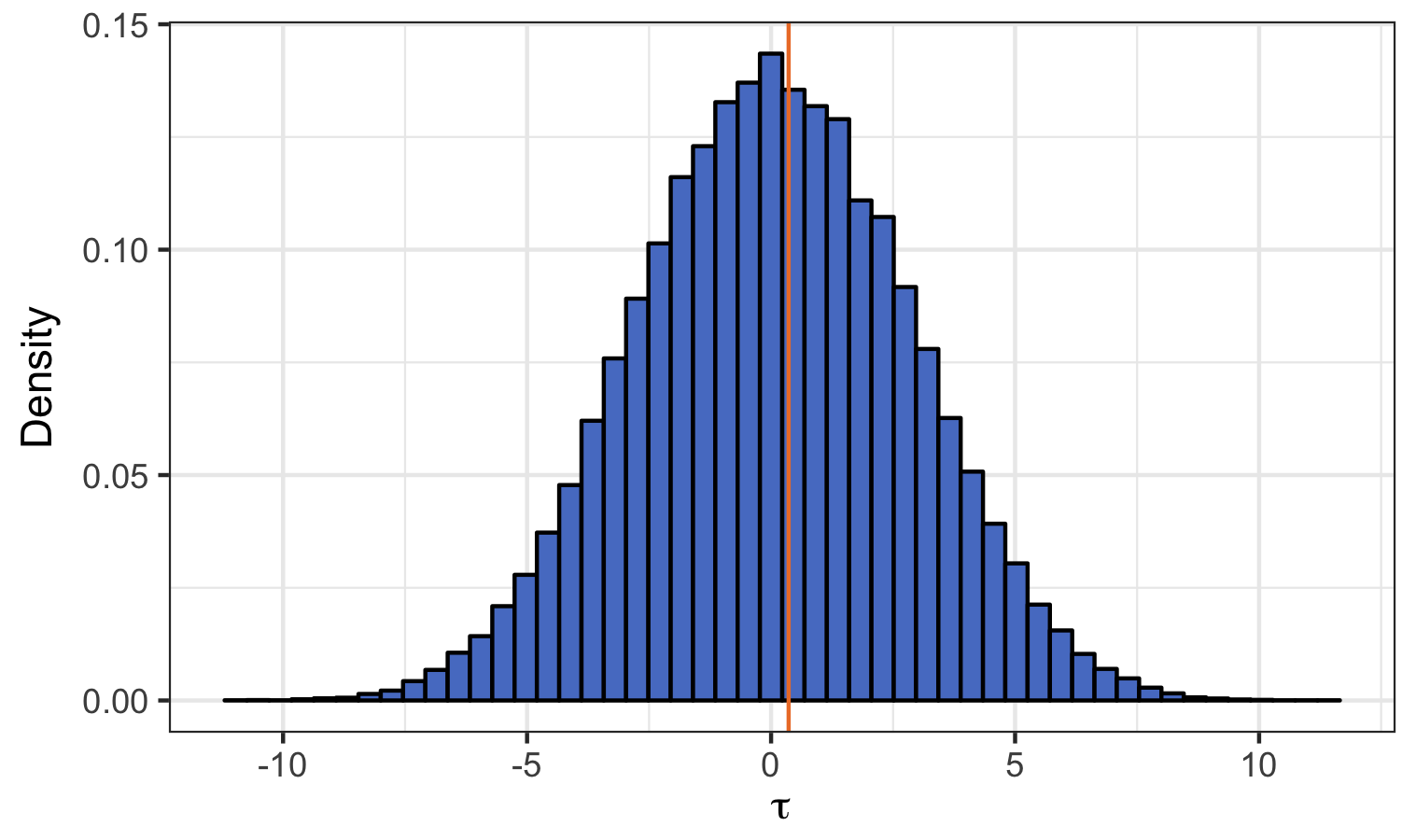}
Market 4
\end{minipage}
\begin{minipage}[t]{0.23\textwidth}
\includegraphics[width=\textwidth]{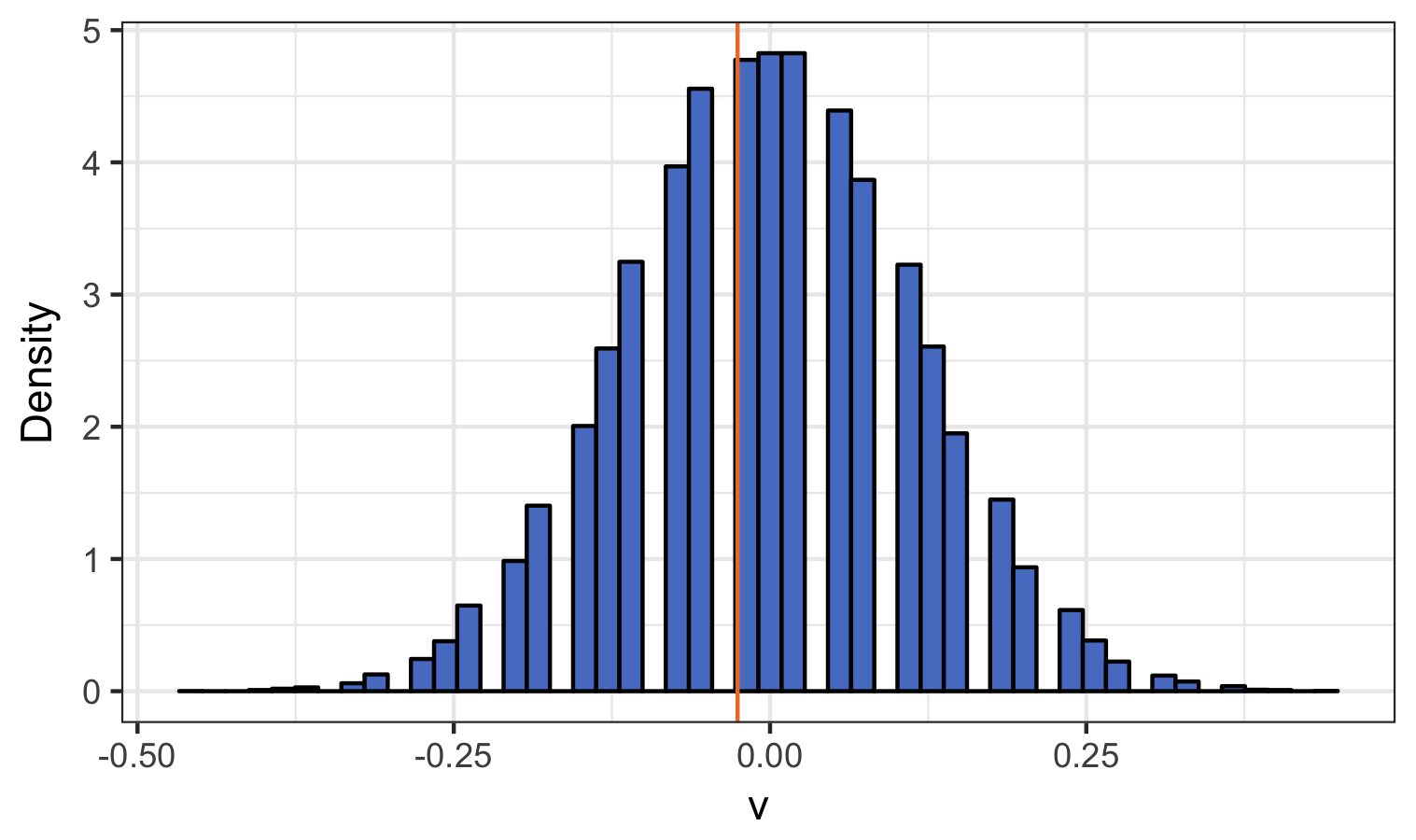}
\end{minipage}
\begin{minipage}[t]{0.23\textwidth}
\includegraphics[width=\textwidth]{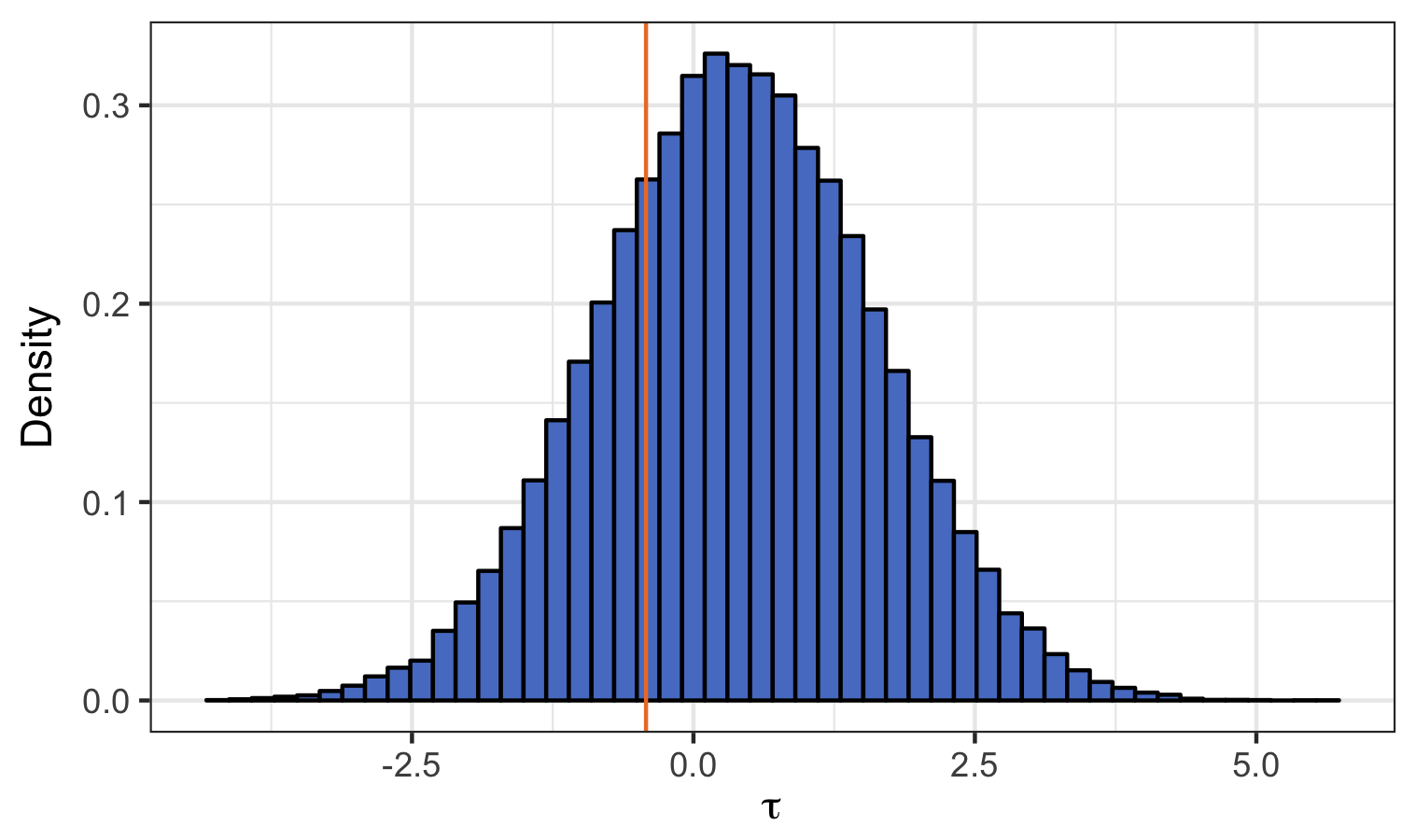}
Market 5
\end{minipage}
\begin{minipage}[t]{0.23\textwidth}
\includegraphics[width=\textwidth]{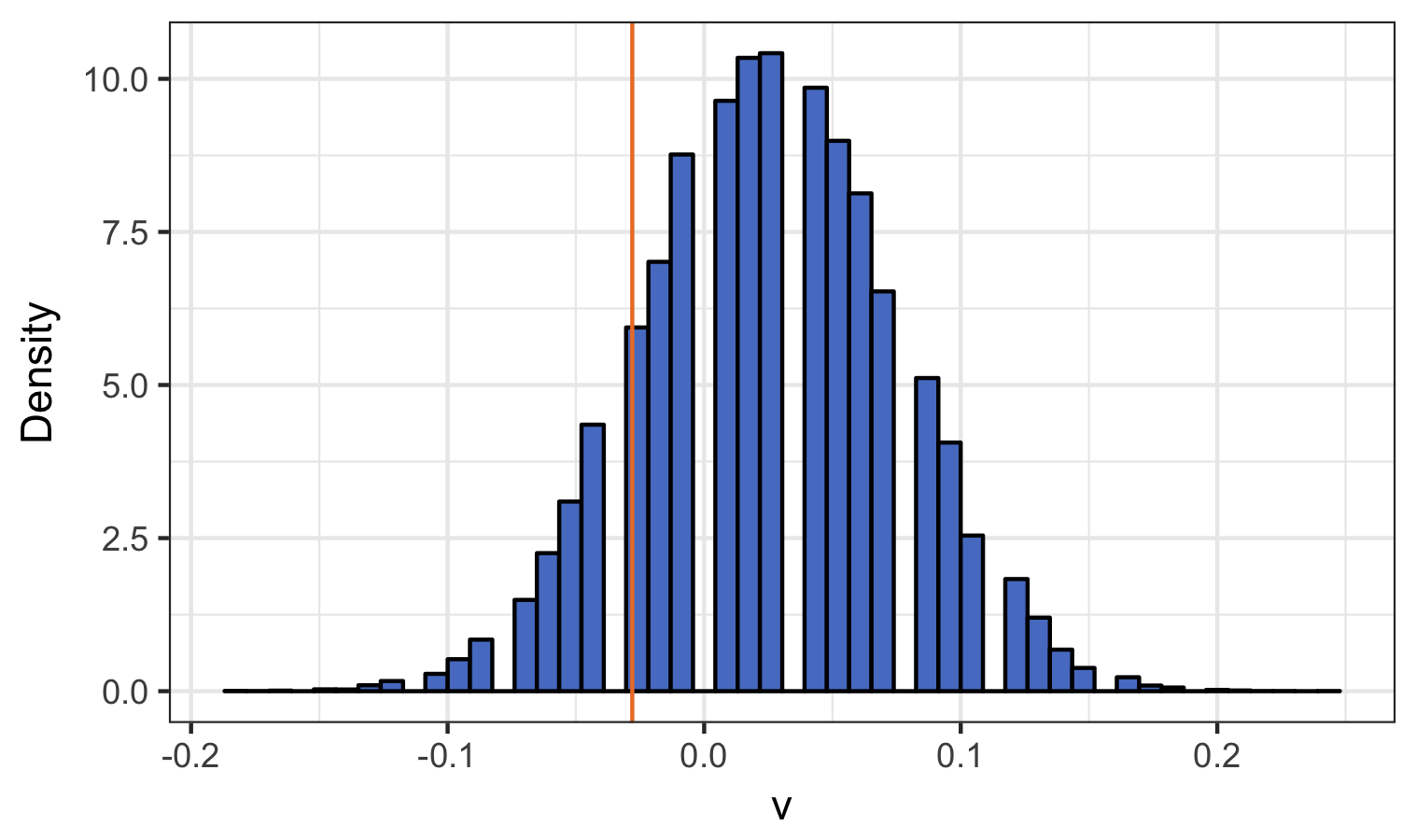}
\end{minipage}
\begin{minipage}[t]{0.23\textwidth}
\includegraphics[width=\textwidth]{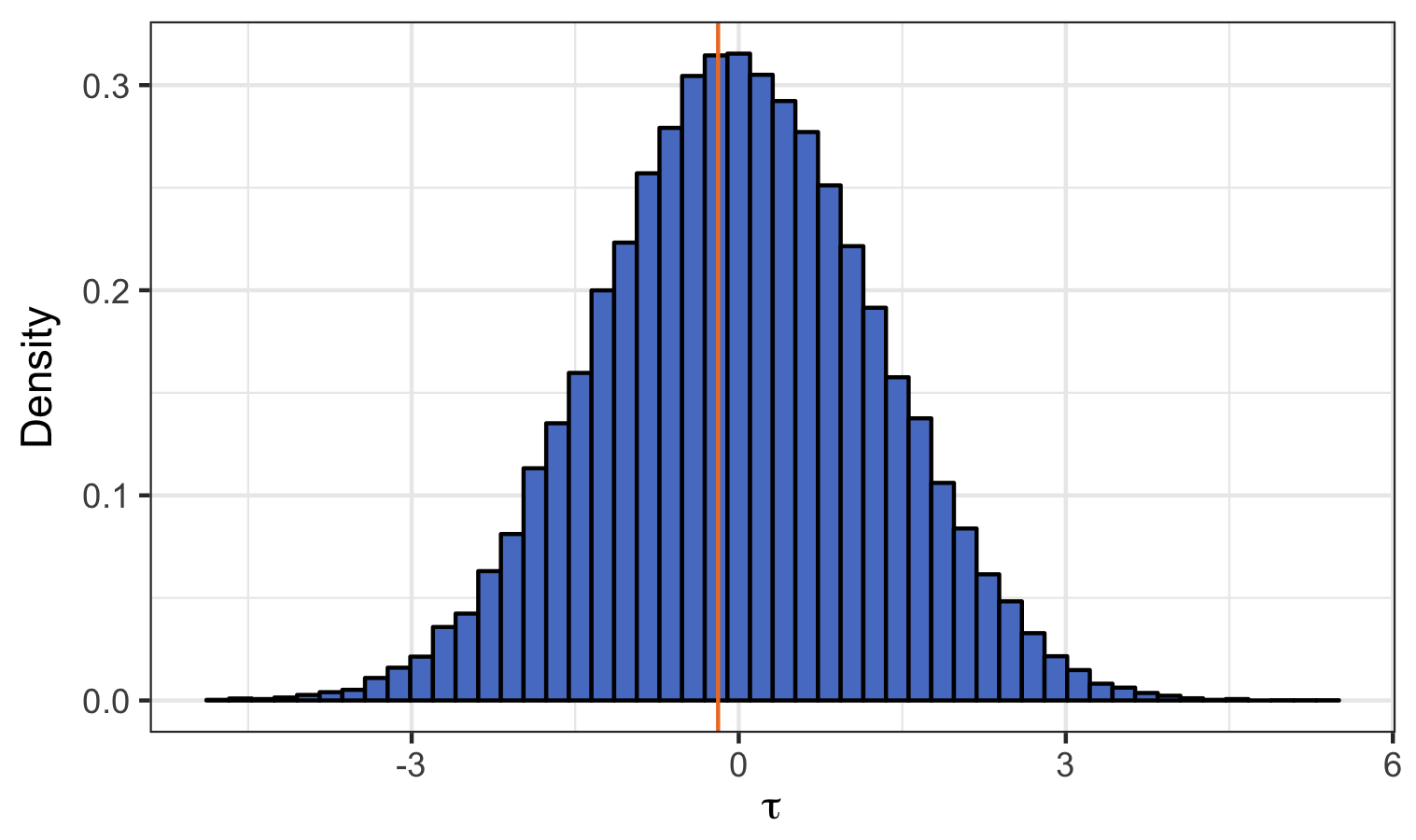}
Market 7
\end{minipage}
\begin{minipage}[t]{0.23\textwidth}
\includegraphics[width=\textwidth]{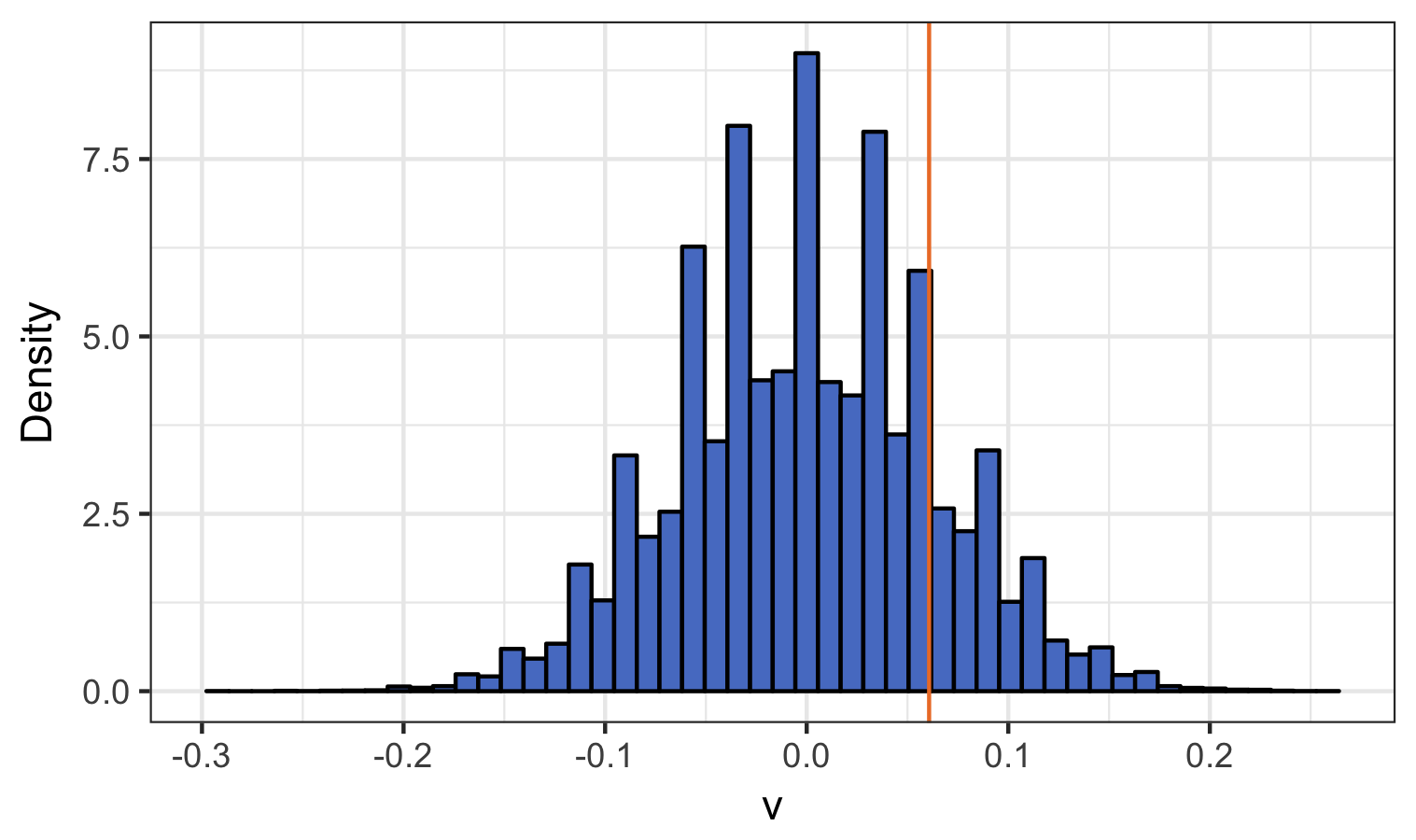}
\end{minipage}
\begin{minipage}[t]{0.23\textwidth}
\includegraphics[width=\textwidth]{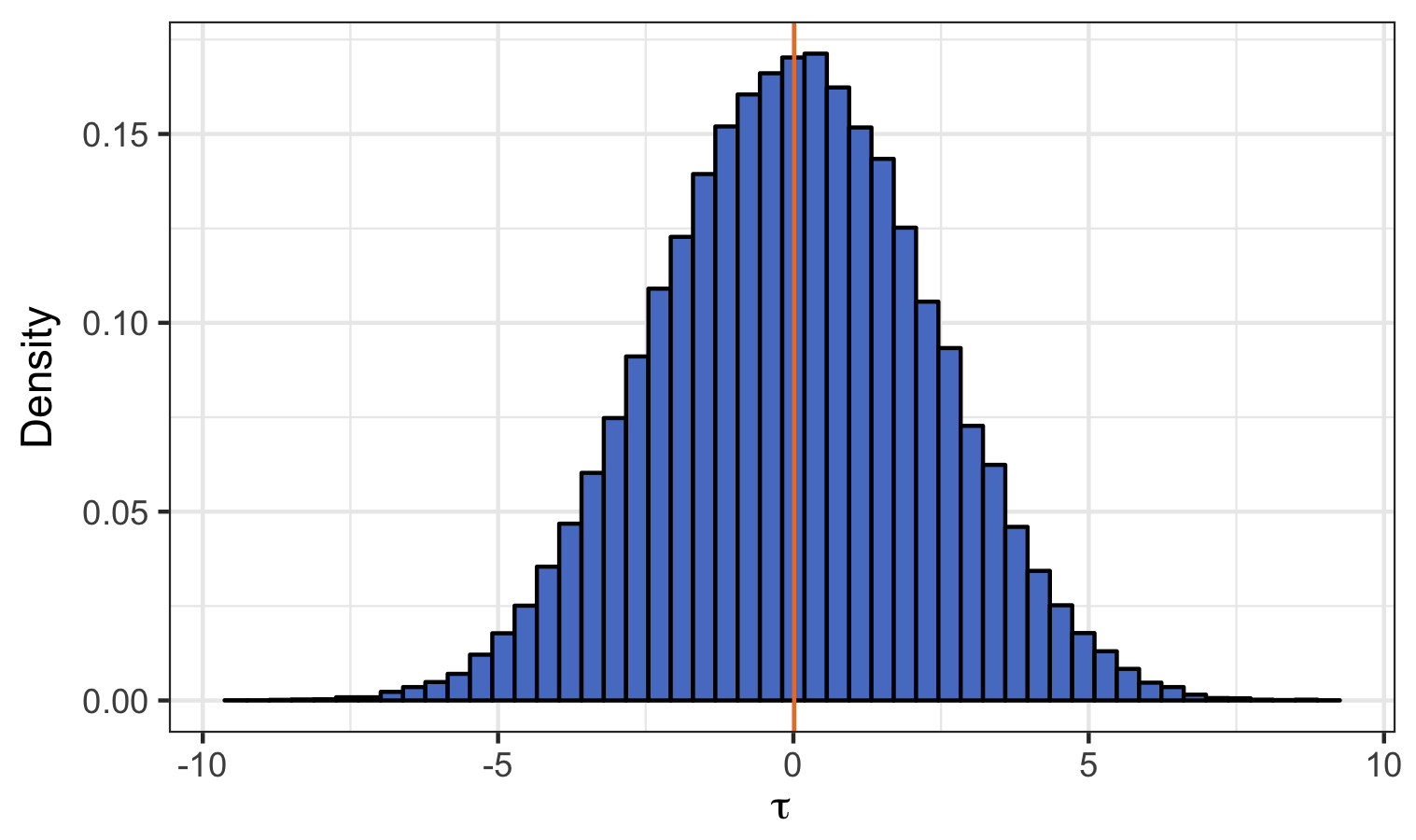}
Market 8
\end{minipage}
\begin{minipage}[t]{0.23\textwidth}
\includegraphics[width=\textwidth]{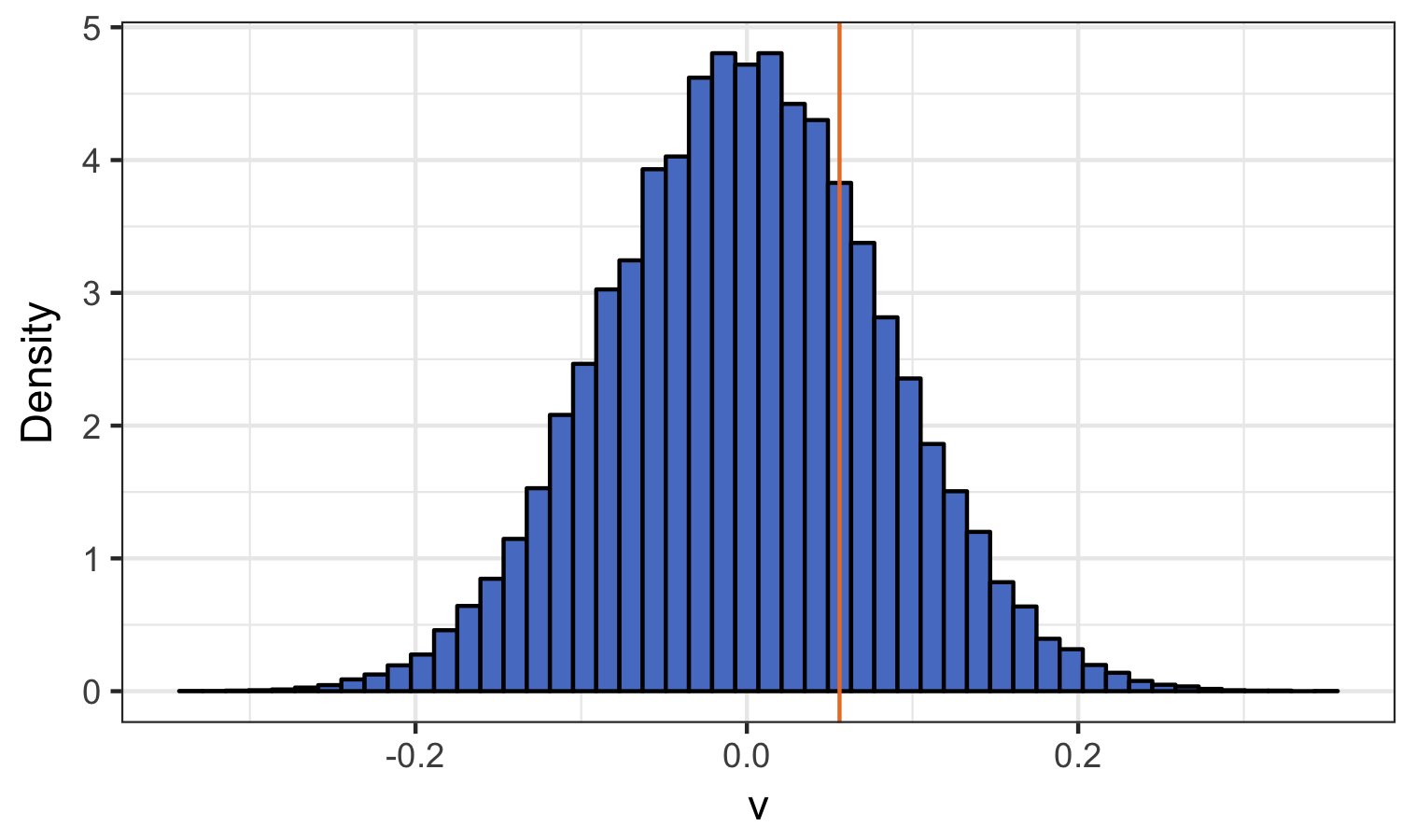}
\end{minipage}
\begin{minipage}[t]{0.23\textwidth}
\includegraphics[width=\textwidth]{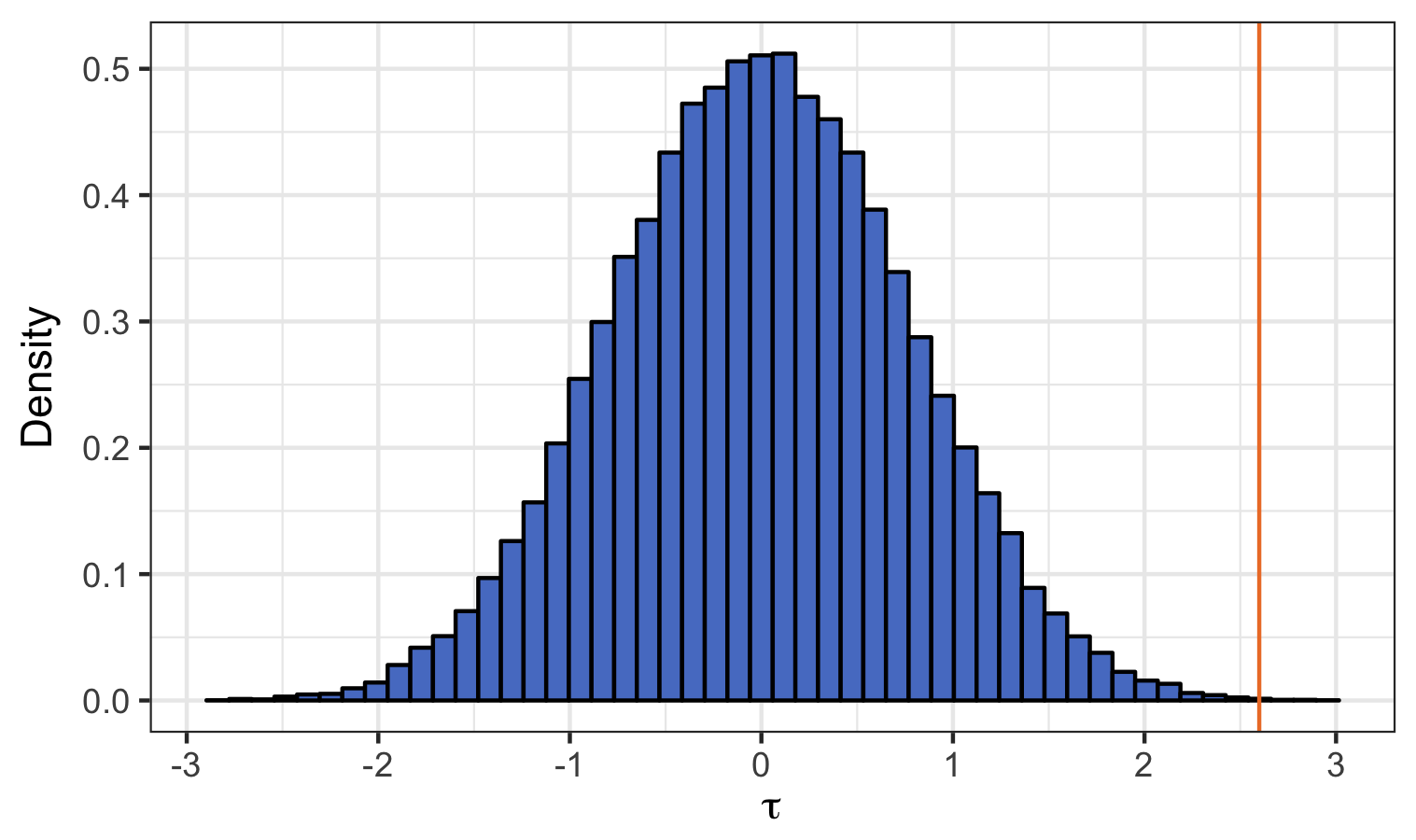}
Market 9
\end{minipage}
\begin{minipage}[t]{0.23\textwidth}
\includegraphics[width=\textwidth]{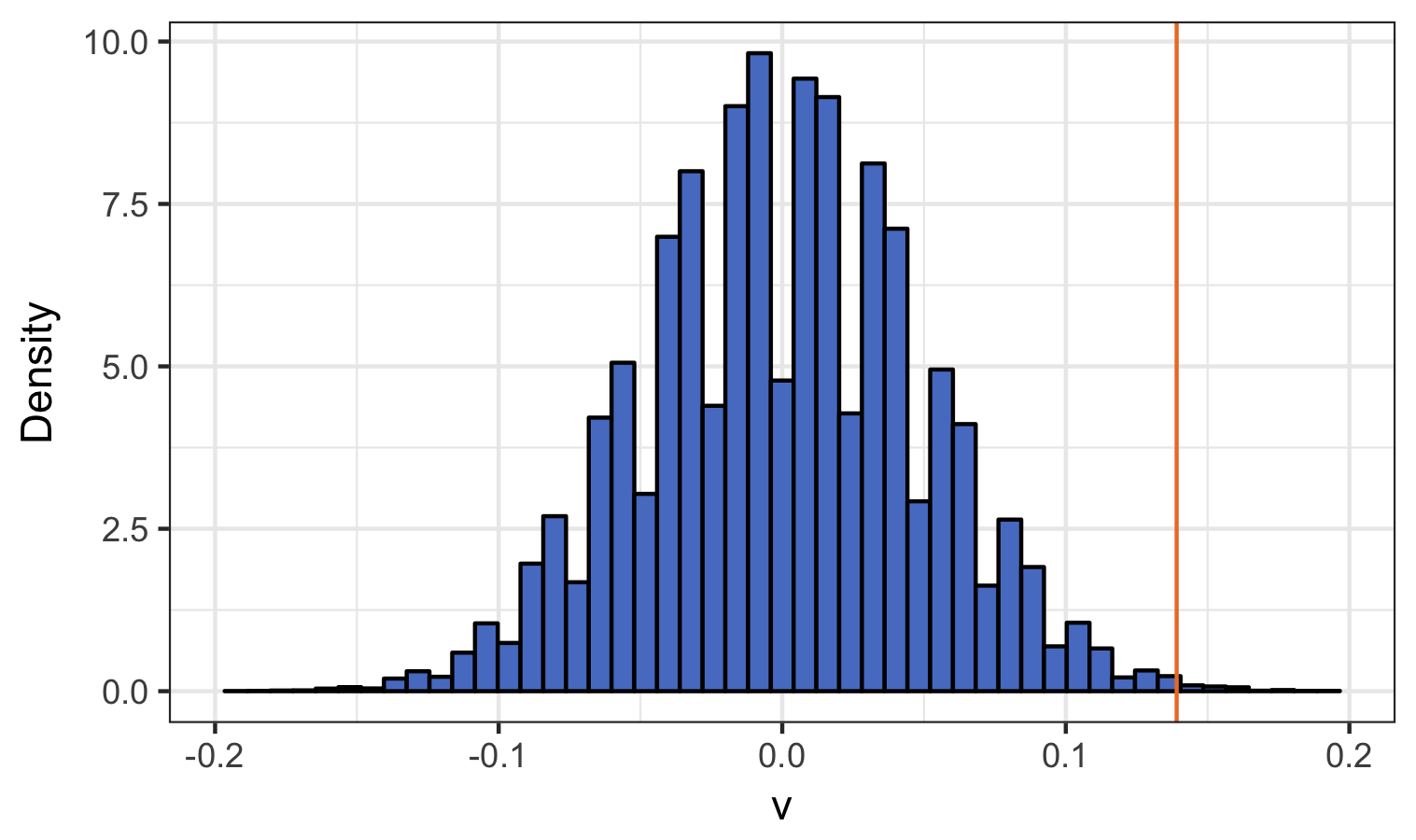}
\end{minipage}  \newline
\caption{Randomization Distribution for Market 2 through 9, left using the $%
\protect\widehat{\bar{\protect\tau}}_{0}$, right using the $\protect\widehat{%
\bar{v}}_{0}$. The orange line indicated the observed value of the
statistic. }
\label{fig:rand_dist1}
\end{figure}

\end{document}